Lidsky V.V. On the two body problem in the classical electrodynamics.

A B S T R A C T

In this paper we propose an approach to the problem of two body motion in classical electrodynamics that takes into account the electromagnetic radiation and the radiation reaction forces. The resulting differential equations are solved numerically.

arXiv:1501.03133 (12-jan-2015)

В.В. Лидский. О задаче двух тел в классической электродинамике.

## А Н Н О Т А Ц И Я

*Предложена классическая теория движения заряженных частиц, учитывающая силы радиационного трения и позволяющая решить задачу о столкновении заряженных частиц, так называемую задачу двух тел в электродинамике.*

1. Если применить к задаче двух тел стандартные уравнения движения релятивистской заряженной частицы в классической электродинамике, то мы столкнемся с парадоксом: кинетическая энергия разлетающихся частиц окажется выше энергии до столкновения. Убедимся в этом на простейшем примере лобового столкновения двух одинаковых частиц. Рассмотрим столкновение в нерелятивистском приближении, а затем учтем релятивистские эффекты, считая их малыми. Как известно, в нерелятивистском приближении сохраняется сумма потенциальной и кинетической энергий частиц, причем в системе центра масс, координата каждой частицы может быть записана как четная функция времени.

Рассмотрим два релятивистских эффекта: первый – запаздывание сигнала, второй – зависимость поля от скорости и ускорения частицы. Величина запаздывания определяется временем, за которое свет проходит расстояние между частицами. Создаваемое движущейся частицей поле задается формулой Лиенара-Вихерта. В одномерном случае единственная отличная от ноля компонента тензора поля имеет вид:

(1.1) $$F^{10} = \frac{e}{\left(\chi^m w_m\right)^3} \cdot \left(\chi^1 w^0 - \chi^0 w^1\right) \cdot \left(1 - \chi \dot{w}_m\right) + \frac{e}{\left(\chi^m w_m\right)^2} \cdot \left(\chi^1 \dot{w}^0 - \chi^0 \dot{w}^1\right),$$

где $\chi^i$ – 4-вектор из «точки излучения» в точку наблюдения, а $w^i, \dot{w}^i$ – 4-скорость и 4-ускорение в «точке излучения» частицы, движущейся к центру масс из $-\infty$. Здесь и далее мы будем считать скорость света равной 1. Поскольку вектор $\chi^i$ светоподобен, в одномерном случае $\chi^0 = \chi^1$ и (1.1) можно упростить:

$F^{10} = e \cdot (\chi^0)^{-2} (w^0 - w^1)^{-2}$. Релятивистское уравнение движения $m \dfrac{du^1}{ds} = eF^{10} u_0$ в трехмерных обозначениях принимает вид:

$$(1.2) \qquad m \frac{\dot{v}}{(1-v^2)^{\frac{3}{2}}} = e^2 \cdot \frac{(1 - w^2(t'))}{(\chi^0)^2 \cdot (1 - w(t'))^2}$$

Здесь $v, \dot{v}$ – скорость и ускорение частицы, движущейся из $+\infty$, $t' = t - \delta t$ – «момент излучения». Ясно, что

$$(1.3) \qquad \delta t = 2x + \delta x,$$

где $x + \delta x$ – координата «точки излучения». Считая скорости малыми, сравнительно со скоростью света, вычислим величины $\delta x, dt$, оставляя члены третьего порядка малости:

$$(1.4) \qquad -\delta x = -w(t)\delta t + \dot{w}(t) \cdot \frac{\delta t^2}{2} - \ddot{w}(t) \cdot \frac{\delta t^3}{6}$$

Из (1.3) и (1.4) находим

$$(1.5) \qquad \delta t = 2x \left( \frac{1 + \dot{w} \cdot (x + wx) - \ddot{w} \cdot \frac{4}{3} x^2}{1 - w + \dot{w} \cdot (2x + wx) - \ddot{w} \cdot 2x^2} \right).$$

Учитывая, что $w(t') = w - \dot{w} \cdot \delta t + \ddot{w} \cdot \delta t^2 / 2$,

из (1.2) находим:

$$(1.6) \qquad m\dot{v} = \frac{e^2}{(2x)^2} \cdot \left( 1 - \frac{5}{2} v^2 - 2 \cdot \dot{v} \cdot (x - vx) - \ddot{v} \cdot \frac{8}{3} x^2 \right),$$

где мы воспользовались тем, что из-за симметрии задачи $w = -v$.

Умножим (1.6) на $v$ и проинтегрируем по времени от $-\infty$ до $+\infty$:

$$\int_{-\infty}^{+\infty} m\dot{v}v \, dt = \int_{-\infty}^{+\infty} \frac{e^2}{(2x)^2} v \, dt + \int_{-\infty}^{+\infty} \frac{e^2}{(2x)^2} \cdot \left( -\frac{5}{2} v^2 - 2 \cdot \dot{v} \cdot (x - vx) - \ddot{v} \cdot \frac{8}{3} x^2 \right) v \, dt.$$

Интеграл в левой части есть изменение кинетической энергии частицы при столкновении. Первый интеграл в правой части – изменение потенциальной энергии, исчезающей при бесконечном удалении частиц. При вычислении второго интеграла в правой части учтем, что в решении нерелятивистского уравнения скорость $v(t)$ оказывается нечетной функцией времени, а $x(t), \dot{v}(t)$ – четные. Интегрируя по частям последнее слагаемое, находим:

$$(1.7) \qquad \Delta\left(\frac{mv^2}{2}\right) = e^2 \cdot \int\limits_{-\infty}^{+\infty} \cdot \left(\frac{\dot{v} \cdot v^2}{2x} + \dot{v}^2 \cdot \frac{2}{3}\right) dt$$

Откуда виден парадокс – кинетическая энергия разлетающихся частиц больше, чем была до столкновения. Компьютерные расчеты показывают, что увеличение кинетической энергии происходит и при нецентральном столкновении.

Физическая причина парадокса понятна – используемые уравнения движения не учитывают излучения неравномерно движущейся частицей электромагнитных волн и потерь энергии на излучение. В уравнения движения необходимо добавить члены, выражающие радиационные потери энергии. Получить лоренц-инвариантные выражения для дополнительных членов можно только следующим образом: построив некоторую фиксированную поверхность $\Sigma$, составить уравнение баланса энергии, приравняв изменение 4-вектора энергии-импульса внутри $\Sigma$ к потоку тензора энергии-импульса через $\Sigma$. При этом, алгоритм построения $\Sigma$ должен быть таким, чтобы в системе отсчета всякого наблюдателя строилась та же самая поверхность $\Sigma$.

Необходимо объяснить безуспешность попыток, исключив поверхность $\Sigma$, получить лоренц-инвариантные уравнения движения частицы, учитывающие реакцию излучения. Причина этого в том, что потери энергии частицей не являются лоренц-инвариантной величиной (см. [1]). Инвариантным является поток тензора энергии импульса через некоторую замкнутую поверхность, охватывающую частицу, только в том случае, если в системе отсчета произвольного наблюдателя вычисляется поток именно через ту самую поверхность. Интеграл потока излучаемой энергии по полному телесному углу по существу является интегралом потока тензора энергии-импульса через сферу достаточно большого радиуса, причем частица находится в центре сферы. Тот же интеграл, вычисленный другим наблюдателем, окажется потоком тензора энергии-импульса через *другую поверхность*, поскольку сфера в его системе отсчета деформируется из-за лоренцева сокращения. Таким образом, единственный способ получить инвариантное выражение для потери энергии, и, следовательно, для «силы» радиационного трения – это выбрать фиксированную поверхность $\Sigma$, указав инвариантную процедуру ее построения, в составить уравнение баланса энергии-импульса, приравняв изменение энергии внутри $\Sigma$ потоку тензора энергии-импульса через $\Sigma$.

Расчет по указанной схеме выполнен нами ранее в [2], где предложена следующая процедура построения поверхности $\Sigma$. Для каждой точки пространства $x^i$ ищем «момент излучения» – точку пересечения светового конуса абсолютного прошлого точки $x^i$ с мировой линией частицы $z^i(\tau)$. Далее в системе отсчета наблюдателя, для которого частица в момент $\tau$ неподвижна, строится сфера фиксированного радиуса $\rho = \dfrac{2e^2}{3m}$. Принимается, что $x^i$ лежит внутри «чулка» $\Sigma$, если она оказалась внутри сферы. Уравнение поверхности $\Sigma$ может быть записано в инвариантной форме

(1.8) $$(x^i - z^i(\tau)) \cdot u_i(\tau) = \rho.$$

Из требования баланса энергии-импульса на $\Sigma$ получено уравнение движения частицы (см. [2], eq.(46)):

(1.9) $$m a^i = e \cdot F^{ik} w_k + \frac{2e^3}{3m} \cdot F^{ik} a_k + \frac{2e^2}{3} \cdot w^i \cdot a^n a_n.$$

При выводе (1.9) предполагалось, что внешнее поле много меньше, чем собственное поле электрона на расстоянии $\rho$, а также, что внешнее поле однородно на масштабах порядка $\rho$:

(1.10) $$\left|F^{ik}\right| \ll \frac{e^2}{\rho^2},\ \left|\partial_l F^{ik}\right| \ll \frac{\left|F^{ik}\right|}{\rho}$$

Уравнение (1.9) отличается от известного уравнения Лоренца-Абрагама-Дирака (LAD) [3] вторым слагаемым в правой части. Дальнейшее исследование уравнения (1.9) показало, что оно не допускает решений, нарушающих закон сохранения энергии [2]. Подчеркнем, что при построении «чулка» $\Sigma$ не предполагается никаких физических особенностей на поверхности $\Sigma$. Частица предполагается точечной. И поля, и частицы могут беспрепятственно проникать внутрь $\Sigma$. Величина $\rho$, которую мы будем называть электромагнитным радиусом частицы, в данном рассуждении играет чисто инструментальную роль, однако, как будет показано в п.5., в задаче двух тел эта величина важна, так как при сближении частиц на расстояния $\Box\, 2 \cdot \rho$ качественно меняется характер движения.

Для решения задачи о столкновении двух частиц сближающихся на расстояния сравнимые с $\rho$, мы, очевидно, не можем пользоваться (1.9) поскольку оба условия (1.10) не выполняются. В данной работе мы выполним расчет по программе, предложенной в [2], с той разницей, что не будем делать никаких упрощающих предположений, а точно вычислим поток энергии импульса поля, создаваемого одной частицей через поверхность $\Sigma$ другой. Затем, приравняв полученные величины изменению энергии-импульса частицы $mu^i$, получим уравнения движения, учитывающие реакцию излучения. Эти уравнения мы проинтегрируем численно и обсудим результаты.

2. Будем обозначать частицы буквами **u** и **w**. Соответственно их 4-скорости $u^i$ и $w^i$, а мировые линии — $z^i(\tau)$ и $y^i(\sigma)$. Построим поверхность $\Sigma$ вокруг мировой линии частицы **u**. Выберем на $z^i(\tau)$ две близкие точки $\tau, \bar{\tau}$. Построим для каждой из них световой конус абсолютного будущего. Эти конусы вырезают на 3-поверхности $\Sigma$ 3-цилиндр, который мы обозначим $\delta\Sigma$. Мы получим уравнение движения частицы приравняв поток тензора энергии импульса поля через поверхность $\delta\Sigma$ разности импульсов частицы в моменты $\tau, \bar{\tau}$:

(2.1) $$P^i(\bar{\tau}) - P^i(\tau) = -\int_{\delta\Sigma} T^{ik} dV_k$$

где $dV_k = e_{klmn} dx^l dx^m dx^n$ — элемент интегрирования по 3-поверхности, направленный вдоль внешней нормали к $\delta\Sigma$.

Выражение для нормали к $\delta\Sigma$ получим дифференцированием (1.8):

(2.2) $$\left( \left( \delta_k^i - u^i(\tau) \cdot \partial_k \tau \right) \cdot u_i(\tau) + (x^i - z^i(\tau)) \cdot \dot{u}_i(\tau) \cdot \partial_k \tau \right) dx^k = 0,$$

где $\dot{u}_k = \partial_\tau u_k$. Производные $\partial_k \tau$ получим, дифференцируя по $\partial_k$ тождество $\left( x^i - z^i(\tau) \right)\left( x_i - z_i(\tau) \right) = 0$. Учитывая (1.8), находим: $\partial_k \tau = \dfrac{(x_k - z_k(\tau))}{\rho}$. Для нормали к поверхности $\delta\Sigma$ получаем выражение:

(2.3) $$n_k = \frac{1}{\sqrt{1 - 2\dot{u}_m \cdot \left(x^m - z^m(\tau)\right)}} \cdot \left( u_k + \left( -1 + \dot{u}_m \cdot \left(x^m - z^m(\tau)\right) \right) \cdot \frac{(x_k - z_k(\tau))}{\rho} \right)$$

Нормировочный множитель мы выбрали из условия $n_k n^k = -1$.

Пусть световой конус абсолютного будущего точки $z^i(\tau)$ вырезает на «чулке» $\Sigma$ 2-сферу **S**. Перейдем в (2.1) от интегрирования по 3-объему $dV_k$ к интегрированию по поверхности **S**. Выберем на **S** некоторую точку $x^i$ и построим ортонормированный базис $\xi^i, \eta^i, \zeta^i$ в 3-пространстве, касательном к $\Sigma$ в точке $x^i$.

Все три вектора $\xi^i, \eta^i, \zeta^i$ ортогональны нормали $n_k$. Из (2.3) видно, что вектор $n_k$ лежит в 2-плоскости **U**, образованной векторами $u_k$ и $\chi_k = x_k - z_k(\tau)$. Выберем базисный вектор $\xi^i$ также лежащим в плоскости **U**, а векторы $\eta^i, \zeta^i$ ортогональными **U**. Легко видеть, что

$$(2.4) \qquad \xi^i = \frac{1}{\sqrt{1 - 2 \cdot \dot{u}_l \chi^l}} \cdot \left( u^i - \dot{u}_l \chi^l \cdot \frac{\chi^i}{\rho} \right)$$

Выберем в качестве элемента интегрирования в (2.1) параллелепипед со сторонами, параллельными векторам $\xi^i, \eta^i, \zeta^i$. Вычислим длину *l* ребра параллельного вектору $\xi^i$, заключенного между поверхностью **S** и поверхностью $\overline{S}$, образованной пересечением светового конуса точки $z^i(\overline{\tau})$ с поверхностью $\Sigma$. Из условия $\left( x^i + l\xi^i - z^i(\overline{\tau}) \right)\left( x_i + l\xi_i - z_i(\overline{\tau}) \right) = 0$, полагая $z^i(\overline{\tau}) = z^i(\tau) + u^i \cdot \delta\tau$, находим

$$(2.5) \qquad l = \delta\tau \cdot \sqrt{1 - 2 \cdot \dot{u}_l \chi^l}$$

Откуда видно, что интеграл (2.1) можно переписать в виде:

$$(2.6) \qquad \delta P^i = -\delta\tau \cdot \int_S T^{ik} \left( u_k + \left(-1 + \dot{u}_m \cdot \chi^m \right) \cdot \frac{\chi_k}{\rho} \right) dS.$$

3. Теперь преобразуем тензор энергии-импульса, для которого мы примем формулу Максвелла-Хевисайда (см. [4], §33):

$$(3.1) \qquad T^{ik} = \frac{1}{4\pi} \cdot \left( -F^{il} F^k{}_l + \frac{1}{4} g^{ik} F_{lm} F^{lm} \right)$$

здесь $F^{ik}$ — 4-тензор электромагнитного поля, $g^{ik}$ — метрический тензор.

Поскольку поле на сфере **S** создается двумя частицами, в (3.1) необходимо подставить сумму $F^{ik} + G^{ik}$, где $G^{ik}, F^{ik}$ – поля частиц **u** и **w**, соответственно. Результат распадается на три слагаемых

$$(3.2) \qquad T^{ik} = T^{ik}_{FF} + T^{ik}_{FG} + T^{ik}_{GG},$$

содержащих соответствующие произведения. Поле $G^{ik}$ определяется запаздывающими потенциалами Лиенара-Вихерта:

$$(3.3) \qquad G^{ik} = \frac{e}{\left(\chi^m u_m\right)^3} \cdot \left( \chi^i u^k - \chi^k u^i \right) \cdot \left(1 - \chi^m \dot{u}_m \right) + \frac{e}{\left(\chi^m u_m\right)^2} \cdot \left( \chi^i \dot{u}^k - \chi^k \dot{u}^i \right),$$

причем «точкой излучения» для всей сферы **S** оказывается точка $z^i(\tau)$. Ввиду $\chi^i u_i = \rho$ на поверхности **S** выражение (3.3) принимает особенно простой вид. Подынтегральное выражение в (2.6) получим, свернув тензоры (3.2) с выражением в круглых скобках, которое мы обозначим $\psi_l$:

(3.4) $$\psi_l = u_l + \left(-1 + \chi^m \dot{u}_m\right) \cdot \frac{\chi_l}{\rho}$$

Опуская промежуточные преобразования, выпишем результаты[a]:

(3.5) $$T_{FF}^{ik}\psi_k = \frac{1}{4\pi} \cdot \left(-F^{il}F^k{}_l + \frac{1}{4}g^{ik}F_{lm}F^{lm}\right) \cdot \left(u_k + \left(-1 + \chi^m \dot{u}_m\right) \cdot \frac{\chi_k}{\rho}\right)$$

(3.6) $$T_{FG}^{ik}\psi_k = -\frac{e_u}{4\pi\rho} \cdot F^{il}\left(\frac{u_l}{\rho} \cdot (1 - \chi^m \dot{u}_m) + \dot{u}_l\right)$$
$$+ \frac{e_u}{4\pi\rho^2} \cdot F_{lm}\left(\chi^i \cdot \dot{u}^l u^m + u^i \cdot \chi^l \dot{u}^m + \dot{u}^i \cdot u^l \chi^m\right)$$

(3.7) $$T_{GG}^{ik}\psi_k = \frac{e_u^2}{8\pi\rho^4} \cdot u^i \cdot (1 - 2\chi^m \dot{u}_m) - \frac{e_u^2}{4\pi\rho^3} \cdot \left(\chi^i \cdot \dot{u}^k \dot{u}_k - \dot{u}^i\right)$$
$$- \frac{1}{8\pi} \frac{e_u^2}{\rho^5} \cdot \chi^i \cdot \left(1 - 3\chi^m \dot{u}_m + 2\chi^m \dot{u}_m \chi^n \dot{u}_n\right)$$

4. Вычислим интеграл по сфере **S** от выражения $T_{GG}^{ik}\psi_k$ (см. (3.7)), не зависящего от параметров мировой линии второй частицы. Поместим начало координат и начало отсчета времени в 4-центр сферы **S**. Выберем полярную ось вдоль направления скорости частицы и введем полярную систему координат $t, R, \theta, \varphi$. Тогда для компонент $\chi^i$ мы получим:

(4.1) $$\chi^i = \{t, R \cdot \sin\vartheta \cdot \cos\varphi, R \cdot \sin\vartheta \cdot \sin\varphi, R \cdot \cos\vartheta\}$$

Уравнения 2-поверхности **S** (1.8) и $\chi_i \chi^i = 0$ принимают вид:

(4.2) $$t = R; \quad t - v \cdot R\cos\vartheta = \rho \cdot \sqrt{1 - v^2}$$

Элемент интегрирования по сфере принимает вид:

(4.3) $$d\mathbf{S} = R^2 \cdot \sin\vartheta \cdot d\vartheta \cdot d\varphi = \frac{\rho^2 \cdot (1 - v^2)}{(1 - v\cos\vartheta)^2} \cdot \sin\vartheta \cdot d\vartheta \cdot d\varphi$$

Вычислим интегралы по поверхности **S**[b]:

(4.4) $$\oiint_S d\mathbf{S} = \int_0^{2\pi} d\varphi \int_0^{\pi} \frac{\rho^2(1-v^2)}{(1-v\cdot\cos\vartheta)^2} \sin\vartheta\, d\vartheta = 4\pi\rho^2$$

$$(4.5) \quad \oiint_S \chi^i \cdot d\mathbf{S} = 4\pi\rho^3 \cdot u^i$$

$$(4.6) \quad \oiint_S \chi^i \chi^k \cdot d\mathbf{S} = \frac{4\pi\rho^4}{3} \cdot \left(4u^i u^k - g^{ik}\right)$$

$$(4.7) \quad \oiint_S \chi^i \chi^k \chi^l \cdot d\mathbf{S} = 8\pi\rho^5 \cdot u^i u^k u^l - \frac{4\pi\rho^5}{3} \cdot \left(g^{ik} \cdot u^l + g^{il} \cdot u^k + g^{kl} \cdot u^i\right)$$

$$(4.8) \quad \oiint_S \chi^i \chi^k \chi^l \chi^m \cdot d\mathbf{S} = \frac{64\pi}{5}\rho^6 \cdot u^i u^k u^l u^m + \frac{4\pi}{15}\rho^6 \cdot \left(g^{ik}g^{lm} + g^{il}g^{km} + g^{im}g^{kl}\right)$$
$$- \frac{8\pi}{5}\rho^6 \cdot \left(g^{ik} \cdot u^l u^m + g^{il} \cdot u^k u^m + g^{kl} \cdot u^i u^m + g^{im} \cdot u^k u^l + g^{km} \cdot u^i u^l + g^{lm} \cdot u^i u^k\right)$$

С помощью формул (4.4) - (4.7) вычисляем интеграл выражения (3.7):

$$(4.9) \quad \oiint_S T_{GG}^{ik} \psi_k dS = \frac{1}{2}\frac{e^2}{\rho} \cdot \dot{u}^i - \frac{2e^2}{3} \cdot u^i \cdot \dot{u}^k \dot{u}_k$$

5. При вычислении интеграла от выражений (3.5) и (3.6) необходимо учитывать, что в различных точках сферы $S$ величины $F^{ik}$ зависят от различных «точек излучения» на мировой линии частицы.

Пусть $y^i(\sigma)$ – мировая линия второй частицы. Выберем на $y^i(\sigma)$ некоторую точку и рассечем $S$ световыми конусами: $\left(\chi^i - y^i\right) \cdot \left(\chi_i - y_i\right) = 0$ (начало координат помещено в центр сферы $S$). Учитывая $\chi^i \chi_i = 0$, получаем уравнение 3-плоскости

$$(5.1) \quad \chi^i y_i = \frac{1}{2} y^i y_i$$

Обозначим $O$ окружность, возникающую как пересечение 3-плоскости (5.1) со сферой $S$. Окружность $O$ есть множество точек, для которых выбранная 4-точка $y^i(\sigma)$ является точкой излучения. Построим единичные векторы $\eta^i$ и $\zeta^i$, ортогональные векторам $k^i$ и $\xi^i$, причем $\zeta^i$ - параллелен плоскости (5.1), а $\eta^i$ – ортогонален $\zeta^i$.

Несложно показать, что этим условиям удовлетворяют векторы

$$(5.2) \quad \zeta^i = \frac{1}{\sqrt{-y_p y^p \cdot \left(\rho^2 - \rho \cdot u_n y^n + \frac{1}{4} \cdot y_n y^n\right)}} \cdot e^{iklm} \chi_k u_l y_m$$
$$\eta^i = \frac{1}{\rho} \cdot e^{iklm} \chi_k u_l \zeta_m$$

здесь $e^{iklm}$ – единичный, полностью антисимметричный псевдотензор. Векторы $\xi^i, k^i, \zeta^i, \eta^i$ образуют правую тройку, в чем легко убедиться, вычислив определитель: $e_{iklm}\xi^i k^k \zeta^l \eta^m = -1$. (Мы считаем $e_{0123} = -1$).

В качестве элемента интегрирования в (2.6) выберем прямоугольник, стороны которого параллельны векторам $\eta^i$ и $\zeta^i$.

Рассмотрим световой конус, вершина которого расположена в точке $\bar{y}^i$, лежащей на мировой линии частицы, достаточно близко к $y^i(\sigma)$, так, что

(5.3) $$\bar{y}^i = y^i + w^i \cdot \delta\sigma$$

где $w^i$ – вектор 4-скорости частицы. Подобно первому этот конус пересекается со сферой $S$ по окружности $\bar{O}$. Вычислим ширину полоски сферы $S$, заключенной между $O$ и $\bar{O}$. Для этого надо вычислить длину $d$ стороны прямоугольника, параллельную вектору $\eta^i$, заключенную между $O$ и $\bar{O}$. Поскольку конец вектора $\chi^i + d \cdot \eta^i$ лежит на световом конусе точки $\bar{y}^i$, можно составить уравнение

(5.4) $$\left(\chi^i + d \cdot \eta^i - y^i - w^i \cdot \delta\sigma\right)\cdot\left(\chi_i + d \cdot \eta_i - y_i - w_i \cdot \delta\sigma\right) = 0$$

решая которое, находим:

(5.5) $$d = -\frac{w_i \cdot (\chi^i - y^i)}{\eta_k y^k} \cdot \delta\sigma$$

Скалярное произведение в знаменателе (5.5) легко вычисляется из (5.2), после чего элемент интегрирования в (2.6) приобретает вид

(5.6) $$\delta S = \frac{\rho \cdot (w_i \cdot (\chi^i - y^i)) \cdot \delta\sigma \cdot \delta q}{\sqrt{-y_p y^p \cdot \left(\rho^2 - \rho \cdot u_n y^n + \frac{1}{4} \cdot y_n y^n\right)}}$$

где $\delta q$ – сторона элементарного прямоугольника, параллельная вектору $\zeta^i$.

Окружность $O$ лежит в 2 плоскости, являющейся пересечением 3-плоскостей (1.8) и (5.1) (напомним, $x^i - z^i(\tau) = \chi^i$). Введем в этой плоскости ортонормированный базис $\lambda^i, \mu^i$ ($\mu^i \mu_i = \lambda^i \lambda_i = -1$). Ясно, что эти базисные векторы ортогональны векторам $y^i, u^i$. Обозначим $N^i$ центр окружности $O$. В системе наблюдателя, связанного с частицей $\mathbf{u}$, $N^i$ есть перпендикуляр, опущенный из центра сферы $S$ на плоскость $\mathbf{O}$. В лабораторной системе мы найдем , построив вектор, составленный из $y^i, u^i$, конец которого лежит на плоскости $\mathbf{O}$:

(5.7) $$N^i = -\frac{y_m y^m \cdot (y^i - u^i \cdot y^k u_k)}{2 \cdot r_s^2} + \frac{\rho \cdot (y^i \cdot y^k u_k - y_m y^m \cdot u^i)}{r_s^2}$$

где $r_s$ – расстояние между частицами **u** и **w** в системе наблюдателя **u**:

(5.8) $$r_s = \sqrt{(u_i y^i)^2 - y_i y^i}$$

Вектор $\chi^i$ из центра сферы **S** в какую-либо точку **O** может быть представлен в виде: $\chi^i = N^i + a \cdot \lambda^i + b \cdot \mu^i$. Учитывая, что $\chi^i$ светоподобен, находим: $a^2 + b^2 = N^i N_i$, откуда заключаем, что

(5.9) $$\chi^i = N^i + R_o \cdot (\lambda^i \cdot \cos\varphi + \mu^i \cdot \sin\varphi),$$

где

(5.10) $$R_o = \sqrt{N^i N_i} = \sqrt{-y^p y_p \cdot \frac{y^m y_m - 4\rho \cdot y^m u_m + 4\rho^2}{4 \cdot r_s^2}}.$$

Мы приходим к выражению для величины $\delta q$: $\delta q = R_o \cdot \delta\varphi$. Выражение (5.6) для элемента интегрирования по сфере **S** принимает вид:

(5.11) $$\delta S = \frac{\rho \cdot (w_i \cdot (\chi^i - y^i)) \cdot \delta\sigma \cdot \delta\varphi}{r_s}$$

Таким образом, интеграл по сфере **S** распадается на интеграл по $d\varphi$ вдоль окружности **O** и на интеграл по $d\sigma$ вдоль мировой линии частицы **w**. Пределы интегрирования $\sigma_1$, $\sigma_2$ определим из условия, что конец вектора $N^i$ оказывается на поверхности сферы **S**. Это приводит к уравнению относительно $y^i(\sigma)$: $N^i \cdot N_i = 0$. Согласно (5.10), эта свертка распадается на два сомножителя:

(5.12) $$y^i(\sigma) \cdot y_i(\sigma) = 0 \quad ; \quad (2\rho u^i(\tau) - y^i(\sigma)) \cdot (2\rho u_i(\tau) - y_i(\sigma)) = 0$$

Таким образом, пределы интегрирования $\sigma_1$, $\sigma_2$ определяются точками пересечения световых конусов точек $z^i(\tau), z^i(\tau) + 2\rho \cdot u^i(\tau)$ с мировой линией частицы **w**.

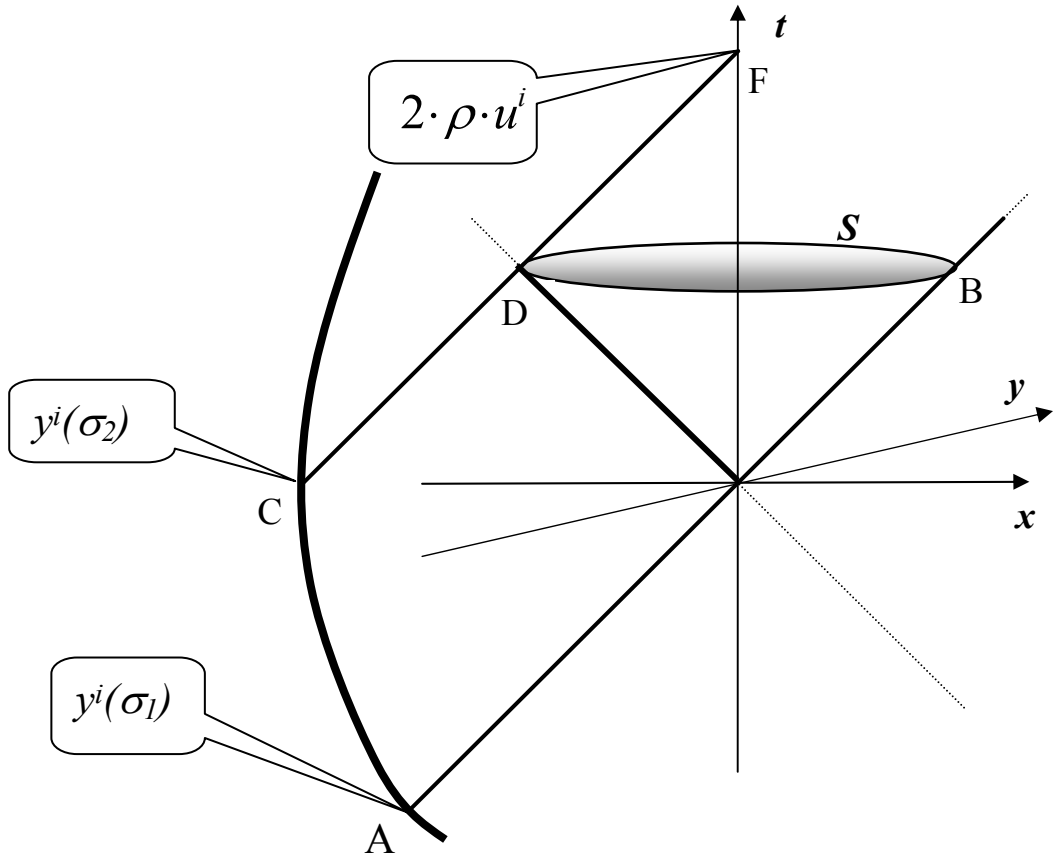

Рис.5.1.

Из физических соображений ясно, что следует выбрать интервал интегрирования между точками, вырезаемыми световыми конусами абсолютного прошлого. Условия (5.12) имеют ясный геометрический смысл (см. рис. 5.1). Луч света из 4-точки $y^i(\sigma_1)$ (луч AB) достигает самой удаленной от частицы точки сферы **S**. Тогда как луч света из точки $y_i(\sigma_2)$ (луч CD) достигает сферу **S** в ближайшей к частице точке. Луч AB лежит на световом конусе абсолютного прошлого 4-центра сферы **S**, то есть точки $z^i(\tau)$. Луч CD лежит на световом конусе абсолютного прошлого 4-точки $z^i(\tau)+2\rho \cdot u^i(\tau)$. Таким образом, на участке интегрирования AC справедливы следующие неравенства:

(5.13)
$$\begin{aligned}&y^k y_k < 0 \\ &\left(y^k - 2\rho u^k\right)\left(y_k - 2\rho u_k\right) > 0 \\ &y^k u_k < 0\end{aligned}$$

Интеграл (2.6) преобразуем к виду:

$$\text{(5.14)} \qquad \delta P^i = -\delta\tau \cdot \int\limits_{\sigma_1}^{\sigma_2} \frac{\rho}{r_s} d\sigma \int\limits_0^{2\pi} T^{ik}\psi_k \cdot \left(w_i \cdot (\chi^i - y^i)\right) \cdot d\varphi$$

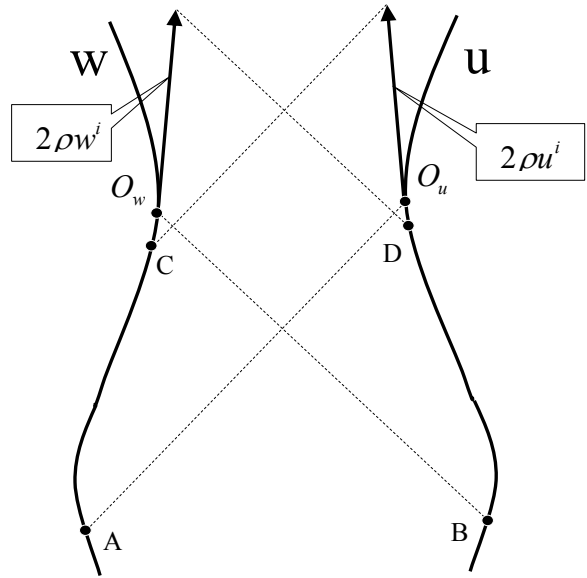

Рис. 5.2.

Характер движения качественно меняется при сближении частиц на расстояния порядка электромагнитного радиуса частиц

$$\text{(5.15)} \qquad \rho = \frac{2e^2}{3m}.$$

Ускорение частицы **u** в точке $O_u$ (см. рис. 5.2) определяется полем частицы **w** на участке мировой линии AC. Если расстояние между частицами велико, то точка C имеет в системе центра масс временную координату меньшую, чем точка $O_u$, следовательно движение частицы **w** на участке AC к этому моменту вполне определено, и для ответа на вопрос об ускорении частицы **u** в т. $O_u$ достаточно решить стандартную задачу о движении в заданном внешнем поле. Однако, если частицы сближаются на расстояния, сравнимые с электромагнитным радиусом, то временная координата 4-точки C может оказаться больше временной координаты точки $O_w$. В этом случае движение частицы **w** на участке AC окажется зависящим от движения частицы **u** на участке BD. Если при этом точка D выше т. $O_u$, то, как мы видим, характер движения меняется качественно – ускорение частицы **u** в выбранной точке $O_u$ зависит от участка ее собственной мировой линии, лежащего в конусе абсолютного будущего рассматриваемой точки $O_u$. Предлагаемая в данной работе теория не позволяет ответить на вопрос о движении частиц при таком сближении. Как видно из рис. 5.2. критическое расстояние оказывается зависящим от компоненты $u^0$ 4-скорости частицы:

(5.16) $$d_m \Box 2 \cdot \rho \cdot u^0.$$

6. Величины $F^{ik}$, входящие в (3.5) и (3.6) определяются выражением Леонара-Вихерта:

(6.1) $$F^{ik} = \frac{e}{\left(\beta^m w_m\right)^3} \cdot \left(\beta^i w^k - \beta^k w^i\right) \cdot \left(1 - \beta^m \dot{w}_m\right) + \frac{e}{\left(\beta^m w_m\right)^2} \cdot \left(\beta^i \dot{w}^k - \beta^k \dot{w}^i\right),$$

где $\beta^i = -y^i + \chi^i$ – 4-вектор из «точки излучения» $y^i(\tau)$ на мировой линии частицы **w** в точку $\chi^i$ на окружности **O**, $w^i, \dot{w}^i$ – скорость и ускорение частицы **w** в момент $\tau$.

Для вычисления интегралов от величин (3.5) и (3.6) необходимо выполнить ряд довольно громоздких преобразований. Для упрощения записи формул введем следующие обозначения:

(6.2) $$T_{FG}^{ik} \cdot \psi_k = B_u u^i + B_a \dot{u}^i + B_\chi \chi^i + B_y y^i + B_w w^i + B \dot{w}^i$$

Тогда, преобразуя (3.6), находим[c]:

(6.3) $$\begin{aligned} B_u = &-\frac{e_u}{4\pi\rho^2} \cdot \frac{e_w}{\left(\beta^m w_m\right)^3} \cdot \left(\tfrac{1}{2} Y \cdot w^l \dot{u}_l + \beta^l \dot{u}_l \cdot y_k w^k\right) \cdot \left(1 - \beta^m \dot{w}_m\right) \\ &-\frac{e_u}{4\pi\rho^2} \cdot \frac{e_w}{\left(\beta^m w_m\right)^2} \cdot \left(\tfrac{1}{2} Y \cdot \dot{u}_l \dot{w}^l + \beta^l \dot{u}_l \cdot \left(1 + y^m \dot{w}_m\right)\right) \end{aligned}$$

(6.4) $$\begin{aligned} B_a = &\frac{e_u}{4\pi\rho^2} \times \frac{e_w}{\left(\beta^m w_m\right)^3} \cdot \left(\tfrac{1}{2} Y \cdot u_l w^l + \left(\rho + U\right) \cdot y_k w^k\right) \cdot \left(1 - \beta^m \dot{w}_m\right) \\ &+\frac{e_u}{4\pi\rho^2} \times \frac{e_w}{\left(\beta^m w_m\right)^2} \cdot \left(\tfrac{1}{2} Y \cdot u_l \dot{w}^l + \left(\rho + U\right) \cdot \left(1 + y^m \dot{w}_m\right)\right) \end{aligned}$$

(6.5) $$\begin{aligned} B_\chi = &-\frac{e_u}{4\pi\rho^2} \cdot \left(-y_k u^k + 2\rho\right) \cdot \left(\dot{u}_l \cdot w^l \cdot \frac{e_w}{\left(\beta^m w_m\right)^3} \cdot \left(1 - \beta^m \dot{w}_m\right) + \dot{u}_l \cdot \dot{w}^l \cdot \frac{e_w}{\left(\beta^m w_m\right)^2}\right) \\ &-\frac{e_u}{4\pi\rho^2} \cdot \left(1 - y^m \dot{u}_m - 2 \cdot \beta^m \dot{u}_m\right) \times \\ &\left(\frac{e_w}{\left(\beta^m w_m\right)^3} \cdot u_k \cdot w^k \cdot \left(1 - \beta^m \dot{w}_m\right) + \frac{e_w}{\left(\beta^m w_m\right)^2} \cdot u_k \cdot \dot{w}^k\right) \end{aligned}$$

$$B_y = \frac{e_u}{4\pi\rho^2} \cdot \left(\rho \cdot \dot{u}_l + u_l \cdot \left(1 - \chi^m \dot{u}_m\right)\right) \times$$

(6.6)
$$\left(\frac{e_w}{\left(\beta^m w_m\right)^3} \cdot w^l \cdot \left(1 - \beta^m \dot{w}_m\right) + \frac{e_w}{\left(\beta^m w_m\right)^2} \cdot \dot{w}^l\right)$$

(6.7) $$B_w = \frac{e_u}{4\pi\rho^2} \cdot \left(\rho \cdot \dot{u}_l + u_l \cdot \left(1 - \chi^m \dot{u}_m\right)\right) \times \frac{e_w}{\left(\beta^m w_m\right)^3} \cdot \beta^l \cdot \left(1 - \beta^m \dot{w}_m\right)$$

(6.8) $$B_b = \frac{e_u}{4\pi\rho^2} \cdot \left(\rho \cdot \dot{u}_l + u_l \cdot \left(1 - \chi^m \dot{u}_m\right)\right) \times \frac{e_w}{\left(\beta^m w_m\right)^2} \cdot \beta^l$$

Величины *B*... должны быть проинтегрированы по окружности **O**.

7. Рассмотрим следующие интегралы:

(7.1) $L_1 = \frac{1}{r_s} \cdot \int_o \frac{d\varphi}{\beta^m \xi_m}; \quad L_2 = \frac{1}{r_s} \cdot \int_o \frac{d\varphi}{\left(\beta^m \xi_m\right)^2}; \quad L_3 = \frac{1}{r_s} \cdot \int_o \frac{d\varphi}{\left(\beta^m \xi_m\right)^3}; \quad L_4 = \frac{1}{r_s} \cdot \int_o \frac{d\varphi}{\left(\beta^m \xi_m\right)^4};$

(7.2) $M_1^i = \frac{1}{r_s} \cdot \int_o \frac{\beta^i \cdot d\varphi}{\beta^m \xi_m}; \quad M_2^i = \frac{1}{r_s} \cdot \int_o \frac{\beta^i \cdot d\varphi}{\left(\beta^m \xi_m\right)^2}; \quad M_3^i = \frac{1}{r_s} \cdot \int_o \frac{\beta^i \cdot d\varphi}{\left(\beta^m \xi_m\right)^3};$

(7.3) $M_4^i = \frac{1}{r_s} \cdot \int_o \frac{\beta^i \cdot d\varphi}{\left(\beta^m \xi_m\right)^4}; \quad M_5^i = \frac{1}{r_s} \cdot \int_o \frac{\beta^i \cdot d\varphi}{\left(\beta^m \xi_m\right)^5};$

(7.4) $N_1^{ik} = \frac{1}{r_s} \cdot \int_o \frac{\beta^i \beta^k \cdot d\varphi}{\left(\beta^m \xi_m\right)}; \quad N_2^{ik} = \frac{1}{r_s} \cdot \int_o \frac{\beta^i \beta^k \cdot d\varphi}{\left(\beta^m \xi_m\right)^2}; \quad N_3^{ik} = \frac{1}{r_s} \cdot \int_o \frac{\beta^i \beta^k \cdot d\varphi}{\left(\beta^m \xi_m\right)^3};$

(7.5) $N_4^{ik} = \frac{1}{r_s} \cdot \int_o \frac{\beta^i \beta^k \cdot d\varphi}{\left(\beta^m \xi_m\right)^4}; \quad N_5^{ik} = \frac{1}{r_s} \cdot \int_o \frac{\beta^i \beta^k \cdot d\varphi}{\left(\beta^m \xi_m\right)^5}$

(7.6) $P_2^{ikl} = \frac{1}{r_s} \cdot \int_o \frac{\beta^i \beta^k \beta^l \cdot d\varphi}{\left(\beta^m \xi_m\right)^2}; \quad P_3^{ikl} = \frac{1}{r_s} \cdot \int_o \frac{\beta^i \beta^k \beta^l \cdot d\varphi}{\left(\beta^m \xi_m\right)^3};$

(7.7) $P_4^{ikl} = \frac{1}{r_s} \cdot \int_o \frac{\beta^i \beta^k \beta^l \cdot d\varphi}{\left(\beta^m \xi_m\right)^4}; \quad P_5^{ikl} = \frac{1}{r_s} \cdot \int_o \frac{\beta^i \beta^k \beta^l \cdot d\varphi}{\left(\beta^m \xi_m\right)^5};$

(7.8) $Q_5^{ikmn} = \frac{1}{r_s} \cdot \int_o \frac{\beta^i \beta^k \beta^m \beta^n \cdot d\varphi}{\left(\beta^m \xi_m\right)^5}$

Интегралы от величин (6.3)-(6.8) могут быть выражены через эти интегралы (7.1)-(7.8), если в последних заменить $\xi_m \to w_m$. Введем обозначения:

$$(7.9) \quad \hat{B}_u = \frac{1}{r_s} \int_0^{2\pi} B_u \cdot \beta^m w_m d\varphi$$

$$(7.10) \quad \hat{B}_a = \frac{1}{r_s} \int_0^{2\pi} B_a \cdot \beta^m w_m d\varphi$$

$$(7.11) \quad \hat{B}_\beta^i = \frac{1}{r_s} \int_0^{2\pi} B_\chi \cdot \beta^i \cdot \beta^m w_m d\varphi$$

$$(7.12) \quad \hat{B}_y = \frac{1}{r_s} \int_0^{2\pi} \left( B_y + B_\chi \right) \cdot \beta^m w_m d\varphi$$

$$(7.13) \quad \hat{B}_w = \frac{1}{r_s} \int_0^{2\pi} B_w \cdot \beta^m w_m d\varphi$$

$$(7.14) \quad \hat{B}_b = \frac{1}{r_s} \int_0^{2\pi} B_b \cdot \beta^m w_m d\varphi$$

Так, что

$$(7.15) \quad \frac{dP_{FG}^i}{d\tau} = -\rho \cdot \int_{\sigma_1}^{\sigma_2} \left( \hat{B}_u u^i + \hat{B}_a \dot{u}^i + \hat{B}_\beta^i + \hat{B}_y y^i + \hat{B}_w w^i + \hat{B}_b \dot{w}^i \right) d\sigma \cdot$$

С помощью несложных преобразований из можно найти (см. **Приложение 3**):

$$(7.16) \quad \begin{aligned} \hat{B}_\beta^i &= -\frac{e_u e_w}{4\pi \rho^2} \cdot \left( (U + 2\rho) \cdot \dot{u}_l + (1 - y^m \dot{u}_m) \cdot u_l \right) \left( \dot{w}^l \cdot \left( M_2^i - N_2^{im} \dot{w}_m \right) + \dot{w}^l \cdot M_1^i \right) \\ &+ \frac{e_u e_w}{4\pi \rho^2} \cdot 2 \cdot \dot{u}_m \times \left( u_k \cdot w^k \cdot \left( N_2^{im} - P_2^{ipm} \dot{w}_p \right) + N_1^{im} \cdot u_k \cdot \dot{w}^k \right) \end{aligned}$$

$$(7.17) \quad \begin{aligned} \hat{B}_u &= -\frac{e_u e_w}{4\pi \rho^2} \cdot \left( \tfrac{1}{2} Y \cdot w^l \dot{u}_l \cdot L_2 + M_2^l \dot{u}_l \cdot y_k w^k \right) \\ &+ \frac{e_u e_w}{4\pi \rho^2} \cdot \left( \tfrac{1}{2} Y \cdot w^l \dot{u}_l \cdot M_2^m + N_2^{ml} \dot{u}_l \cdot y_k w^k \right) \cdot \dot{w}_m \\ &- \frac{e_u e_w}{4\pi \rho^2} \cdot \left( \tfrac{1}{2} Y \cdot \dot{u}_l \dot{w}^l L_1 + M_1^l \dot{u}_l \cdot \left( 1 + y^m \dot{w}_m \right) \right) \end{aligned}$$

$$(7.18) \quad \begin{aligned} \hat{B}_a &= \frac{e_u e_w}{4\pi \rho^2} \cdot \left( \tfrac{1}{2} Y \cdot u_l w^l + (\rho + U) \cdot y_k w^k \right) \cdot \left( L_2 - M_2^m \dot{w}_m \right) \\ &+ \frac{e_u e_w}{4\pi \rho^2} \cdot \left( \tfrac{1}{2} Y \cdot u_l \dot{w}^l + (\rho + U) \cdot \left( 1 + y^m \dot{w}_m \right) \right) \cdot L_1 \end{aligned}$$

$$(7.19) \quad \begin{aligned} \hat{B}_y &= -\frac{e_u e_w}{4\pi \rho^2} \cdot (U + \rho) \cdot \left( \dot{u}_l \cdot w^l \cdot \left( L_2 - M_2^m \dot{w}_m \right) + \dot{u}_l \cdot \dot{w}^l \cdot L_1 \right) \\ &+ \frac{e_u e_w}{4\pi \rho^2} \cdot \dot{u}_m \times \left( u_k \cdot w^k \cdot \left( M_2^m - N_2^{nm} \dot{w}_n \right) + M_1^m \cdot u_k \cdot \dot{w}^k \right) \end{aligned}$$

$$(7.20) \quad \hat{B}_w = \frac{e_u e_w}{4\pi\rho^2} \cdot (\rho + U)(1 - y^m \dot{u}_m) \cdot (L_2 - M_2^m \dot{w}_m)$$

$$- \frac{e_u e_w}{4\pi\rho^2} \cdot U \cdot \dot{u}_l \cdot (M_2^l - N_2^{lm} \dot{w}_m)$$

$$(7.21) \quad \hat{B}_b = \frac{e_u e_w}{4\pi\rho^2} \cdot \left(-U \cdot \dot{u}_l M_1^l + (\rho + U) \cdot (1 - y^m \dot{u}_m) \cdot L_1\right)$$

Где введены обозначения

$$(7.22) \quad U = -y_k u^k; \quad Y = y_k y^k$$

8. Рассмотрим выражение (3.5). Тензор энергии-импульса приведем к виду[d]:

$$(8.1) \quad T_{FF}^{ip} = \frac{e_w^2}{4\pi} \cdot \left(\frac{1}{(\beta^m w_m)^5} \cdot (\beta^p w^i + \beta^i w^p) \cdot (1 - \beta^m b_m) + \frac{1}{(\beta^m w_m)^4} \cdot (\beta^p b^i + \beta^i b^p)\right)$$

$$+ \frac{e_w^2}{4\pi} \cdot \left(-\frac{1}{(\beta^m w_m)^6} \cdot \beta^i \beta^p \cdot (1 - \beta^m b_m)^2 - \frac{1}{(\beta^m w_m)^4} \cdot \beta^i \beta^p \cdot b^k b_k - \frac{1}{2} \cdot \frac{g^{ip}}{(\beta^m w_m)^4}\right)$$

Интеграл (5.14) по сфере $S$ распадается на четыре слагаемых, которые мы вычислим независимо. Каждое из слагаемых представляет собой двойной интеграл по мировой линии частицы 2 и по окружности $O$:

$$(8.2) \quad \frac{dP_{FF}^i}{d\tau} = -\rho \cdot \int_{\sigma_1}^{\sigma_2} d\sigma \cdot \left(X^i + X_w \cdot w^i + X_b \cdot \dot{w}^i + X_\psi^{\ i}\right)$$

где

$$(8.3) \quad X_\beta^{\ i} = \frac{e_w^2}{4\pi} \cdot \int_0^{2\pi} d\varphi \cdot \frac{\beta^i}{r_s} \cdot \left(\frac{w^p \cdot (1 - \beta^m b_m)}{(\beta^m w_m)^4} + \frac{\dot{w}^p - \beta^p \cdot \dot{w}^k \dot{w}_k}{(\beta^m w_m)^3} - \frac{\beta^p \cdot (1 - \beta^m b_m)^2}{(\beta^m w_m)^5}\right) \psi_p$$

$$(8.4) \quad X_w = \frac{e_w^2}{4\pi} \cdot \int_0^{2\pi} d\varphi \cdot \frac{1}{r_s} \cdot \frac{\beta^p}{(\beta^m w_m)^4} \cdot (1 - \beta^m \dot{w}_m) \cdot \psi_p$$

$$(8.5) \quad X_b = \frac{e_w^2}{4\pi} \cdot \int_0^{2\pi} d\varphi \cdot \frac{1}{r_s} \cdot \frac{\beta^p}{(\beta^m w_m)^3} \psi_p$$

$$(8.6) \quad X_\psi^{\ i} = -\frac{1}{2} \cdot \frac{e_w^2}{4\pi} \cdot \int_0^{2\pi} d\varphi \cdot \frac{1}{r_s} \cdot \frac{g^{ip}}{(\beta^m w_m)^3} \psi_p$$

Каждое из выражений (8.3)-(8.6) разобьем на три в соответствии с тремя слагаемыми величины $\psi_p$:

$$(8.7) \quad X_\beta^{\ i} = \frac{e_w^2}{4\pi} \cdot \left(X_{\beta u}^i + X_{\beta \rho}^i + X_{\beta a,m}^i a^m\right); \quad X_w = \frac{e_w^2}{4\pi} \cdot \left(X_{wu} + X_{w\rho} + X_{wa,m} a^m\right);$$

$$X_\psi^{\ i} = \frac{e_w^2}{4\pi} \cdot \left(X_{\psi u}^i + X_{\psi \rho}^i + X_{\psi a,m}^i a^m\right); \quad X_b = \frac{e_w^2}{4\pi} \cdot \left(X_{bu} + X_{b\rho} + X_{ba,m} a^m\right);$$

Преобразуя (8.3), найдем[e]:

(8.8)
$$X_{\beta u}{}^i = w^p u_p \cdot \left(M_4^i - N_4^{im}\dot{w}_m\right) + M_3^i \cdot \dot{w}^p u_p$$
$$- \left(M_5^i - 2\cdot N_5^{im}\dot{w}_m + P_5^{imn}\dot{w}_m\dot{w}_n + \dot{w}^k\dot{w}_k \cdot M_3^i\right)\cdot\left(\rho - y^p u_p\right)$$

(8.9)
$$X_{\beta\rho}{}^i = -\left(w^p y_p \cdot \left(M_4^i - N_4^{im}\dot{w}_m\right) + \left(1 + \dot{w}^p y_p\right)\cdot M_3^i\right)\frac{1}{\rho}$$
$$- \left(M_5^i - 2\cdot N_5^{im}\dot{w}_m + P_5^{imn}\dot{w}_m\dot{w}_n + \dot{w}^k\dot{w}_k \cdot M_3^i\right)\cdot\frac{y^p y_p}{2\rho}$$

(8.10)
$$X_{\beta a}{}^{im} = \frac{w^p y_p}{\rho}\cdot\left(N_4^{im} + M_4^i y^m + \left(P_4^{imq} + y^m N_4^{iq}\right)\dot{w}_q\right) + \frac{1+\dot{w}^p y_p}{\rho}\cdot\left(N_3^{im} + y^m M_3^i\right)$$
$$+ \left(N_5^{im} + y^m M_5^i - 2\cdot\left(P_5^{imq} + y^m N_5^{iq}\right)\dot{w}_q\right)\cdot\frac{y_p y^p}{2\rho}$$
$$+ \left(\left(Q_5^{imqk} + y^m P_5^{iqk}\right)\dot{w}_q\dot{w}_k + \left(N_3^{im} + y^m M_3^i\right)\cdot\dot{w}^k\dot{w}_k\right)\cdot\frac{y_p y^p}{2\rho}$$

Подобным образом можно найти[f]:

(8.11)
$$X_{wu} = \left(\rho - y^l u_l\right)\cdot\left(L_4 - M_4^m\dot{w}_m\right)$$

(8.12)
$$X_{w\rho} = \left(L_4 - M_4^m\dot{w}_m\right)\cdot\frac{y^l y_l}{2\rho}$$

(8.13)
$$X_{wa}^m = -\left(\left(M_4^m + y^m\cdot L_4\right) - \left(N_4^{mp} + y^m M_4^p\right)\dot{w}_p\right)\cdot\frac{y^l y_l}{2\rho}$$

(8.14)
$$X_{wu} = \left(\rho - y^l u_l\right)\cdot L_3$$

(8.15)
$$X_{b\rho} = \frac{y^l y_l}{2\rho}\cdot L_3$$

(8.16)
$$X_{ba}^m = -\left(y^m\cdot L_3 + M_3^m\right)\cdot\frac{y^l y_l}{2\rho}$$

(8.17)
$$X_{\psi u}{}^i = -\frac{1}{2}\cdot L_3\cdot u^i$$

(8.18)
$$X_{\psi\rho}{}^i = \frac{1}{2\rho}\cdot\left(M_3^i + y^i\cdot L_3\right)$$

(8.19)
$$X_{\psi a}{}^i = -\frac{1}{2\rho}\cdot\left(N_3^{mi} + y^m M_3^i + y^i M_3^m + y^m y^i L_3\right)$$

9. Вычислим интегралы (7.1)-(7.8). Рассмотрим производящую функцию

(9.1)
$$L_1 = \frac{1}{r_s}\cdot\int_o \frac{d\varphi}{\beta^m \xi_m}$$

интеграл в (9.1) берется по окружности **O**, по которой скользит конец 4-вектора $\beta^i$:

(9.2) $$\beta^i = \chi^i - y^i = \tilde{N}^i + R_o \cdot \left( \lambda^i \cos\varphi + \mu^i \sin\varphi \right),$$

где $\tilde{N}^i = N^i - y^i$ – 4-вектор из положения частицы **w** в центр окружности.

Выберем базис $\left( \lambda^i, \mu^i \right)$ в плоскости **O** так, чтобы $\mu^i \xi_i = 0$:

(9.3) $$\mu^i = A \cdot e^{iklm} u_k y_l \xi_m; \quad \lambda^i = \frac{1}{\sqrt{(y^k u_k)^2 - y^p y_p}} \cdot e^{iklm} u_k y_l \mu_m$$

здесь нормировочный множитель[g] $A = \dfrac{1}{\sqrt{-e^{iklm} u_k y_l \xi_m \cdot e_{ipqr} u^p y^q \xi^r}}$.

Или

(9.4) $$A = \frac{1}{\sqrt{\tilde{\Theta}^{nm} \xi_n \xi_m}}$$

Где

(9.5) $$\tilde{\Theta}^{nm} = -\left( (u_k y^k)^2 - y_k y^k \right) \cdot g^{nm} - \left( y^n \cdot y^m - u_k y^k \cdot (u^m y^n + u^n y^m) + y_k y^k \cdot u^m u^n \right)$$

Заметим, что[h]

(9.6) $$\tilde{\Theta}^{ik} = r_s^2 \cdot \left( \lambda^i \lambda^k + \mu^i \mu^k \right)$$

откуда[i]

(9.7) $$\tilde{\Theta}^{nm} \xi_m = r_s \cdot \sqrt{\tilde{\Theta}^{nm} \xi_n \xi_m} \cdot \lambda^i$$

Интеграл (9.1) может быть вычислен:

(9.8) $$L_1 = \frac{1}{r_s} \cdot \int_0^{2\pi} \frac{d\varphi}{a + b\cos\varphi} = \frac{1}{r_s} \cdot \frac{2\pi}{\sqrt{a^2 - b^2}},$$

где $a = \tilde{N}^i \cdot \xi_i$; $b = R_o \cdot \lambda^i \xi_i = \dfrac{R_o}{r_s} \cdot e^{iklm} u_k y_l \mu_m \xi_i = -R_o \cdot \dfrac{\mu^i \mu_i}{r_s \cdot A} = \dfrac{R_o}{r_s \cdot A}$. Откуда находим:

(9.9) $$L_1 = \frac{2\pi}{\sqrt{\Theta^{nm} \xi_m \xi_n}},$$

где

(9.10) $$\Theta^{nm} = \tilde{N}^n \tilde{N}^m \cdot r_s^2 + R_o^2 \cdot e^{iklm} u_k y_l \cdot e_{ipqr} u^p y^q g^{rn}$$

Или

(9.11) $$\Theta^{nm} = \tilde{N}^n \tilde{N}^m \cdot r_s^2 - R_o^2 \cdot \tilde{\Theta}^{nm}$$

Выражение (9.10) можно упростить, если принять во внимание (5.7) и (5.10). В **Приложении 1** получен следующий результат:

$$(9.12) \quad \begin{aligned} \Theta^{nm} &= \left(\left(\rho - y^p u_p\right)\cdot y^n + \tfrac{1}{2} y^p y_p u^n\right)\cdot\left(\left(\rho - y^q u_q\right)\cdot y^m + \tfrac{1}{2} y^q y_q u^m\right) \\ &\quad - y^p y_p \cdot \left(\tfrac{1}{4} y^q y_q - \rho\cdot y^q u_q + \rho^2\right)\cdot g^{nm} \end{aligned}$$

В **Приложении 2** показано, как вычисляются интегралы (7.1)-(7.6). Выпишем результаты вычислений:

$$(9.13) \quad M_2^i = 2\pi \cdot r_s^2 \cdot \frac{\tilde{N}^i \tilde{N}^m \xi_m - R_o^2 \cdot \lambda^i \lambda^m \xi_m}{\left(\Theta^{nm}\xi_m \xi_n\right)^{\frac{3}{2}}}$$

$$(9.14) \quad L_2 = 2\pi \cdot r_s^2 \cdot \frac{\tilde{N}^m \xi_m}{\left(\Theta^{nm}\xi_m \xi_n\right)^{\frac{3}{2}}}$$

$$(9.15) \quad M_1^i = \tilde{N}^i \cdot L_1 + X \cdot \lambda^i, \quad \text{where } X = \frac{2\pi}{\lambda^i \xi_i}\cdot\left(\frac{1}{r_s} - \frac{\tilde{N}^i \xi_i}{\sqrt{\Theta^{nm}\xi_n \xi_m}}\right)$$

$$(9.16) \quad N_1^{ik} = \tilde{N}^i \tilde{N}^k \cdot L_1 + \left(\tilde{N}^i \lambda^k + \tilde{N}^k \lambda^i\right)\cdot X + R_o^2 \cdot\left(\lambda^i \lambda^k \cdot (L_1 - X_1) + \mu^i \mu^k \cdot X_1\right)$$

где
$$X_1 = \frac{\tilde{N}^k \xi_k \cdot X}{R_o^2 \cdot \lambda^k \xi_k} + L_1$$

$$(9.17) \quad N_2^{ik} = \tilde{N}^i \tilde{N}^k \cdot L_2 + \left(\tilde{N}^i \lambda^k + \tilde{N}^k \lambda^i\right)\cdot X_2 + X_\lambda \cdot \lambda^i \lambda^k + X_\mu \cdot \mu^i \mu^k$$

где

$$(9.18) \quad X_2 = -2\pi \cdot r_s^2 \cdot \frac{R_o^2 \cdot \lambda^m \xi_m}{\left(\Theta^{nm}\xi_m \xi_n\right)^{\frac{3}{2}}}; \quad X_\mu = -\frac{X}{\left(\lambda^k \xi_k\right)}; \quad X_\lambda = R_o^2 \cdot L_2 - X_\mu$$

$$(9.19) \quad \begin{aligned} P_2^{ikl} &= \tilde{N}^i \cdot \tilde{N}^k \cdot \tilde{N}^l \cdot L_2 + \tilde{N}^i \cdot\left(N_2^{kl} - \tilde{N}^l \cdot M_2^k\right) + \tilde{N}^l \cdot\left(N_2^{ik} - \tilde{N}^k \cdot M_2^i\right) \\ &\quad + \tilde{N}^k \cdot\left(N_2^{il} - \tilde{N}^i \cdot M_2^l\right) + X_3 \cdot \lambda^i \lambda^k \lambda^l + X_4 \cdot\left(\lambda^i \mu^k \mu^l + \mu^i \lambda^k \mu^l + \mu^i \mu^k \lambda^l\right) \end{aligned}$$

где

$$(9.20) \quad X_4 = \frac{R_o^2 \cdot X_1 - \tilde{N}^l \cdot X_\mu \cdot \xi_l}{\lambda^l \cdot \xi_l}; \quad X_3 = \tilde{N}_k \cdot \tilde{N}^k \cdot X_2 - X_4$$

$$(9.21) \quad N_3^{ik} = -\frac{1}{2}\frac{\partial}{\partial \xi_k} M_2^i = \frac{3}{2}\cdot\frac{2\pi \cdot \Theta^{im}\xi_m \cdot \Theta^{kn}\xi_n}{\left(\Theta^{nm}\xi_m \xi_n\right)^{\frac{5}{2}}} - \frac{1}{2}\frac{2\pi \cdot \Theta^{ik}}{\left(\Theta^{nm}\xi_m \xi_n\right)^{\frac{3}{2}}}$$

$$(9.22) \quad P_4^{ikl} = -\frac{1}{2}\cdot\frac{2\pi \cdot\left(\Theta^{il}\Theta^{kn} + \Theta^{kl}\Theta^{in} + \Theta^{ik}\Theta^{ln}\right)\cdot \xi_n}{\left(\Theta^{nm}\xi_m \xi_n\right)^{\frac{5}{2}}} + \frac{5}{2}\frac{2\pi \cdot \Theta^{im}\xi_m \Theta^{kn}\xi_n \Theta^{lp}\xi_p}{\left(\Theta^{nm}\xi_m \xi_n\right)^{\frac{7}{2}}}$$

$$(9.23) \quad Q_5^{iklm} = 2\pi \cdot \left( \frac{1}{8} \cdot \frac{\left(\Theta^{il}\Theta^{km} + \Theta^{kl}\Theta^{im} + \Theta^{ik}\Theta^{lm}\right)}{\left(\Theta^{nm}\xi_m\xi_n\right)^{\frac{5}{2}}} + \frac{35}{8} \frac{\Theta^{iq}\xi_q\Theta^{kn}\xi_n\Theta^{lp}\xi_p\Theta^{ms}\xi_s}{\left(\Theta^{nm}\xi_m\xi_n\right)^{\frac{9}{2}}} \right)$$

$$-\frac{5\pi}{4} \frac{\left(\Theta^{im}\Theta^{kq}\Theta^{lp} + \Theta^{iq}\Theta^{km}\Theta^{lp} + \Theta^{il}\Theta^{kp}\Theta^{mq} + \Theta^{kl}\Theta^{ip}\Theta^{mq} + \Theta^{iq}\Theta^{kp}\Theta^{lm} + \Theta^{ik}\Theta^{lp}\Theta^{mq}\right)\xi_q\xi_p}{\left(\Theta^{nm}\xi_m\xi_n\right)^{\frac{7}{2}}}$$

Свертка тензора $\Theta^{nm}$ с $u_m$ имеет вид:

$$(9.24) \quad \Theta^{nm}u_m = \left(\rho - y^p u_p\right) \cdot \Upsilon^n$$

Где через $\Upsilon^n$ обозначено следующее выражение в:

$$(9.25) \quad \Upsilon^n = y^n \cdot \left(\left(\rho - y^q u_q\right) \cdot y^m u_m + \tfrac{1}{2} y^q y_q\right) + u^n \cdot y^p y_p \cdot \left(\tfrac{1}{2} y^m u_m - \rho\right)$$

Заметим, что

$$(9.26) \quad \Upsilon^n u_n = \left(\rho - y^q u_q\right) \cdot r_s^2.$$

С помощью (9.24) и (9.26), учитывая $\beta^i u_i = \rho - y^i u_i$, находим[j]:

$$(9.27) \quad P_5^{ikl} = 2\pi \cdot \left( \frac{1}{8} \cdot \frac{\left(\Theta^{il}\Upsilon^k + \Theta^{kl}\Upsilon^i + \Theta^{ik}\Upsilon^l\right)}{\left(\Theta^{nm}\xi_m\xi_n\right)^{\frac{5}{2}}} + \frac{35}{8} \frac{\Theta^{iq}\xi_q\Theta^{kn}\xi_n\Theta^{lp}\xi_p\Upsilon^s\xi_s}{\left(\Theta^{nm}\xi_m\xi_n\right)^{\frac{9}{2}}} \right)$$

$$-\frac{5\pi}{4} \frac{\left(\Upsilon^i\Theta^{kq}\Theta^{lp} + \Theta^{iq}\Upsilon^k\Theta^{lp} + \Theta^{iq}\Theta^{kp}\Upsilon^l + \Theta^{kl}\Theta^{ip}\Upsilon^q + \Theta^{il}\Theta^{kp}\Upsilon^q + \Theta^{ik}\Theta^{lp}\Upsilon^q\right)\xi_q\xi_p}{\left(\Theta^{nm}\xi_m\xi_n\right)^{\frac{7}{2}}}$$

$$(9.28) \quad N_5^{ik} = 2\pi \cdot \left( \frac{1}{8} \cdot \frac{\left(\Upsilon^i\Upsilon^k + \Upsilon^k\Upsilon^i + \Theta^{ik} \cdot r_s^2\right)}{\left(\Theta^{nm}\xi_m\xi_n\right)^{\frac{5}{2}}} + \frac{35}{8} \frac{\Theta^{iq}\xi_q\Theta^{kn}\xi_n\Upsilon^p\xi_p\Upsilon^s\xi_s}{\left(\Theta^{nm}\xi_m\xi_n\right)^{\frac{9}{2}}} \right)$$

$$-\frac{5\pi}{4} \frac{\left(2 \cdot \Upsilon^i\Theta^{kq}\Upsilon^p + 2 \cdot \Upsilon^k\Theta^{ip}\Upsilon^q + \Theta^{iq}\Theta^{kp} \cdot r_s^2 + \Theta^{ik}\Upsilon^p\Upsilon^q\right)\xi_q\xi_p}{\left(\Theta^{nm}\xi_m\xi_n\right)^{\frac{7}{2}}}$$

$$(9.29) \quad M_5^i = \frac{\pi}{4} \cdot \left( \frac{3 \cdot \Upsilon^i \cdot r_s^2}{\left(\Theta^{nm}\xi_m\xi_n\right)^{\frac{5}{2}}} - \frac{15 \cdot \left(\Upsilon^i\Upsilon^q\Upsilon^p + \Theta^{ip}\Upsilon^q \cdot r_s^2\right)\xi_q\xi_p}{\left(\Theta^{nm}\xi_m\xi_n\right)^{\frac{7}{2}}} + \frac{35 \cdot \Theta^{iq}\xi_q\left(\Upsilon^n\xi_n\right)^3}{\left(\Theta^{nm}\xi_m\xi_n\right)^{\frac{9}{2}}} \right)$$

$$(9.30) \quad N_4^{ik} = -\frac{1}{2} \cdot \frac{2\pi \cdot \left(\Upsilon^i\Theta^{kn} + \Upsilon^k\Theta^{in} + \Theta^{ik}\Upsilon^n\right) \cdot \xi_n}{\left(\Theta^{nm}\xi_m\xi_n\right)^{\frac{5}{2}}} + \frac{5}{2} \frac{2\pi \cdot \Theta^{im}\xi_m\Theta^{kn}\xi_n \cdot \Upsilon^p\xi_p}{\left(\Theta^{nm}\xi_m\xi_n\right)^{\frac{7}{2}}}$$

$$(9.31) \quad M_4^i = -\frac{1}{2} \cdot \frac{2\pi \cdot \left(2 \cdot \Upsilon^i \Upsilon^n + \Theta^{in} \cdot r_s^2\right) \cdot \xi_n}{\left(\Theta^{nm}\xi_m\xi_n\right)^{\frac{5}{2}}} + \frac{5}{2} \frac{2\pi \cdot \Theta^{im}\xi_m \cdot \left(\Upsilon^p \xi_p\right)^2}{\left(\Theta^{nm}\xi_m\xi_n\right)^{\frac{7}{2}}}$$

$$(9.32) \quad L_4 = -3\pi \cdot \frac{\Upsilon^n \xi_n \cdot r_s^2}{\left(\Theta^{nm}\xi_m\xi_n\right)^{\frac{5}{2}}} + 5\pi \cdot \frac{\left(\Upsilon^p \xi_p\right)^3}{\left(\Theta^{nm}\xi_m\xi_n\right)^{\frac{7}{2}}}$$

$$(9.33) \quad M_3^i = 3\pi \cdot \frac{\Theta^{im}\xi_m \cdot \Upsilon^n \xi_n}{\left(\Theta^{nm}\xi_m\xi_n\right)^{\frac{5}{2}}} - \pi \cdot \frac{\Upsilon^i}{\left(\Theta^{nm}\xi_m\xi_n\right)^{\frac{3}{2}}}$$

$$(9.34) \quad L_3 = 3\pi \cdot \frac{\left(\Upsilon^m \xi_m\right)^2}{\left(\Theta^{nm}\xi_m\xi_n\right)^{\frac{5}{2}}} - \pi \cdot \frac{r_s^2}{\left(\Theta^{nm}\xi_m\xi_n\right)^{\frac{3}{2}}}$$

## 10. Результаты.

Ниже представлены результаты расчетов столкновений двух частиц с зарядами противоположных знаков при различных значениях начальной энергии и прицельного расстояния. Через $u$ обозначена абсолютная величина пространственной компоненты 4-скорости каждой из частиц в системе центра масс. Через $d$ обозначено прицельное расстояние на бесконечности. Массы сталкивающихся частиц приняты равными единице. Заряды – $\pm 1$.

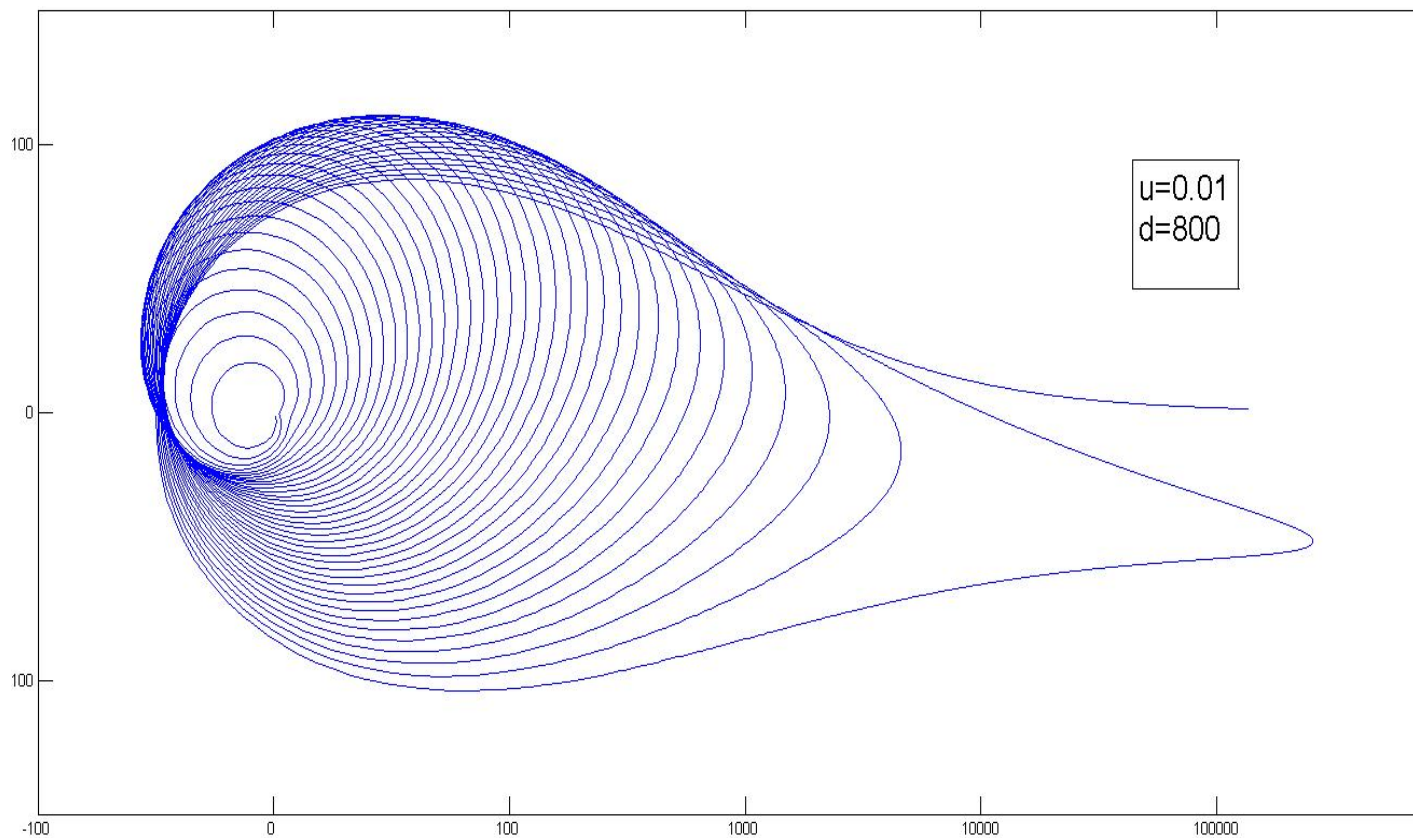

Рис. 10.1. Траектория захвата частицы. При столкновении с малыми энергиями частицы совершают множество оборотов вокруг центра масс, постепенно сближаясь из-за радиационных потерь энергии. Неэллиптическая форма траектории связна с выбранной логарифмической шкалой расстояний.

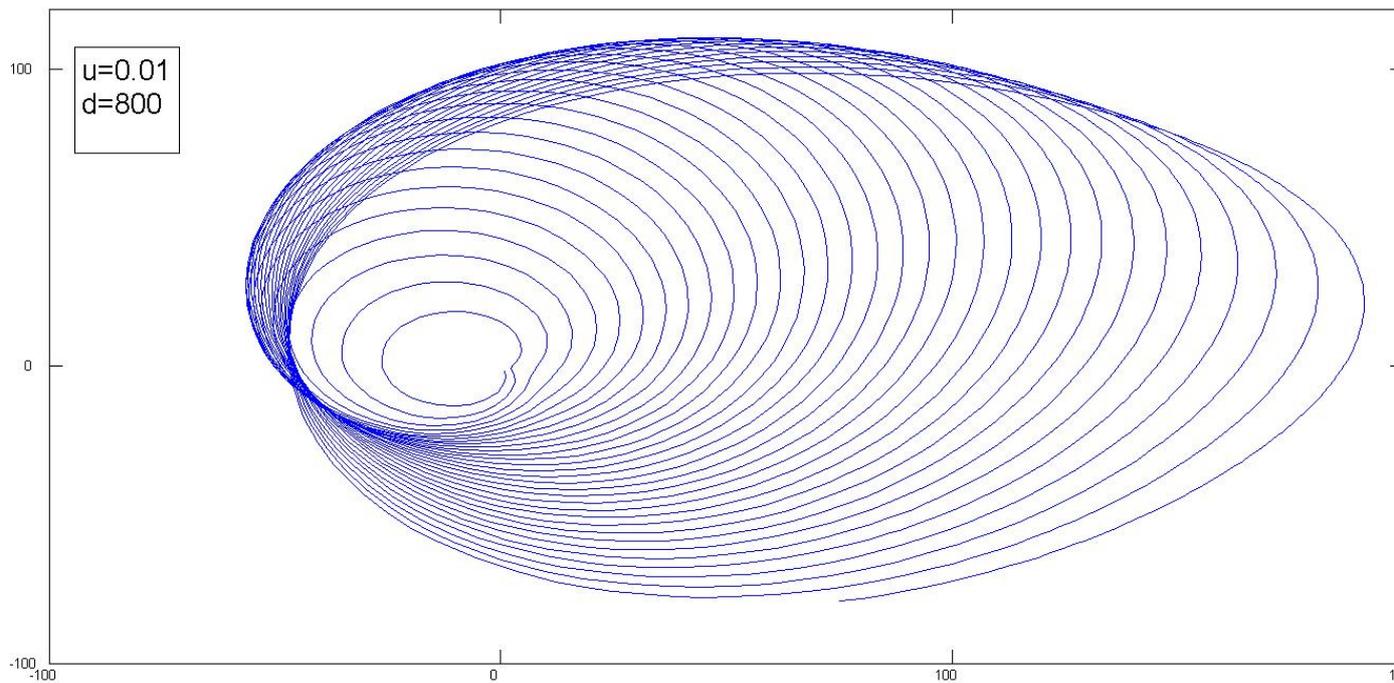

Рис. 10.2. Финальная стадия захвата частицы при тех же параметрах.

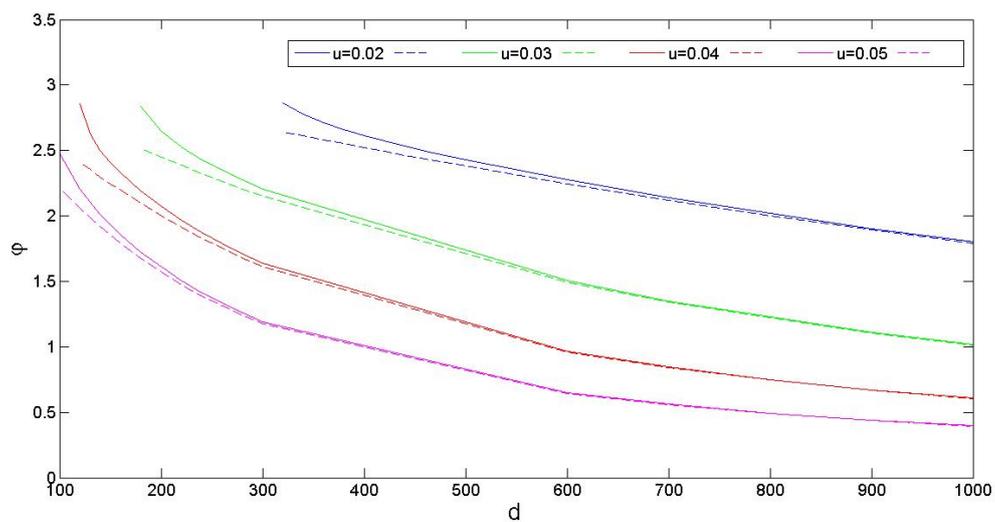

Рис. 10.3. Зависимость угла отклонения частицы в системе центра масс от прицельного расстояния при значениях 4-скорости $u_x = 0.02; 0.03; 0.04; 0.05$. Пунктиром показана соответствующая кривая, вычисленная по формуле Резерфорда.

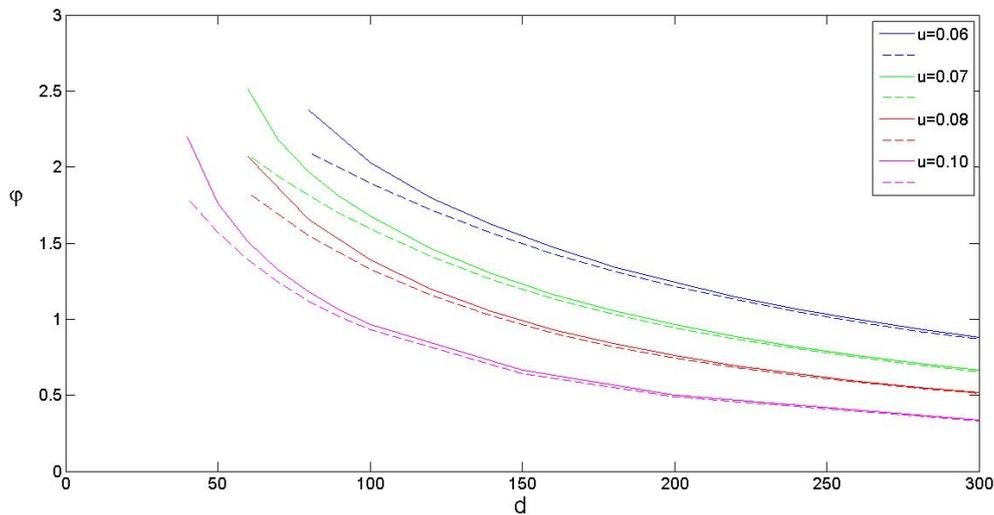

Рис. 10.4. Зависимость угла отклонения частицы в системе центра масс от прицельного расстояния при значениях 4-скорости $u_x = 0.06; 0.07; 0.08; 0.10$. Пунктиром показана соответствующая кривая, вычисленная по формуле Резерфорда.

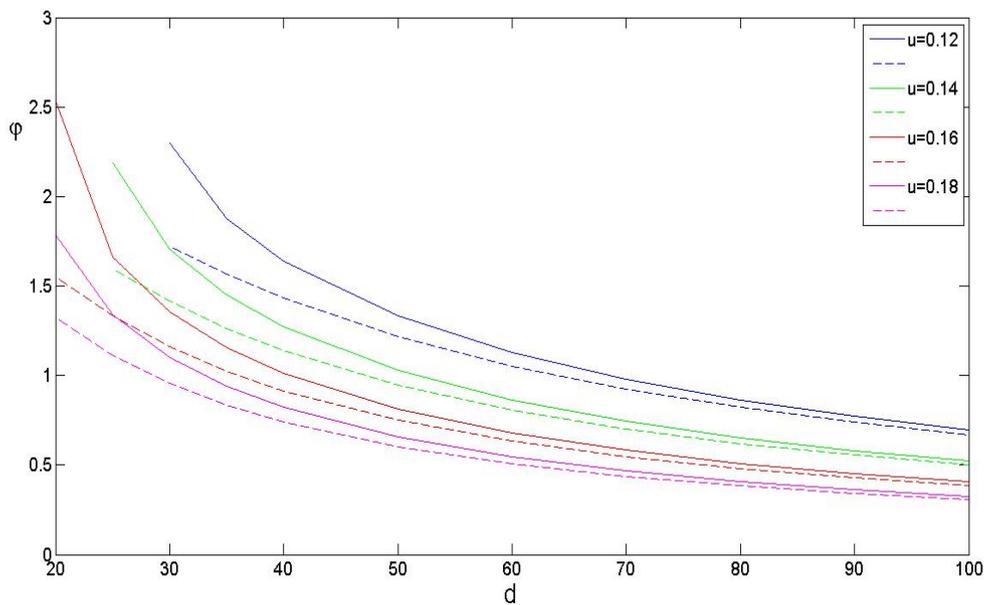

Рис. 10.5. Зависимость угла отклонения частицы в системе центра масс от прицельного расстояния при значениях 4-скорости $u_x = 0.12; 0.14; 0.16; 0.18$. Пунктиром показана соответствующая кривая, вычисленная по формуле Резерфорда.

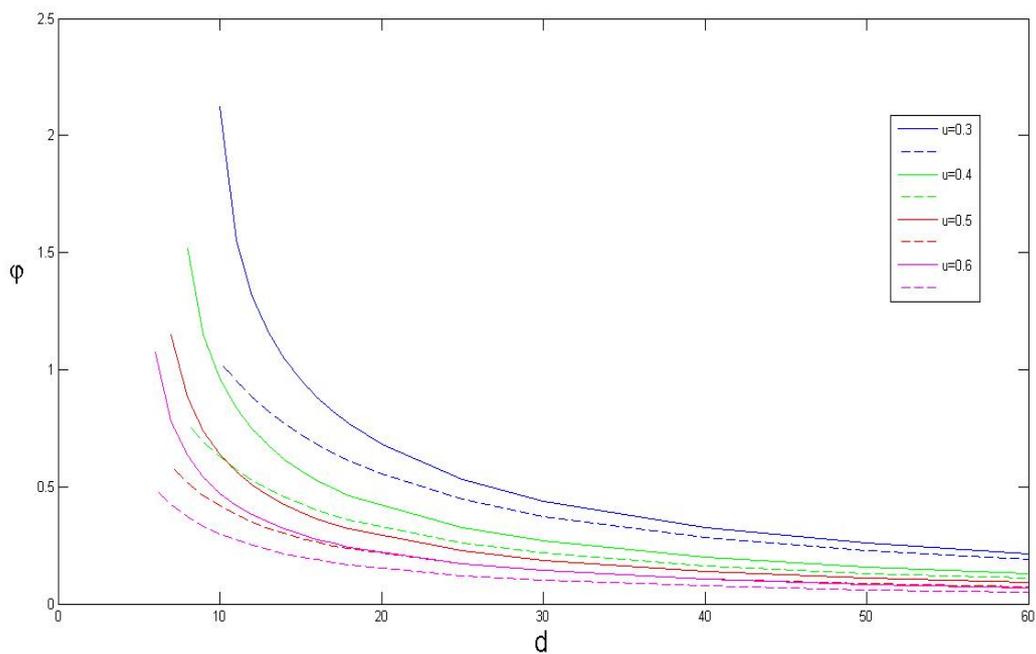

Рис. 10.6. Зависимость угла отклонения частицы в системе центра масс от прицельного расстояния при значениях 4-скорости $u_x = 0.3; 0.4; 0.5; 0.6$. Пунктиром показана соответствующая кривая, вычисленная по формуле Резерфорда.

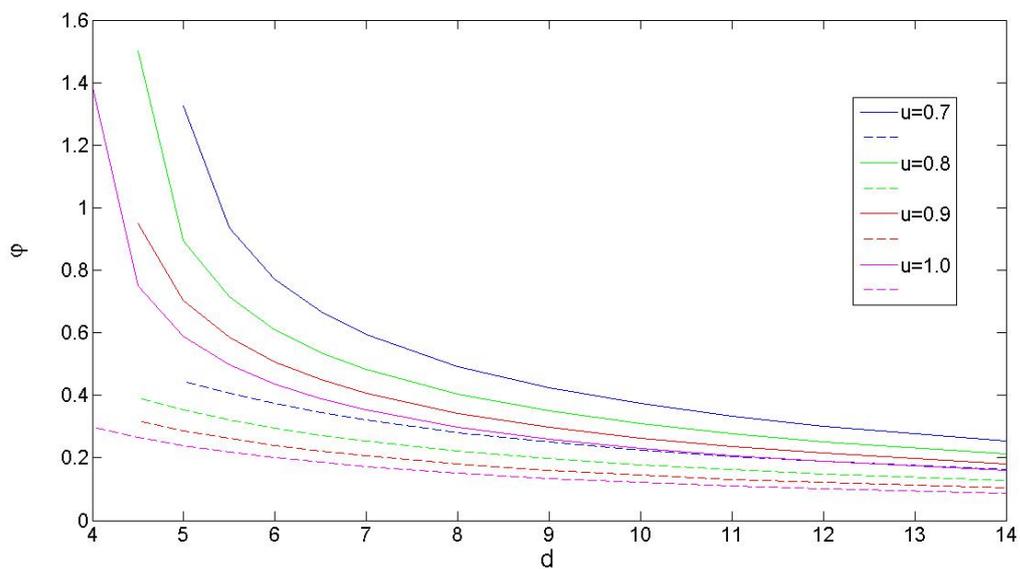

Рис. 10.7. Зависимость угла отклонения частицы от прицельного расстояния при значениях 4-скорости $u_x = 0.7; 0.8; 0.9; 1.0$. Пунктиром показана соответствующая кривая, вычисленная по формуле Резерфорда.

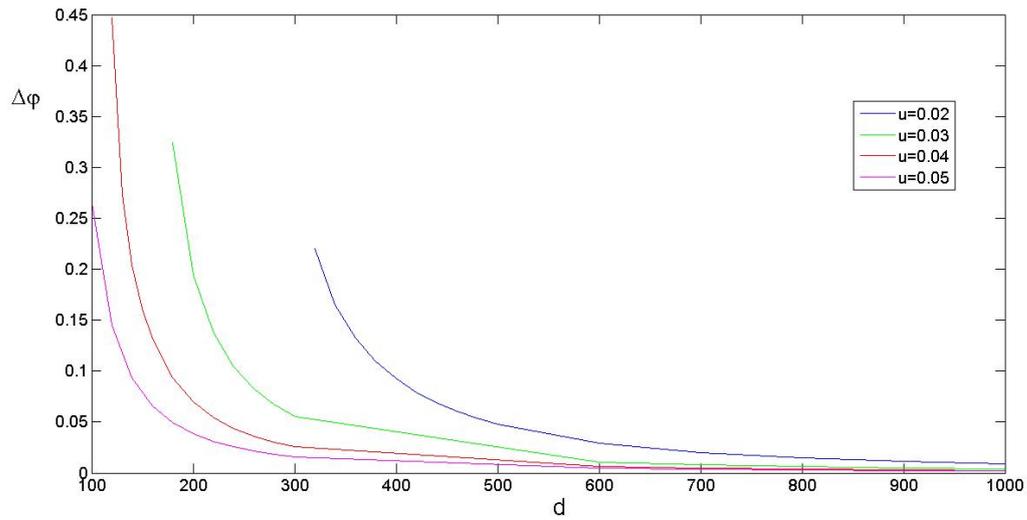

Рис. 10.8. Разность между рассчитанным углом отклонения и формулой Резерфорда в зависимости от прицельного расстояния при четырех значениях начальной скорости.

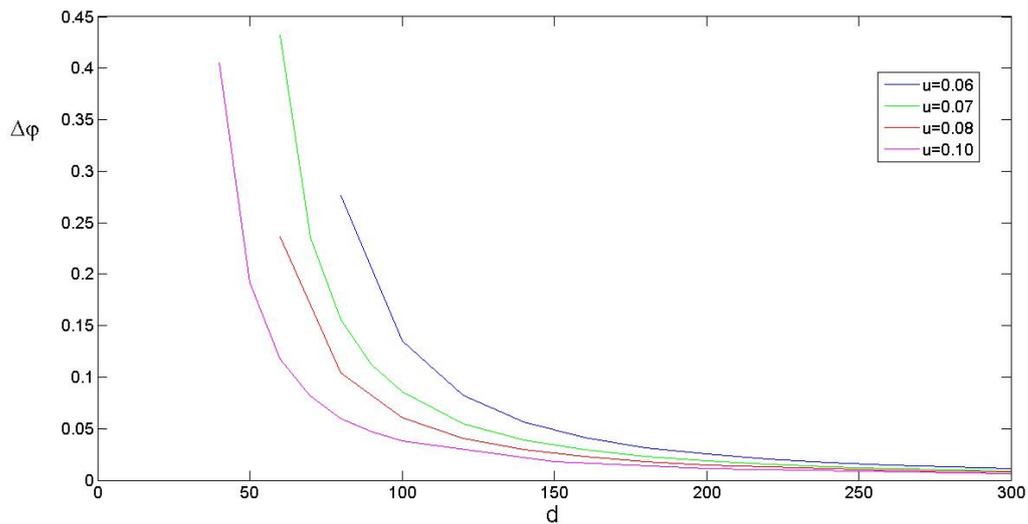

Рис. 10.9. Разность между рассчитанным углом отклонения и формулой Резерфорда в зависимости от прицельного расстояния при четырех значениях начальной скорости.

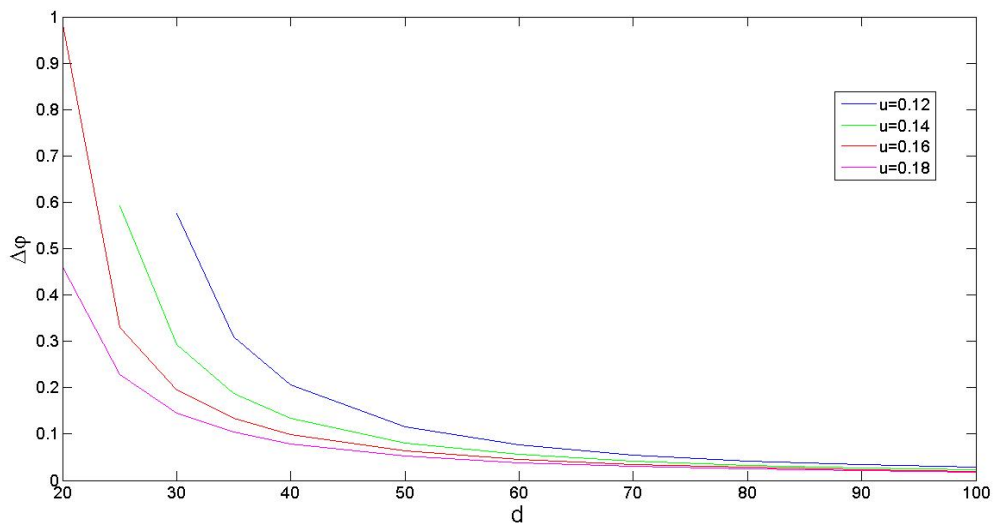

Рис. 10.10. Разность между рассчитанным углом отклонения и формулой Резерфорда в зависимости от прицельного расстояния при четырех значениях начальной скорости.

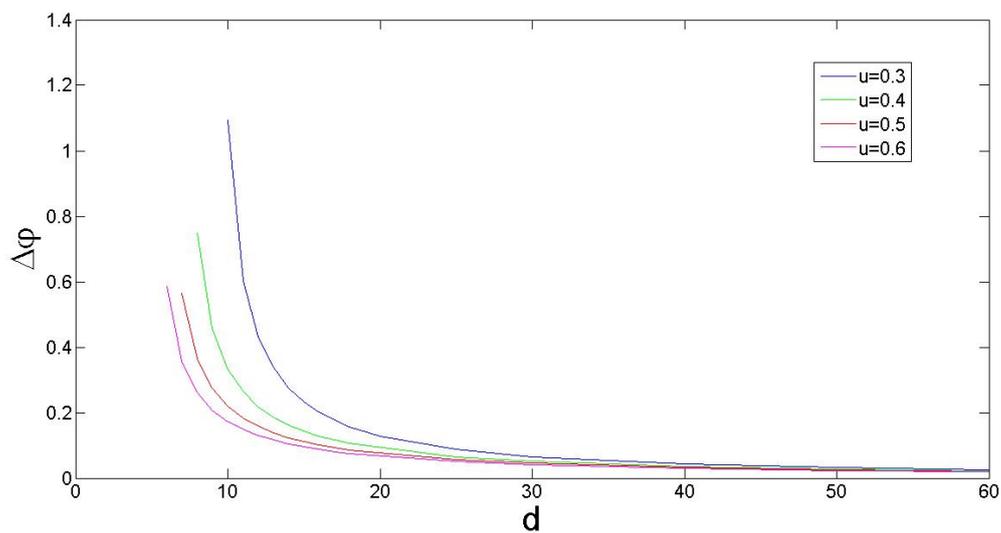

Рис. 10.11. Разность между рассчитанным углом отклонения и формулой Резерфорда в зависимости от прицельного расстояния при четырех значениях начальной скорости.

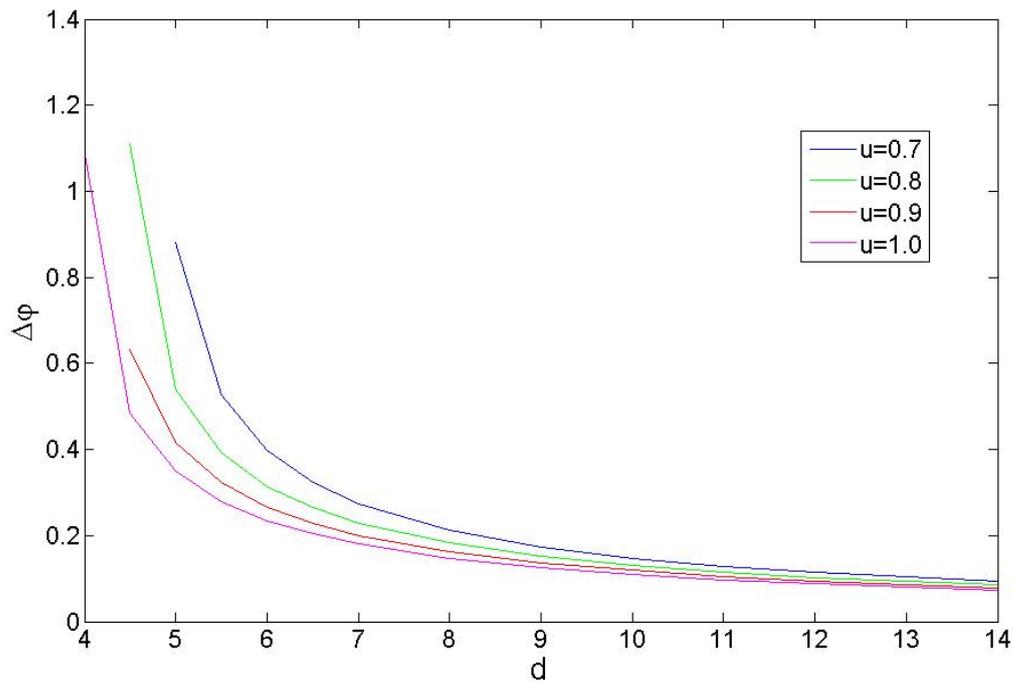

Рис. 10.12. Разность между рассчитанным углом отклонения и формулой Резерфорда в зависимости от прицельного расстояния при четырех значениях начальной скорости.

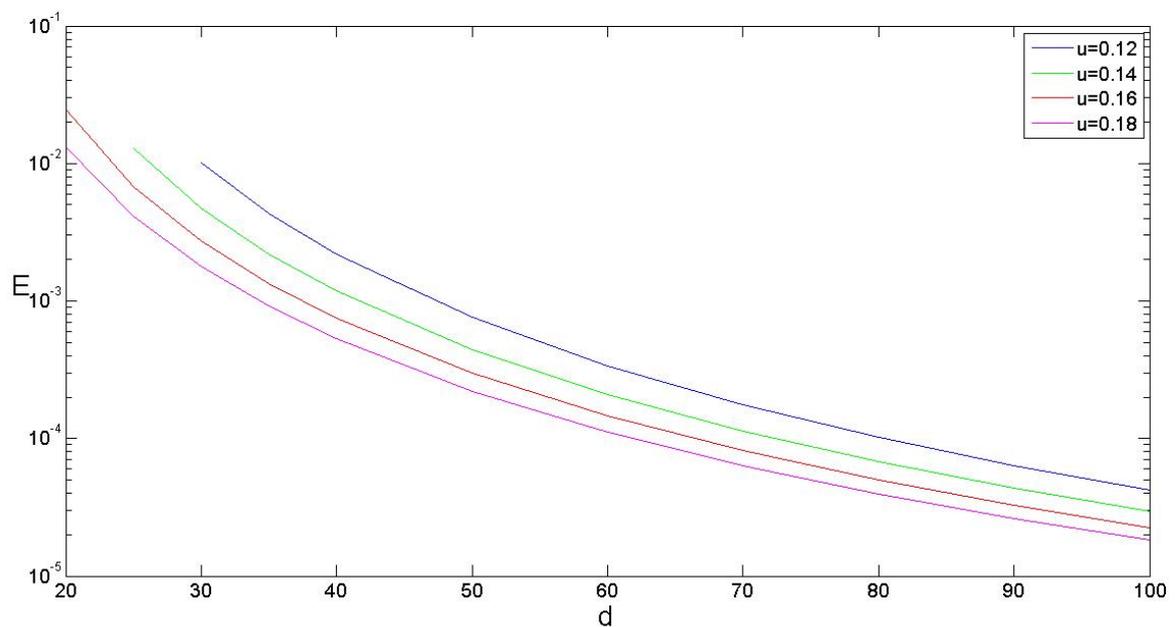

Рис. 10.13. Зависимость потерь энергии от прицельного расстояния при различных значениях начальной энергии.

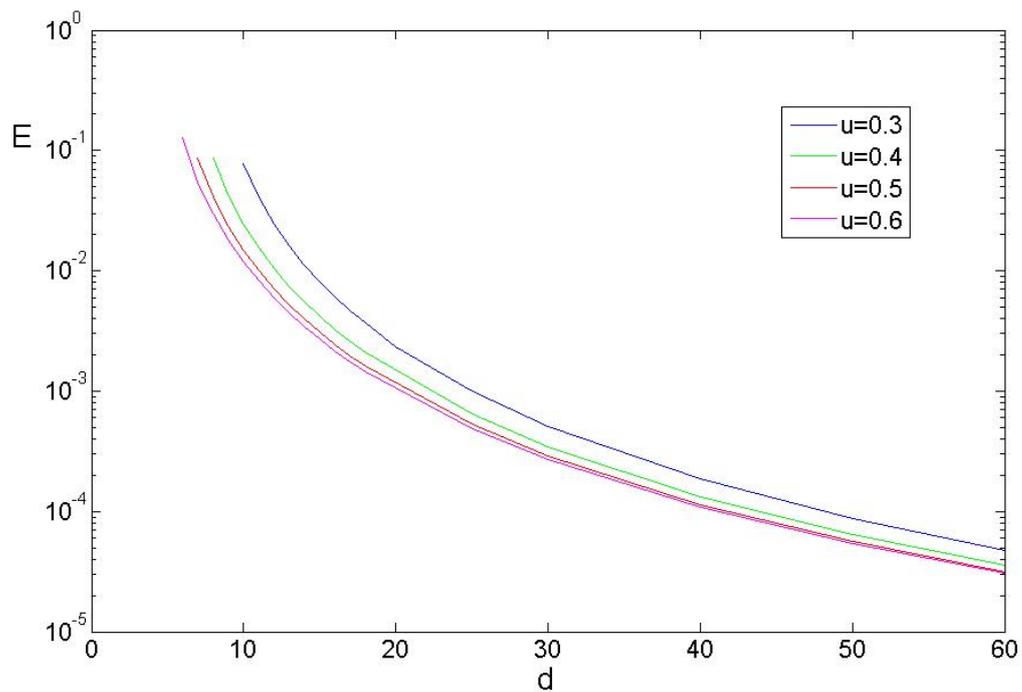

Рис. 10.14. Зависимость потерь энергии от прицельного расстояния при различных значениях начальной энергии.

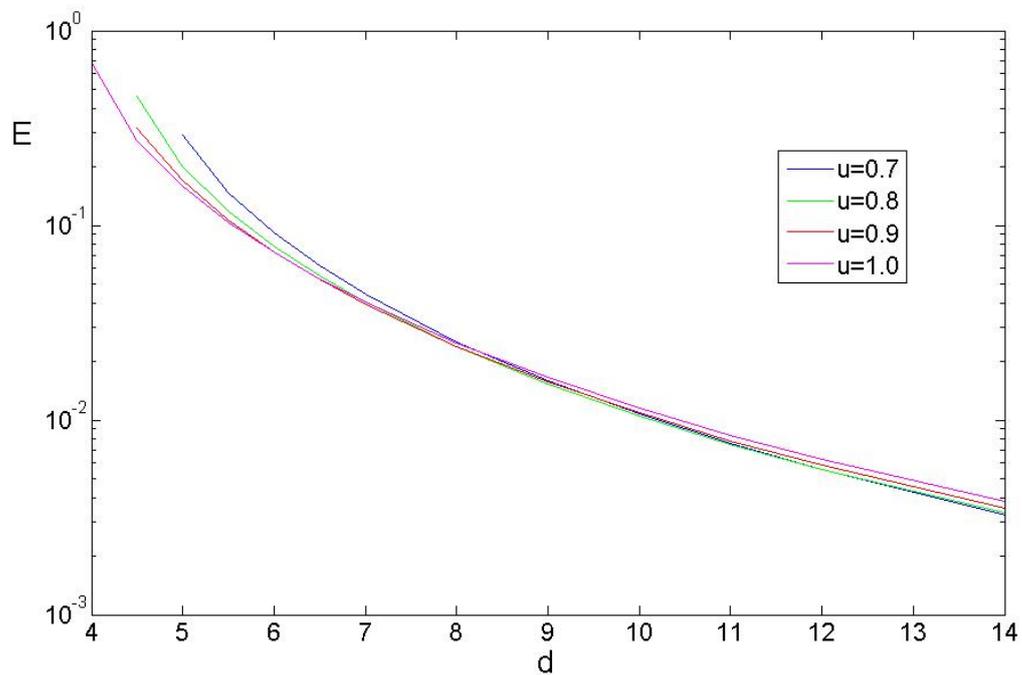

Рис. 10.15. Зависимость потерь энергии от прицельного расстояния при различных значениях начальной энергии.

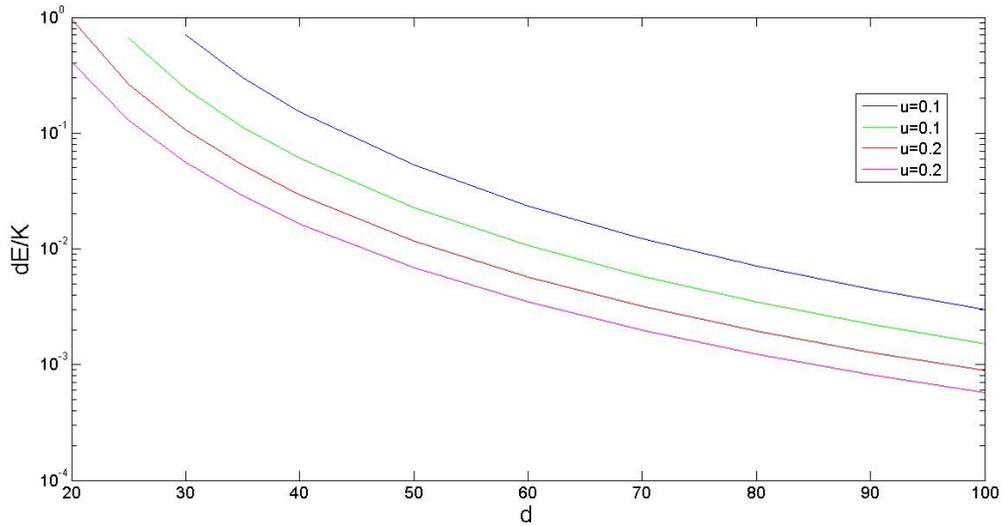

Рис. 10.16. Зависимость относительных потерь энергии от прицельного расстояния при различных значениях начальной энергии.

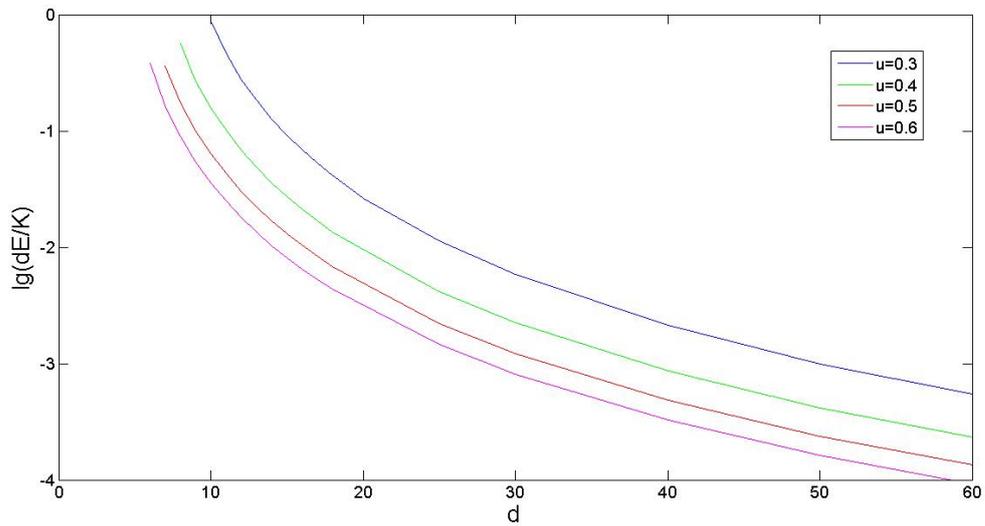

Рис. 10.17. Зависимость относительных потерь энергии от прицельного расстояния при различных значениях начальной энергии.

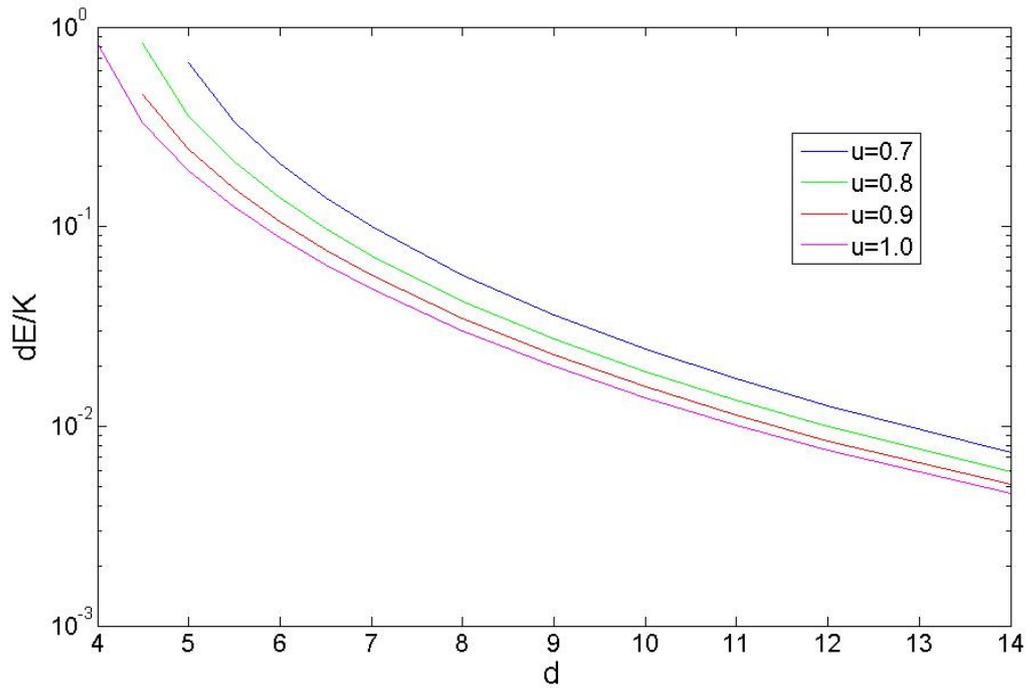

Рис. 10.18. Зависимость относительных потерь энергии от прицельного расстояния при различных значениях начальной энергии.

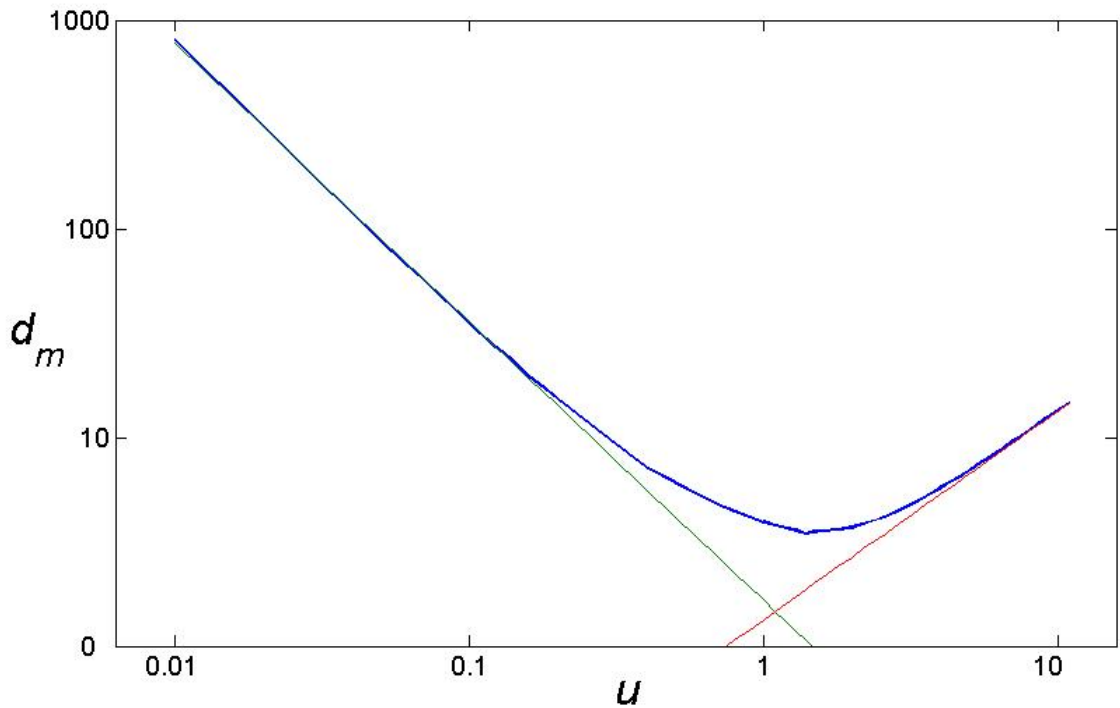

Рис. 10.19. Зависимость критического прицельного расстояния $d_m$, при котором происходит захват частицы, от начальной скорости $u$. По осям выбраны логарифмические шкалы. Асимптотические прямые задаются уравнениями –
$y = 0.222 - \frac{4}{3}x$ (зеленая прямая) и $y = x + 0.125$ (красная прямая).

Из рисунка[k] мы видим, что при малых скоростях сечение захвата частицы падает с ростом начальной скорости, а при достижении релятивистских скоростей начинает расти. При малых скоростях зависимость сечения от скорости может быть задана формулой:

(10.1) $$d_m = \frac{5}{3} \cdot u^{-\frac{4}{3}}.$$

На этом участке захват частицы определяется соотношением величин излучаемой энергии и исходной кинетической энергии сталкивающихся частиц. Значение (10.1) соответствует равенству этих величин.

Рост критического прицельного расстояния на релятивистском участке связан с уходом частицы под электромагнитный радиус $\rho$ (см. (5.15)) . При этом частицы уходят под электромагнитный радиус с положительными значениями полной энергии. Зависимость критического прицельного расстояния от скорости в системе центра масс задается приближенной формулой:

(10.2) $$d_m = \frac{4}{3}u.$$

**ЛИТЕРАТУРА**

**Приложение 1.**

$$e^{ikls} \cdot e_{pqrs} = -\delta_p^i \cdot \left(\delta_q^k \delta_r^l - \delta_q^l \delta_r^k\right) + \delta_p^k \cdot \left(\delta_q^i \delta_r^l - \delta_q^l \delta_r^i\right) + \delta_p^l \cdot \left(\delta_q^k \delta_r^i - \delta_q^i \delta_r^k\right)$$

$$\Theta^{nm} = \left(N^n - y^n\right) \cdot \left(N^m - y^m\right) \cdot \left(\left(y^k u_k\right)^2 - y^p y_p\right) + R_o^2 \cdot e^{iklm} u_k y_l \cdot e_{ipqr} u^p y^q g^{rn}$$

$$XE_r^m = e^{iklm} u_k y_l \cdot e_{ipqr} u^p y^q = e^{ikms} u_i y_k \cdot e_{pqrs} u^p y^q$$

$$XE_r^m = u_i y_k \cdot u^p y^q \cdot \left(-\delta_p^i \cdot \left(\delta_q^k \delta_r^m - \delta_q^m \delta_r^k\right) + \delta_p^k \cdot \left(\delta_q^i \delta_r^m - \delta_q^m \delta_r^i\right) + \delta_p^m \cdot \left(\delta_q^k \delta_r^i - \delta_q^i \delta_r^k\right)\right)$$

$$XE_r^m = -\left(y^q y_q \delta_r^m - y^m y_r\right) + u^p y_p \cdot \left(y^q u_q \delta_r^m - y^m u_r\right) + u^m \cdot \left(y^q y_q u_r - y^q u_q y_r\right)$$

$$XE_r^m = \delta_r^m \cdot \left(\left(y^q u_q\right)^2 - y^q y_q\right) + \left(y^m y_r - u^p y_p \cdot y^m u_r + u^m \cdot \left(y^q y_q u_r - y^q u_q y_r\right)\right)$$

$$XE_r^m = \delta_r^m \cdot \left(\left(y^q u_q\right)^2 - y^q y_q\right) + y^m \left(y_r - u^p y_p \cdot u_r\right) + u^m \cdot y^q \left(y_q u_r - u_q y_r\right)$$

$$XE_r^m g^{rn} = g^{mn} \cdot \left(\left(y^q u_q\right)^2 - y^q y_q\right) + y^m \left(y^n - u^p y_p \cdot u^n\right) + u^m \cdot y^q \left(y_q u^n - u_q y^n\right)$$

$$XE_r^m g^{rn} = g^{mn} \cdot \left(\left(y^q u_q\right)^2 - y^q y_q\right) + y^m y^n - u^p y_p \cdot \left(u^n y^m + y^n u^m\right) + u^m u^n \cdot y^q y_q$$

$$\Theta^{nm} = \left(N^n - y^n\right) \cdot \left(N^m - y^m\right) \cdot \left(\left(y^k u_k\right)^2 - y^p y_p\right) + R_o^2 \cdot \left(\left(y^k u_k\right)^2 - y^p y_p\right) \cdot g^{nm}$$
$$+ R_o^2 \cdot \left(y^m y^n - u^p y_p \cdot \left(u^n y^m + y^n u^m\right) + u^m u^n \cdot y^q y_q\right)$$

$$N^i = -\frac{y_m y^m \cdot \left(y^i - u^i \cdot y^k u_k\right)}{2 \cdot \left(\left(u_i y^i\right)^2 - y^i y_i\right)} + \frac{\rho \cdot \left(y^i \cdot y^k u_k - y_m y^m \cdot u^i\right)}{\left(\left(u_i y^i\right)^2 - y^i y_i\right)}$$

$$R_o = \sqrt{N^i N_i} = \sqrt{-y^p y_p \cdot \frac{y^m y_m - 4\rho \cdot y^m u_m + 4\rho^2}{4 \cdot \left(\left(y^k u_k\right)^2 - y^k y_k\right)}}$$

$$R_o^2 = N^i N_i = -y^p y_p \cdot \frac{y^m y_m - 4\rho \cdot y^m u_m + 4\rho^2}{4 \cdot \left(\left(y^k u_k\right)^2 - y^k y_k\right)}$$

$$N^i = \frac{-y_p y^p \cdot \left(y^i - u^i \cdot y^q u_q\right) + 2\rho \cdot \left(y^i \cdot y^q u_q - y_p y^p \cdot u^i\right)}{2 \cdot \left(\left(u_i y^i\right)^2 - y^i y_i\right)}$$

$$\Theta^{nm} = \left(XYY \cdot y^n y^m + XUY \cdot \left(u^n y^m + u^m y^n\right) + XUU \cdot u^n u^m + R_o^2 \cdot g^{nm}\right) \cdot \left(\left(u_i y^i\right)^2 - y^i y_i\right)$$

$$U = -y^i u_i \quad Y = y^i y_i$$

$$XYY = \left(\frac{-y_p y^p + 2\rho \cdot y^q u_q}{2 \cdot \left((u_i y^i)^2 - y^i y_i\right)} - 1\right) \cdot \left(\frac{-y_p y^p + 2\rho \cdot y^q u_q}{2 \cdot \left((u_i y^i)^2 - y^i y_i\right)} - 1\right) + R_o^2 \frac{1}{(y^k u_k)^2 - y^p y_p}$$

$$XYY = \left(\frac{Y + 2\rho \cdot U}{2 \cdot (U^2 - Y)} + 1\right) \cdot \left(\frac{Y + 2\rho \cdot U}{2 \cdot (U^2 - Y)} + 1\right) - Y \cdot \frac{Y + 4\rho \cdot U + 4\rho^2}{4 \cdot (U^2 - Y)^2}$$

$$XYY = \frac{Y^2 + 4\rho \cdot UY + 4\rho^2 U^2}{4 \cdot (U^2 - Y)^2} + \frac{Y + 2\rho \cdot U}{(U^2 - Y)} + 1 - Y \cdot \frac{Y + 4\rho \cdot U + 4\rho^2}{4 \cdot (U^2 - Y)^2}$$

$$XYY = \frac{4\rho^2 U^2 - 4\rho^2 Y}{4 \cdot (U^2 - Y)^2} + \frac{Y + 2\rho \cdot U}{(U^2 - Y)} + 1 = \frac{Y + 2\rho \cdot U + \rho^2}{(U^2 - Y)} + 1 = \frac{(\rho + U)^2}{(U^2 - Y)}$$

$$XUY \cdot (y^n u^m + y^m u^n) = \left(\frac{-y_p y^p \cdot y^n + 2\rho \cdot y^n \cdot y^q u_q}{2 \cdot \left((u_i y^i)^2 - y^i y_i\right)} - y^n\right) \cdot \left(\frac{y_p y^p \cdot u^m \cdot y^q u_q - 2\rho \cdot y_p y^p \cdot u^m}{2 \cdot \left((u_i y^i)^2 - y^i y_i\right)}\right)$$

$$+ \left(\frac{-y_p y^p \cdot y^m + 2\rho \cdot y^m \cdot y^q u_q}{2 \cdot \left((u_i y^i)^2 - y^i y_i\right)} - y^m\right) \cdot \left(\frac{y_p y^p \cdot u^n \cdot y^q u_q - 2\rho \cdot y_p y^p \cdot u^n}{2 \cdot \left((u_i y^i)^2 - y^i y_i\right)}\right)$$

$$+ R_o^2 \frac{-u^p y_p \cdot (u^n y^m + y^n u^m)}{(y^k u_k)^2 - y^p y_p}$$

$$XUY = \left(\frac{-y_p y^p + 2\rho \cdot y^q u_q}{2 \cdot \left((u_i y^i)^2 - y^i y_i\right)} - 1\right) \cdot \left(\frac{y_p y^p \cdot y^q u_q - 2\rho \cdot y_p y^p}{2 \cdot \left((u_i y^i)^2 - y^i y_i\right)}\right)$$

$$- y^p y_p \cdot \frac{y^m y_m - 4\rho \cdot y^m u_m + 4\rho^2}{4 \cdot \left((y^k u_k)^2 - y^k y_k\right)} \cdot \frac{-u^p y_p}{(y^k u_k)^2 - y^p y_p}$$

$$XUY = \left(\frac{-Y - 2\rho \cdot U}{2 \cdot (U^2 - Y)} - 1\right) \cdot \left(\frac{-YU - 2\rho \cdot Y}{2 \cdot (U^2 - Y)}\right) - YU \cdot \frac{Y + 4\rho \cdot U + 4\rho^2}{4 \cdot (U^2 - Y)^2}$$

$$XUY = \left(\frac{Y + 2\rho \cdot U}{2 \cdot (U^2 - Y)} + 1\right) \cdot \left(\frac{U + 2\rho}{2 \cdot (U^2 - Y)}\right) \cdot Y - YU \cdot \frac{Y + 4\rho \cdot U + 4\rho^2}{4 \cdot (U^2 - Y)^2}$$

$$XUY = \frac{YU + 2\rho \cdot U^2 + 2\rho Y + 4\rho^2 U}{4 \cdot (U^2 - Y)^2} Y + \frac{U + 2\rho}{2 \cdot (U^2 - Y)} Y - YU \cdot \frac{Y + 4\rho \cdot U + 4\rho^2}{4 \cdot (U^2 - Y)^2}$$

$$XUY = \frac{-2\rho \cdot U^2 + 2\rho Y}{4 \cdot (U^2 - Y)^2} Y + \frac{U + 2\rho}{2 \cdot (U^2 - Y)} Y$$

$$XUY = -\frac{\rho}{2 \cdot (U^2 - Y)} Y + \frac{U + 2\rho}{2 \cdot (U^2 - Y)} Y$$

$$XUY = \frac{(U + \rho) \cdot Y}{2 \cdot (U^2 - Y)}$$

$$N^i = \frac{-y_p y^p \cdot (y^i - u^i \cdot y^q u_q) + 2\rho \cdot (y^i \cdot y^q u_q - y_p y^p \cdot u^i)}{2 \cdot ((u_i y^i)^2 - y^i y_i)}$$

$$XUU = \left(\frac{y_p y^p \cdot y^q u_q - 2\rho \cdot y_p y^p}{2 \cdot ((u_i y^i)^2 - y^i y_i)}\right)^2 + R_o^2 \frac{y^q y_q}{(y^k u_k)^2 - y^p y_p}$$

$$XUU = \left(\frac{y_p y^p \cdot y^q u_q - 2\rho \cdot y_p y^p}{2 \cdot ((u_i y^i)^2 - y^i y_i)}\right)^2 - y^p y_p \cdot \frac{y^m y_m - 4\rho \cdot y^m u_m + 4\rho^2}{4 \cdot ((y^k u_k)^2 - y^k y_k)} \frac{y^q y_q}{(y^k u_k)^2 - y^p y_p}$$

$$XUU = Y^2 \frac{(U + 2\rho)^2}{4 \cdot (U^2 - Y)^2} - Y^2 \cdot \frac{Y + 4\rho \cdot U + 4\rho^2}{4 \cdot (U^2 - Y)^2}$$

$$XUU = Y^2 \frac{U^2 - Y}{4 \cdot (U^2 - Y)^2} = \frac{Y^2}{4 \cdot (U^2 - Y)}$$

$$\Theta^{nm} = \left(\frac{(\rho + U)^2}{(U^2 - Y)} \cdot y^n y^m + \frac{(U + \rho) \cdot Y}{2 \cdot (U^2 - Y)} \cdot (u^n y^m + u^m y^n) + \frac{Y^2}{4 \cdot (U^2 - Y)} \cdot u^n u^m + R_o^2 \cdot g^{nm}\right) \cdot (U^2 - Y)$$

$$\Theta^{nm} = (\rho + U)^2 \cdot y^n y^m + \frac{(U + \rho) \cdot Y}{2} \cdot (u^n y^m + u^m y^n) + \frac{Y^2}{4} \cdot u^n u^m + R_o^2 \cdot g^{nm} \cdot (U^2 - Y)$$

$$\Theta^{nm} = ((\rho + U) \cdot y^n + \tfrac{1}{2} Y u^n) \cdot ((\rho + U) \cdot y^m + \tfrac{1}{2} Y u^m) + R_o^2 \cdot g^{nm} \cdot (U^2 - Y)$$

$$\Theta^{nm} = ((\rho + U) \cdot y^n + \tfrac{1}{2} Y u^n) \cdot ((\rho + U) \cdot y^m + \tfrac{1}{2} Y u^m) - Y \cdot (\tfrac{1}{4} Y + \rho \cdot U + \rho^2) \cdot g^{nm}$$

$$\boxed{\begin{aligned}\Theta^{nm} &= ((\rho - y^p u_p) \cdot y^n + \tfrac{1}{2} y^p y_p u^n) \cdot ((\rho - y^q u_q) \cdot y^m + \tfrac{1}{2} y^q y_q u^m) \\ &\quad - y^p y_p \cdot (\tfrac{1}{4} y^q y_q - \rho \cdot y^q u_q + \rho^2) \cdot g^{nm}\end{aligned}}$$

# Приложение 2.

Величина $M_2^i$ вычисляется дифференцированием $L_1$:

(10.3) $$M_2^i = -\frac{\partial}{\partial \xi_i} L_1 = \frac{2\pi \cdot \Theta^{im} \xi_m}{\left(\Theta^{nm} \xi_m \xi_n\right)^{\frac{3}{2}}}$$

(10.4) $$M_2^i = 2\pi \cdot r_s^2 \cdot \frac{\tilde{N}^i \tilde{N}^m \xi_m - R_o^2 \cdot \lambda^i \lambda^m \xi_m}{\left(\Theta^{nm} \xi_m \xi_n\right)^{\frac{3}{2}}}$$

Для вычисления $L_2$ свернем $M_2^i$ с $u_i$ и поделим на $(\rho + U)$. Вычисления приводят к выражению[1]:

(10.5) $$L_2 = -2\pi \cdot \frac{\left(U^2 + \rho \cdot U - \frac{1}{2}Y\right) \cdot y^m \xi_m + \frac{1}{2}(U+2\rho) \cdot Y \cdot u^m \xi_m}{\left(\Theta^{nm} \xi_m \xi_n\right)^{\frac{3}{2}}}$$

(10.6) $$L_2 = 2\pi \cdot r_s^2 \cdot \frac{\tilde{N}^m \xi_m}{\left(\Theta^{nm} \xi_m \xi_n\right)^{\frac{3}{2}}}$$

Для вычисления $M_1^i$ подставим (9.2) в (7.2):

$$M_1^i = \frac{1}{r_s} \cdot \int_o \frac{\left(\tilde{N}^i + R_o \cdot \left(\lambda^i \cos\varphi + \mu^i \sin\varphi\right)\right) \cdot d\varphi}{\beta^m \xi_m}$$

Откуда видно, что

$$M_1^i = \tilde{N}^i \cdot L_1 + X \cdot \lambda^i$$

где величину $X$ определим из условия $M_1^i \xi_i = \frac{2\pi}{r_s}$:

$$X = \frac{1}{\lambda^i \xi_i} \cdot \left(\frac{2\pi}{r_s} - \tilde{N}^i \xi_i \cdot L_1\right) = A \cdot r_s \cdot \left(\frac{2\pi}{r_s} - \tilde{N}^i \xi_i \cdot L_1\right)$$

$$X = \frac{2\pi \cdot r_s}{\sqrt{\tilde{\Theta}^{nm} \xi_n \xi_m}} \cdot \left(\frac{1}{r_s} - \frac{\tilde{N}^i \xi_i}{\sqrt{\Theta^{nm} \xi_m \xi_n}}\right)$$

(10.7) $$X = \frac{2\pi}{\lambda^i \xi_i} \cdot \left(\frac{1}{r_s} - \frac{\tilde{N}^i \xi_i}{\sqrt{\Theta^{nm} \xi_n \xi_m}}\right)$$

$$X = \frac{R_o}{r_s} \cdot \int_o \frac{\cos\varphi \cdot d\varphi}{a + b \cdot \cos\varphi}$$

$$M_1^i = \frac{2\pi \cdot \tilde{N}^i}{\sqrt{\Theta^{nm} \xi_m \xi_n}} + \frac{2\pi}{\lambda^k \xi_k} \cdot \left(\frac{1}{r_s} - \frac{\tilde{N}^k \xi_k}{\sqrt{\Theta^{nm} \xi_m \xi_n}}\right) \cdot \lambda^i$$

$$M_1^i = \tilde{N}^i \cdot L_1 + \lambda^i \cdot X$$

Для вычисления $N_1^{ik} = \dfrac{1}{r_s} \cdot \displaystyle\int_o \dfrac{\beta^i \beta^k \cdot d\varphi}{(\beta^m \xi_m)}$, заметим

$$N_1^{ik} = \frac{1}{r_s} \cdot \int_o \frac{\left(\tilde{N}^i + R_o \cdot (\lambda^i \cos\varphi + \mu^i \sin\varphi)\right)\left(\tilde{N}^k + R_o \cdot (\lambda^k \cos\varphi + \mu^k \sin\varphi)\right) \cdot d\varphi}{(a + b \cdot \cos\varphi)}$$

$$N_1^{ik} = \frac{1}{r_s} \cdot \int_o \frac{\tilde{N}^i \tilde{N}^k \cdot d\varphi}{(a + b \cdot \cos\varphi)} + \frac{1}{r_s} \cdot \int_o \frac{\left(\tilde{N}^i \lambda^k + \tilde{N}^k \lambda^i\right) R_o \cdot \cos\varphi \cdot d\varphi}{(a + b \cdot \cos\varphi)}$$

$$+ \frac{1}{r_s} \cdot \int_o \frac{\left(R_o^2 \cdot (\lambda^i \lambda^k \cos^2\varphi + \mu^i \mu^k \sin^2\varphi)\right) \cdot d\varphi}{(a + b \cdot \cos\varphi)}$$

(10.8) $$N_1^{ik} = \tilde{N}^i \tilde{N}^k \cdot L_1 + \left(\tilde{N}^i \lambda^k + \tilde{N}^k \lambda^i\right) \cdot X + R_o^2 \cdot \left(\lambda^i \lambda^k \cdot (L_1 - X_1) + \mu^i \mu^k \cdot X_1\right)$$

Величина $X_1$ может быть определена из условия: $N_1^{ik} \xi_k = \dfrac{2\pi}{r_s} \cdot \tilde{N}^i$:

$$\frac{2\pi}{r_s} \cdot \tilde{N}^i = \tilde{N}^i \tilde{N}^k \xi_k \cdot L_1 + \left(\tilde{N}^i \lambda^k \xi_k + \tilde{N}^k \xi_k \lambda^i\right) \cdot X + R_o^2 \cdot \lambda^i \lambda^k \xi_k \cdot (L_1 - X_1)$$

Откуда

(10.9) $$X_1 = \frac{\tilde{N}^k \xi_k \cdot X}{R_o^2 \cdot \lambda^k \xi_k} + L_1$$

$$X_1 = 2\pi \cdot \frac{\tilde{N}^k \xi_k}{R_o^2 \cdot (\lambda^k \xi_k)^2} \cdot \left(\frac{1}{r_s} - \frac{\tilde{N}^i \xi_i}{\sqrt{\Theta^{nm} \xi_n \xi_m}}\right) + \frac{2\pi}{\sqrt{\Theta^{nm} \xi_n \xi_m}}$$

(10.10) $$\mu^i = \frac{e^{iklm} u_k y_l \xi_m}{\sqrt{\tilde{\Theta}^{nm} \xi_n \xi_m}}; \quad \lambda^i = \frac{e^{iklm} u_k y_l e_{mpqr} u^p y^q \xi^r}{r_s \cdot \sqrt{\tilde{\Theta}^{nm} \xi_n \xi_m}}$$

Для вычисления $N_2^{ik}$, заметим

$$N_2^{ik} = \frac{1}{r_s} \cdot \int_o \frac{\left(\tilde{N}^i + R_o \cdot (\lambda^i \cos\varphi + \mu^i \sin\varphi)\right)\left(\tilde{N}^k + R_o \cdot (\lambda^k \cos\varphi + \mu^k \sin\varphi)\right) \cdot d\varphi}{(a + b \cdot \cos\varphi)^2}$$

$$N_2^{ik} = \frac{1}{r_s} \cdot \int_o \frac{\tilde{N}^i \tilde{N}^k \cdot d\varphi}{(a + b \cdot \cos\varphi)^2} + \frac{R_o}{r_s} \cdot \int_o \frac{\left(\tilde{N}^i \lambda^k + \tilde{N}^k \lambda^i\right) \cdot \cos\varphi \cdot d\varphi}{(a + b \cdot \cos\varphi)^2} +$$

$$+ \frac{R_o^2}{r_s} \cdot \int_o \frac{\left(\lambda^i \lambda^k \cos^2\varphi + \mu^i \mu^k \sin^2\varphi\right) \cdot d\varphi}{(a + b \cdot \cos\varphi)^2}$$

$$N_2^{ik} = \tilde{N}^i \tilde{N}^k \cdot L_2 + \left(\tilde{N}^i \lambda^k + \tilde{N}^k \lambda^i\right) \cdot X_2 + X_\lambda \cdot \lambda^i \lambda^k + X_\mu \cdot \mu^i \mu^k$$

$$N_2^{ik} u_k = (\rho + U) \cdot M_2^i$$

$$(\rho + U) \cdot M_1^i = \tilde{N}^i \tilde{N}^k u_k \cdot L_2 + \tilde{N}^k u_k \lambda^i \cdot X_2$$

$$M_2^i = \tilde{N}^i \cdot L_2 + \lambda^i \cdot X_2$$

$$M_2^i = 2\pi \cdot r_s^2 \cdot \frac{\tilde{N}^i \tilde{N}^m \xi_m - R_o^2 \cdot \lambda^i \lambda^m \xi_m}{\left(\Theta^{nm} \xi_m \xi_n\right)^{\frac{3}{2}}}$$

$$X_2 = -2\pi \cdot r_s^2 \cdot \frac{R_o^2 \cdot \lambda^m \xi_m}{\left(\Theta^{nm} \xi_m \xi_n\right)^{\frac{3}{2}}}$$

$$N_2^{ik} \xi_k = M_1^i$$

$$M_1^i = \frac{2\pi \cdot \tilde{N}^i}{\sqrt{\Theta^{nm} \xi_m \xi_n}} + \frac{2\pi}{\lambda^k \xi_k} \cdot \left(\frac{1}{r_s} - \frac{\tilde{N}^k \xi_k}{\sqrt{\Theta^{nm} \xi_m \xi_n}}\right) \cdot \lambda^i$$

$$M_1^i = \tilde{N}^i \tilde{N}^k \xi_k \cdot L_2 + \left(\tilde{N}^i \lambda^k \xi_k + \tilde{N}^k \xi_k \lambda^i\right) \cdot X_2 + X_\lambda \cdot \lambda^i \lambda^k \xi_k$$

$$\tilde{N}^k \xi_k \cdot X_2 + X_\lambda \cdot \lambda^k \xi_k = \frac{2\pi}{\lambda^k \xi_k} \cdot \left(\frac{1}{r_s} - \frac{\tilde{N}^k \xi_k}{\sqrt{\Theta^{nm} \xi_m \xi_n}}\right)$$

$$-2\pi \cdot \frac{r_s^2 R_o^2 \cdot \lambda^m \xi_m}{\left(\Theta^{nm} \xi_m \xi_n\right)^{\frac{3}{2}}} \cdot \tilde{N}^k \xi_k + X_\lambda \cdot \lambda^k \xi_k = \frac{2\pi}{\lambda^k \xi_k} \cdot \left(\frac{1}{r_s} - \frac{\tilde{N}^k \xi_k}{\sqrt{\Theta^{nm} \xi_m \xi_n}}\right)$$

$$X_\lambda = \frac{2\pi}{\left(\lambda^k \xi_k\right)^2} \cdot \left(\frac{1}{r_s} - \frac{\tilde{N}^k \xi_k}{\sqrt{\Theta^{nm} \xi_m \xi_n}}\right) + 2\pi \cdot \frac{r_s^2 R_o^2}{\left(\Theta^{nm} \xi_m \xi_n\right)^{\frac{3}{2}}} \cdot \tilde{N}^k \xi_k$$

$$X_\lambda + X_\mu = R_o^2 \cdot L_2$$

$$X_\lambda = \frac{2\pi}{\left(\lambda^k \xi_k\right)^2} \cdot \left(\frac{1}{r_s} - \frac{\tilde{N}^k \xi_k}{\sqrt{\Theta^{nm} \xi_m \xi_n}}\right) + 2\pi \cdot \frac{r_s^2 R_o^2}{\left(\Theta^{nm} \xi_m \xi_n\right)^{\frac{3}{2}}} \cdot \tilde{N}^k \xi_k$$

$$L_2 = 2\pi \cdot r_s^2 \cdot \frac{\tilde{N}^m \xi_m}{\left(\Theta^{nm} \xi_m \xi_n\right)^{\frac{3}{2}}}$$

$$\frac{2\pi}{\left(\lambda^k \xi_k\right)^2} \cdot \left(\frac{1}{r_s} - \frac{\tilde{N}^k \xi_k}{\sqrt{\Theta^{nm} \xi_m \xi_n}}\right) + 2\pi \cdot \frac{r_s^2 R_o^2}{\left(\Theta^{nm} \xi_m \xi_n\right)^{\frac{3}{2}}} \cdot \tilde{N}^k \xi_k + X_\mu = R_o^2 \cdot 2\pi \cdot r_s^2 \cdot \frac{\tilde{N}^m \xi_m}{\left(\Theta^{nm} \xi_m \xi_n\right)^{\frac{3}{2}}}$$

$$X_\mu = -\frac{2\pi}{\left(\lambda^k \xi_k\right)^2} \cdot \left(\frac{1}{r_s} - \frac{\tilde{N}^k \xi_k}{\sqrt{\Theta^{nm} \xi_m \xi_n}}\right)$$

$$X = \frac{2\pi}{\lambda^i \xi_i} \cdot \left(\frac{1}{r_s} - \frac{\tilde{N}^i \xi_i}{\sqrt{\Theta^{nm} \xi_n \xi_m}}\right)$$

$$X_\mu = -\frac{X}{\left(\lambda^k \xi_k\right)}$$

$$X_\lambda = R_o^2 \cdot L_2 - X_\mu$$

$$\Theta^{nm} = \tilde{N}^n \tilde{N}^m \cdot r_s^2 - R_o^2 \cdot \tilde{\Theta}^{nm}$$

$$P_2^{ikl} = \frac{1}{r_s} \cdot \int_o \frac{\begin{pmatrix} \left(\tilde{N}^i + R_o \cdot \left(\lambda^i \cos\varphi + \mu^i \sin\varphi\right)\right)\left(\tilde{N}^k + R_o \cdot \left(\lambda^k \cos\varphi + \mu^k \sin\varphi\right)\right) \\ \times \left(\tilde{N}^l + R_o \cdot \left(\lambda^l \cos\varphi + \mu^l \sin\varphi\right)\right) \end{pmatrix} \cdot d\varphi}{\left(a + b \cdot \cos\varphi\right)^2}$$

$$P_2^{ikl} = \frac{1}{r_s} \cdot \int_o \frac{\tilde{N}^i \cdot \tilde{N}^k \cdot \tilde{N}^l \cdot d\varphi}{\left(a + b \cdot \cos\varphi\right)^2} + \frac{1}{r_s} \cdot \int_o \frac{\tilde{N}^i \cdot \tilde{N}^k \cdot \left(\beta^l - \tilde{N}^l\right) \cdot d\varphi}{\left(a + b \cdot \cos\varphi\right)^2} +$$

$$+\frac{1}{r_s} \cdot \int_o \frac{\tilde{N}^i \cdot \left(\beta^k - \tilde{N}^k\right) \cdot \tilde{N}^l \cdot d\varphi}{\left(a + b \cdot \cos\varphi\right)^2} + \frac{1}{r_s} \cdot \int_o \frac{\left(\beta^i - \tilde{N}^i\right) \cdot \tilde{N}^k \cdot \tilde{N}^l \cdot d\varphi}{\left(a + b \cdot \cos\varphi\right)^2}$$

$$+\frac{1}{r_s} \cdot \int_o \frac{\left(\beta^i - \tilde{N}^i\right) \cdot \left(\beta^k - \tilde{N}^k\right) \cdot \tilde{N}^l \cdot d\varphi}{\left(a + b \cdot \cos\varphi\right)^2} + \frac{1}{r_s} \cdot \int_o \frac{\left(\beta^i - \tilde{N}^i\right) \cdot \tilde{N}^k \cdot \left(\beta^l - \tilde{N}^l\right) \cdot d\varphi}{\left(a + b \cdot \cos\varphi\right)^2}$$

$$+\frac{1}{r_s} \cdot \int_o \frac{\tilde{N}^i \cdot \left(\beta^k - \tilde{N}^k\right) \cdot \left(\beta^l - \tilde{N}^l\right) \cdot d\varphi}{\left(a + b \cdot \cos\varphi\right)^2}$$

$$+\frac{1}{r_s} \cdot \int_o \frac{R_o^3 \cdot \left(\lambda^i \cos\varphi + \mu^i \sin\varphi\right) \cdot \left(\lambda^k \cos\varphi + \mu^k \sin\varphi\right) \cdot \left(\lambda^l \cos\varphi + \mu^l \sin\varphi\right) \cdot d\varphi}{\left(a + b \cdot \cos\varphi\right)^2}$$

$$P_2^{ikl} = \frac{1}{r_s} \cdot \int_o \frac{\tilde{N}^i \cdot \tilde{N}^k \cdot \tilde{N}^l \cdot d\varphi}{(a+b\cdot\cos\varphi)^2} + \frac{1}{r_s} \cdot \int_o \frac{\tilde{N}^i \cdot \beta^k \cdot (\beta^l - \tilde{N}^l) \cdot d\varphi}{(a+b\cdot\cos\varphi)^2} +$$

$$+ \frac{1}{r_s} \cdot \int_o \frac{\beta^i \cdot (\beta^k - \tilde{N}^k) \cdot \tilde{N}^l \cdot d\varphi}{(a+b\cdot\cos\varphi)^2} + \frac{1}{r_s} \cdot \int_o \frac{(\beta^i - \tilde{N}^i) \cdot \tilde{N}^k \cdot \beta^l \cdot d\varphi}{(a+b\cdot\cos\varphi)^2}$$

$$+ \frac{1}{r_s} \cdot \int_o \frac{R_o^3 \cdot (\lambda^i \cos\varphi + \mu^i \sin\varphi) \cdot (\lambda^k \cos\varphi + \mu^k \sin\varphi) \cdot (\lambda^l \cos\varphi + \mu^l \sin\varphi) \cdot d\varphi}{(a+b\cdot\cos\varphi)^2}$$

$$\boxed{P_2^{ikl} = \tilde{N}^i \cdot \tilde{N}^k \cdot \tilde{N}^l \cdot L_2 + \tilde{N}^i \cdot (N_2^{kl} - \tilde{N}^l \cdot M_2^k) + \tilde{N}^l \cdot (N_2^{ik} - \tilde{N}^k \cdot M_2^i) + \tilde{N}^k \cdot (N_2^{il} - \tilde{N}^i \cdot M_2^l) \\ + X_3 \cdot \lambda^i \lambda^k \lambda^l + X_4 \cdot (\lambda^i \mu^k \mu^l + \mu^i \lambda^k \mu^l + \mu^i \mu^k \lambda^l)}$$

Определим величины $X_3, X_4$

$$P_2^{ikl} g_{kl} = 0$$

$$P_2^{ikl} g_{kl} = \tilde{N}^i \cdot \tilde{N}^k \tilde{N}_k \cdot L_2 + \tilde{N}^i \cdot (-\tilde{N}_k \cdot M_2^k) + \tilde{N}_k \cdot (N_2^{ik} - \tilde{N}^k \cdot M_2^i) + \tilde{N}_l \cdot (N_2^{il} - \tilde{N}^i \cdot M_2^l)$$
$$- X_3 \cdot \lambda^i + X_4 \cdot (-\lambda^i)$$

$$P_2^{ikl} g_{kl} = \tilde{N}^i \cdot \tilde{N}^k \tilde{N}_k \cdot L_2 - 2 \cdot \tilde{N}^i \cdot \tilde{N}_k \cdot M_2^k - \tilde{N}_k \cdot \tilde{N}^k \cdot M_2^i + 2 \cdot \tilde{N}_l \cdot N_2^{il} - (X_3 + X_4) \lambda^i$$

$$(X_3 + X_4) = \tilde{N}_k \cdot \tilde{N}^k \cdot M_2^i \cdot \lambda_i - 2 \cdot \tilde{N}_l \cdot N_2^{il} \cdot \lambda_i$$

$$\tilde{N}_l N_2^{il} \cdot \lambda_i = -(\tilde{N}^l \tilde{N}_l) \cdot X_2$$

$$M_2^i = 2\pi \cdot r_s^2 \cdot \frac{\tilde{N}^i \tilde{N}^m \xi_m - R_o^2 \cdot \lambda^i \lambda^m \xi_m}{(\Theta^{nm} \xi_m \xi_n)^{\frac{3}{2}}}$$

$$M_2^i \lambda_i = 2\pi \cdot r_s^2 \cdot \frac{R_o^2 \cdot \lambda^m \xi_m}{(\Theta^{nm} \xi_m \xi_n)^{\frac{3}{2}}}$$

$$(X_3 + X_4) = \tilde{N}_k \cdot \tilde{N}^k \cdot \left( 2\pi \cdot r_s^2 \cdot \frac{R_o^2 \cdot \lambda^m \xi_m}{(\Theta^{nm} \xi_m \xi_n)^{\frac{3}{2}}} + 2 \cdot X_2 \right)$$

$$X_2 = -2\pi \cdot r_s^2 \cdot \frac{R_o^2 \cdot \lambda^m \xi_m}{(\Theta^{nm} \xi_m \xi_n)^{\frac{3}{2}}}$$

$$\boxed{(X_3 + X_4) = \tilde{N}_k \cdot \tilde{N}^k \cdot X_2}$$

$$P_2^{ikl} \xi_l = N_1^{ik}$$

$$P_2^{ikl} \cdot \xi_l = \tilde{N}^i \cdot \tilde{N}^k \cdot \tilde{N}^l \cdot \xi_l \cdot L_2 + \tilde{N}^i \cdot \left(N_2^{kl} - \tilde{N}^l \cdot M_2^k\right) \cdot \xi_l + \tilde{N}^l \cdot \left(N_2^{ik} - \tilde{N}^k \cdot M_2^i\right) \cdot \xi_l$$
$$+ \tilde{N}^k \cdot \left(N_2^{il} - \tilde{N}^i \cdot M_2^l\right) \cdot \xi_l + X_3 \cdot \lambda^i \lambda^k \lambda^l \cdot \xi_l + X_4 \cdot \mu^i \mu^k \lambda^l \cdot \xi_l$$

$$N_1^{ik} = \tilde{N}^i \tilde{N}^k \cdot L_1 + \left(\tilde{N}^i \lambda^k + \tilde{N}^k \lambda^i\right) \cdot X + R_o^2 \cdot \left(\lambda^i \lambda^k \cdot (L_1 - X_1) + \mu^i \mu^k \cdot X_1\right)$$

$$P_2^{ikl} \cdot \xi_l \mu_k = \tilde{N}^l \cdot N_2^{ik} \mu_k \cdot \xi_l - X_4 \cdot \mu^i \lambda^l \cdot \xi_l$$
$$P_2^{ikl} \cdot \xi_l \mu_k = -\tilde{N}^l \cdot X_\mu \cdot \mu^i \cdot \xi_l - X_4 \cdot \mu^i \lambda^l \cdot \xi_l$$
$$-R_o^2 \cdot X_1 = -\tilde{N}^l \cdot X_\mu \cdot \xi_l - X_4 \cdot \lambda^l \cdot \xi_l$$

$$X_4 = \frac{R_o^2 \cdot X_1 - \tilde{N}^l \cdot X_\mu \cdot \xi_l}{\lambda^l \cdot \xi_l}$$

$$X_3 = \tilde{N}_k \cdot \tilde{N}^k \cdot X_2 - X_4$$

Формулы для $N_3^{ik}, P_4^{ikl}, Q_5^{iklm}$ получаются непосредственно из (10.3) многократным дифференцированием по $\xi^i$:

$$N_3^{ik} = -\frac{1}{2}\frac{\partial}{\partial \xi_k} M_2^i = -\frac{1}{2}\left(\frac{2\pi \cdot \Theta^{ik}}{\left(\Theta^{nm}\xi_m\xi_n\right)^{\frac{3}{2}}} - 3 \cdot \frac{2\pi \cdot \Theta^{im}\xi_m \cdot \Theta^{kn}\xi_n}{\left(\Theta^{nm}\xi_m\xi_n\right)^{\frac{5}{2}}}\right)$$

$$N_3^{ik} = -\frac{1}{2}\frac{\partial}{\partial \xi_k} M_2^i = \frac{3}{2} \cdot \frac{2\pi \cdot \Theta^{im}\xi_m \cdot \Theta^{kn}\xi_n}{\left(\Theta^{nm}\xi_m\xi_n\right)^{\frac{5}{2}}} - \frac{1}{2}\frac{2\pi \cdot \Theta^{ik}}{\left(\Theta^{nm}\xi_m\xi_n\right)^{\frac{3}{2}}}$$

$$P_4^{ikl} = -\frac{1}{3}\frac{\partial}{\partial \xi_l} N_3^{ik} = -\frac{1}{3}\left(\frac{3}{2} \cdot \frac{2\pi \cdot \left(\Theta^{il}\Theta^{kn}\xi_n + \Theta^{kl}\Theta^{in}\xi_n + \Theta^{ik}\Theta^{ln}\xi_n\right)}{\left(\Theta^{nm}\xi_m\xi_n\right)^{\frac{5}{2}}} - \frac{15}{2}\frac{2\pi \cdot \Theta^{im}\xi_m \Theta^{kn}\xi_n \Theta^{lp}\xi_p}{\left(\Theta^{nm}\xi_m\xi_n\right)^{\frac{7}{2}}}\right)$$

$$P_4^{ikl} = -\frac{1}{2} \cdot \frac{2\pi \cdot \left(\Theta^{il}\Theta^{kn} + \Theta^{kl}\Theta^{in} + \Theta^{ik}\Theta^{ln}\right) \cdot \xi_n}{\left(\Theta^{nm}\xi_m\xi_n\right)^{\frac{5}{2}}} + \frac{5}{2}\frac{2\pi \cdot \Theta^{im}\xi_m \Theta^{kn}\xi_n \Theta^{lp}\xi_p}{\left(\Theta^{nm}\xi_m\xi_n\right)^{\frac{7}{2}}}$$

$$Q_5^{iklm} = -\frac{1}{4}\frac{\partial}{\partial \xi_m} P_4^{ikl}$$

$$Q_5^{iklm} = -\frac{1}{4} \cdot \left( -\frac{1}{2} \cdot \frac{2\pi \cdot \left(\Theta^{il}\Theta^{km} + \Theta^{kl}\Theta^{im} + \Theta^{ik}\Theta^{lm}\right)}{\left(\Theta^{nm}\xi_m\xi_n\right)^{\frac{5}{2}}} + \frac{5}{2} \cdot \frac{2\pi \cdot \left(\Theta^{il}\Theta^{kn} + \Theta^{kl}\Theta^{in} + \Theta^{ik}\Theta^{ln}\right) \cdot \xi_n \cdot \Theta^{mp}\xi_p}{\left(\Theta^{nm}\xi_m\xi_n\right)^{\frac{7}{2}}} \right)$$

$$-\frac{1}{4} \cdot \left( \frac{5}{2} \frac{2\pi \cdot \left(\Theta^{im}\Theta^{kq}\xi_q\Theta^{lp}\xi_p + \Theta^{iq}\xi_q\Theta^{km}\Theta^{lp}\xi_p + \Theta^{iq}\xi_q\Theta^{kp}\xi_p\Theta^{lm}\right)}{\left(\Theta^{nm}\xi_m\xi_n\right)^{\frac{7}{2}}} - \frac{35}{2} \frac{2\pi \cdot \Theta^{iq}\xi_q\Theta^{kn}\xi_n\Theta^{lp}\xi_p\Theta^{ms}\xi_s}{\left(\Theta^{nm}\xi_m\xi_n\right)^{\frac{9}{2}}} \right)$$

$$Q_5^{iklm} = 2\pi \cdot \left( \frac{1}{8} \cdot \frac{\left(\Theta^{il}\Theta^{km} + \Theta^{kl}\Theta^{im} + \Theta^{ik}\Theta^{lm}\right)}{\left(\Theta^{nm}\xi_m\xi_n\right)^{\frac{5}{2}}} - \frac{5}{8} \cdot \frac{\left(\Theta^{il}\Theta^{kn} + \Theta^{kl}\Theta^{in} + \Theta^{ik}\Theta^{ln}\right) \cdot \xi_n \cdot \Theta^{mp}\xi_p}{\left(\Theta^{nm}\xi_m\xi_n\right)^{\frac{7}{2}}} \right)$$

$$+2\pi \cdot \left( -\frac{5}{8} \frac{\left(\Theta^{im}\Theta^{kq}\Theta^{lp} + \Theta^{iq}\Theta^{km}\Theta^{lp} + \Theta^{iq}\Theta^{kp}\Theta^{lm}\right)\xi_q\xi_p}{\left(\Theta^{nm}\xi_m\xi_n\right)^{\frac{7}{2}}} + \frac{35}{8} \frac{\Theta^{iq}\xi_q\Theta^{kn}\xi_n\Theta^{lp}\xi_p\Theta^{ms}\xi_s}{\left(\Theta^{nm}\xi_m\xi_n\right)^{\frac{9}{2}}} \right)$$

$$\boxed{\begin{aligned}Q_5^{iklm} = 2\pi \cdot &\left( \frac{1}{8} \cdot \frac{\left(\Theta^{il}\Theta^{km} + \Theta^{kl}\Theta^{im} + \Theta^{ik}\Theta^{lm}\right)}{\left(\Theta^{nm}\xi_m\xi_n\right)^{\frac{5}{2}}} + \frac{35}{8} \frac{\Theta^{iq}\xi_q\Theta^{kn}\xi_n\Theta^{lp}\xi_p\Theta^{ms}\xi_s}{\left(\Theta^{nm}\xi_m\xi_n\right)^{\frac{9}{2}}} \right) \\ &-\frac{5\pi}{4} \frac{\left(\Theta^{im}\Theta^{kq}\Theta^{lp} + \Theta^{iq}\Theta^{km}\Theta^{lp} + \Theta^{iq}\Theta^{kp}\Theta^{lm} + \Theta^{il}\Theta^{kp}\Theta^{mq} + \Theta^{kl}\Theta^{ip}\Theta^{mq} + \Theta^{ik}\Theta^{lp}\Theta^{mq}\right)\xi_q\xi_p}{\left(\Theta^{nm}\xi_m\xi_n\right)^{\frac{7}{2}}}\end{aligned}}$$

Уменьшаем количество индексов, сворачивая с $u_i$ и деля на $(\rho + U)$, поскольку $\beta^i u_i = \rho + U$.

$$\Theta^{nm} u_m = \left(\rho - y^p u_p\right) \cdot \Upsilon^n$$

Где через $\Upsilon^n$ обозначено следующее выражение:

$$\Upsilon^n = y^n \cdot \left(\left(\rho - y^q u_q\right) \cdot y^m u_m + \tfrac{1}{2} y^q y_q\right) + u^n \cdot y^p y_p \cdot \left(\tfrac{1}{2} y^m u_m - \rho\right)$$

$$\Upsilon^n u_n = \left(\rho - y^q u_q\right) \cdot r_s^2.$$

$$P_5^{ikl} = 2\pi \cdot \left( \frac{1}{8} \cdot \frac{\left(\Theta^{il}\Upsilon^k + \Theta^{kl}\Upsilon^i + \Theta^{ik}\Upsilon^l\right)}{\left(\Theta^{nm}\xi_m\xi_n\right)^{\frac{5}{2}}} + \frac{35}{8} \frac{\Theta^{iq}\xi_q\Theta^{kn}\xi_n\Theta^{lp}\xi_p\Upsilon^s\xi_s}{\left(\Theta^{nm}\xi_m\xi_n\right)^{\frac{9}{2}}} \right)$$

$$-\frac{5\pi}{4} \frac{\left(\Upsilon^i\Theta^{kq}\Theta^{lp} + \Theta^{iq}\Upsilon^k\Theta^{lp} + \Theta^{iq}\Theta^{kp}\Upsilon^l + \Theta^{il}\Theta^{kp}\Upsilon^q + \Theta^{kl}\Theta^{ip}\Upsilon^q + \Theta^{ik}\Theta^{lp}\Upsilon^q\right)\xi_q\xi_p}{\left(\Theta^{nm}\xi_m\xi_n\right)^{\frac{7}{2}}}$$

$$N_5^{ik} = 2\pi \cdot \left( \frac{1}{8} \cdot \frac{\left(\Upsilon^i \Upsilon^k + \Upsilon^k \Upsilon^i + \Theta^{ik} \cdot r_s^2\right)}{\left(\Theta^{nm}\xi_m\xi_n\right)^{\frac{5}{2}}} + \frac{35}{8} \frac{\Theta^{iq}\xi_q \Theta^{kn}\xi_n \Upsilon^p \xi_p \Upsilon^s \xi_s}{\left(\Theta^{nm}\xi_m\xi_n\right)^{\frac{9}{2}}} \right)$$

$$-\frac{5\pi}{4} \frac{\left(\Upsilon^i \Theta^{kq} \Upsilon^p + \Theta^{iq}\Upsilon^k \Upsilon^p + \Theta^{iq}\Theta^{kp} \cdot r_s^2 + \Upsilon^i \Theta^{kp}\Upsilon^q + \Upsilon^k \Theta^{ip}\Upsilon^q + \Theta^{ik}\Upsilon^p\Upsilon^q\right)\xi_q \xi_p}{\left(\Theta^{nm}\xi_m\xi_n\right)^{\frac{7}{2}}}$$

$$\boxed{N_5^{ik} = 2\pi \cdot \left( \frac{1}{8} \cdot \frac{\left(\Upsilon^i \Upsilon^k + \Upsilon^k \Upsilon^i + \Theta^{ik} \cdot r_s^2\right)}{\left(\Theta^{nm}\xi_m\xi_n\right)^{\frac{5}{2}}} + \frac{35}{8} \frac{\Theta^{iq}\xi_q \Theta^{kn}\xi_n \Upsilon^p \xi_p \Upsilon^s \xi_s}{\left(\Theta^{nm}\xi_m\xi_n\right)^{\frac{9}{2}}} \right)}$$

$$\boxed{-\frac{5\pi}{4} \frac{\left(2\cdot\Upsilon^i \Theta^{kq} \Upsilon^p + 2\cdot\Upsilon^k \Theta^{ip}\Upsilon^q + \Theta^{iq}\Theta^{kp} \cdot r_s^2 + \Theta^{ik}\Upsilon^p\Upsilon^q\right)\xi_q \xi_p}{\left(\Theta^{nm}\xi_m\xi_n\right)^{\frac{7}{2}}}}$$

$$M_5^i = 2\pi \cdot \left( \frac{1}{8} \cdot \frac{\left(\Upsilon^i \cdot r_s^2 + \Upsilon^i \cdot r_s^2 + \Upsilon^i \cdot r_s^2\right)}{\left(\Theta^{nm}\xi_m\xi_n\right)^{\frac{5}{2}}} + \frac{35}{8} \frac{\Theta^{iq}\xi_q \Upsilon^n \xi_n \Upsilon^p \xi_p \Upsilon^s \xi_s}{\left(\Theta^{nm}\xi_m\xi_n\right)^{\frac{9}{2}}} \right)$$

$$-\frac{5\pi}{4} \frac{\left(2\cdot\Upsilon^i \Upsilon^q \Upsilon^p + 2\cdot\Theta^{ip}\Upsilon^q \cdot r_s^2 + \Theta^{iq}\Upsilon^p \cdot r_s^2 + \Upsilon^i \Upsilon^p \Upsilon^q\right)\xi_q \xi_p}{\left(\Theta^{nm}\xi_m\xi_n\right)^{\frac{7}{2}}}$$

$$M_5^i = 2\pi \cdot \left( \frac{3}{8} \cdot \frac{\Upsilon^i \cdot r_s^2}{\left(\Theta^{nm}\xi_m\xi_n\right)^{\frac{5}{2}}} + \frac{35}{8} \frac{\Theta^{iq}\xi_q \Upsilon^n \xi_n \Upsilon^p \xi_p \Upsilon^s \xi_s}{\left(\Theta^{nm}\xi_m\xi_n\right)^{\frac{9}{2}}} \right) - \frac{15\pi}{4} \frac{\left(\Upsilon^i \Upsilon^q \Upsilon^p + \Theta^{ip}\Upsilon^q \cdot r_s^2\right)\xi_q \xi_p}{\left(\Theta^{nm}\xi_m\xi_n\right)^{\frac{7}{2}}}$$

$$M_5^i = \frac{\pi}{4} \cdot \left( \frac{3\cdot\Upsilon^i \cdot r_s^2}{\left(\Theta^{nm}\xi_m\xi_n\right)^{\frac{5}{2}}} - \frac{15\cdot\left(\Upsilon^i \Upsilon^q \Upsilon^p + \Theta^{ip}\Upsilon^q \cdot r_s^2\right)\xi_q \xi_p}{\left(\Theta^{nm}\xi_m\xi_n\right)^{\frac{7}{2}}} + \frac{35\cdot\Theta^{iq}\xi_q \Upsilon^n \xi_n \Upsilon^p \xi_p \Upsilon^s \xi_s}{\left(\Theta^{nm}\xi_m\xi_n\right)^{\frac{9}{2}}} \right)$$

$$\boxed{M_5^i = \frac{\pi}{4} \cdot \left( \frac{3\cdot\Upsilon^i \cdot r_s^2}{\left(\Theta^{nm}\xi_m\xi_n\right)^{\frac{5}{2}}} - \frac{15\cdot\left(\Upsilon^i \Upsilon^q \Upsilon^p + \Theta^{ip}\Upsilon^q \cdot r_s^2\right)\xi_q \xi_p}{\left(\Theta^{nm}\xi_m\xi_n\right)^{\frac{7}{2}}} + \frac{35\cdot\Theta^{iq}\xi_q \left(\Upsilon^n \xi_n\right)^3}{\left(\Theta^{nm}\xi_m\xi_n\right)^{\frac{9}{2}}} \right)}$$

$$\boxed{N_4^{ik} = -\frac{1}{2} \cdot \frac{2\pi \cdot \left(\Upsilon^i \Theta^{kn} + \Upsilon^k \Theta^{in} + \Theta^{ik}\Upsilon^n\right)\cdot\xi_n}{\left(\Theta^{nm}\xi_m\xi_n\right)^{\frac{5}{2}}} + \frac{5}{2} \frac{2\pi \cdot \Theta^{im}\xi_m \Theta^{kn}\xi_n \cdot \Upsilon^p \xi_p}{\left(\Theta^{nm}\xi_m\xi_n\right)^{\frac{7}{2}}}}$$

$$M_4^i = -\frac{1}{2} \cdot \frac{2\pi \cdot \left(\Upsilon^i \Upsilon^n + \Theta^{in}\cdot r_s^2 + \Upsilon^i \Upsilon^n\right)\cdot\xi_n}{\left(\Theta^{nm}\xi_m\xi_n\right)^{\frac{5}{2}}} + \frac{5}{2} \frac{2\pi \cdot \Theta^{im}\xi_m \cdot \left(\Upsilon^p \xi_p\right)^2}{\left(\Theta^{nm}\xi_m\xi_n\right)^{\frac{7}{2}}}$$

$$M_4^i = -\frac{1}{2} \cdot \frac{2\pi \cdot \left(2 \cdot \Upsilon^i \Upsilon^n + \Theta^{in} \cdot r_s^2\right) \cdot \xi_n}{\left(\Theta^{nm}\xi_m\xi_n\right)^{\frac{5}{2}}} + \frac{5}{2} \frac{2\pi \cdot \Theta^{im}\xi_m \cdot \left(\Upsilon^p \xi_p\right)^2}{\left(\Theta^{nm}\xi_m\xi_n\right)^{\frac{7}{2}}}$$

$$L_4 = -\frac{1}{2} \cdot \frac{2\pi \cdot \left(2 \cdot \Upsilon^n \cdot r_s^2 + \Upsilon^n \cdot r_s^2\right) \cdot \xi_n}{\left(\Theta^{nm}\xi_m\xi_n\right)^{\frac{5}{2}}} + \frac{5}{2} \frac{2\pi \cdot \Upsilon^m \xi_m \cdot \left(\Upsilon^p \xi_p\right)^2}{\left(\Theta^{nm}\xi_m\xi_n\right)^{\frac{7}{2}}}$$

$$L_4 = -3\pi \cdot \frac{\Upsilon^n \xi_n \cdot r_s^2}{\left(\Theta^{nm}\xi_m\xi_n\right)^{\frac{5}{2}}} + 5\pi \cdot \frac{\left(\Upsilon^p \xi_p\right)^3}{\left(\Theta^{nm}\xi_m\xi_n\right)^{\frac{7}{2}}}$$

$$M_3^i = \frac{3}{2} \cdot \frac{2\pi \cdot \Theta^{im}\xi_m \cdot \Upsilon^n \xi_n}{\left(\Theta^{nm}\xi_m\xi_n\right)^{\frac{5}{2}}} - \frac{1}{2} \frac{2\pi \cdot \Upsilon^i}{\left(\Theta^{nm}\xi_m\xi_n\right)^{\frac{3}{2}}}$$

$$L_3 = \frac{3}{2} \cdot \frac{2\pi \cdot \left(\Upsilon^m \xi_m\right)^2}{\left(\Theta^{nm}\xi_m\xi_n\right)^{\frac{5}{2}}} - \frac{1}{2} \frac{2\pi \cdot r_s^2}{\left(\Theta^{nm}\xi_m\xi_n\right)^{\frac{3}{2}}}$$

$$L_3 = 3\pi \cdot \frac{\left(\Upsilon^m \xi_m\right)^2}{\left(\Theta^{nm}\xi_m\xi_n\right)^{\frac{5}{2}}} - \pi \cdot \frac{r_s^2}{\left(\Theta^{nm}\xi_m\xi_n\right)^{\frac{3}{2}}}$$

**Приложение 3.**

$$B_\chi = -\frac{e_u}{4\pi\rho^2} \cdot (U + 2\rho) \cdot \left(a_l \cdot w^l \cdot \frac{e_w}{\left(\beta^m w_m\right)^3} \cdot \left(1 - \beta^m b_m\right) + a_l \cdot b^l \cdot \frac{e_w}{\left(\beta^m w_m\right)^2}\right)$$
$$- \frac{e_u}{4\pi\rho^2} \cdot \left(1 - y^m a_m - 2 \cdot \beta^m a_m\right) \times \left(\frac{e_w}{\left(\beta^m w_m\right)^3} \cdot u_k \cdot w^k \cdot \left(1 - \beta^m b_m\right) + \frac{e_w}{\left(\beta^m w_m\right)^2} \cdot u_k \cdot b^k\right)$$

$$B_\chi = -\frac{e_u}{4\pi\rho^2} \cdot (U+2\rho) \cdot \left( a_l \cdot w^l \cdot \frac{e_w}{(\beta^m w_m)^3} \cdot (1-\beta^m b_m) + a_l \cdot b^l \cdot \frac{e_w}{(\beta^m w_m)^2} \right)$$

$$-\frac{e_u}{4\pi\rho^2} \cdot (1-y^m a_m) \times \left( \frac{e_w}{(\beta^m w_m)^3} \cdot u_k \cdot w^k \cdot (1-\beta^m b_m) + \frac{e_w}{(\beta^m w_m)^2} \cdot u_k \cdot b^k \right)$$

$$+\frac{e_u}{4\pi\rho^2} \cdot (2 \cdot \beta^m a_m) \times \left( \frac{e_w}{(\beta^m w_m)^3} \cdot u_k \cdot w^k \cdot (1-\beta^m b_m) + \frac{e_w}{(\beta^m w_m)^2} \cdot u_k \cdot b^k \right)$$

$$\hat{B}^i_\beta = -\frac{e_u e_w}{4\pi\rho^2} \cdot (U+2\rho) \cdot \left( a_l \cdot w^l \cdot (M^i_2 - N^{im}_2 b_m) + a_l \cdot b^l \cdot M^i_1 \right)$$

$$-\frac{e_u e_w}{4\pi\rho^2} \cdot (1-y^m a_m) \times \left( u_k \cdot w^k \cdot (M^i_2 - N^{im}_2 b_m) + M^i_1 \cdot u_k \cdot b^k \right)$$

$$+\frac{e_u e_w}{4\pi\rho^2} \cdot 2 \cdot a_m \times \left( u_k \cdot w^k \cdot (N^{im}_2 - P^{ikm}_2 b_k) + N^{im}_1 \cdot u_k \cdot b^k \right)$$

==**$\hat{B}^i_\beta = -\frac{e_u e_w}{4\pi\rho^2} \cdot \left( (U+2\rho)\cdot a_l + (1-y^m a_m)\cdot u_l \right)\left( w^l \cdot (M^i_2 - N^{im}_2 b_m) + b^l \cdot M^i_1 \right)$**==

==**$+\frac{e_u e_w}{4\pi\rho^2} \cdot 2 \cdot a_m \times \left( u_k \cdot w^k \cdot (N^{im}_2 - P^{ipm}_2 b_p) + N^{im}_1 \cdot u_k \cdot b^k \right)$**==

$$B_u = -\frac{e_u}{4\pi\rho^2} \cdot \frac{e_w}{(\beta^m w_m)^3} \cdot \left( \tfrac{1}{2} Y \cdot w^l a_l + \beta^l a_l \cdot y_k w^k \right) \cdot (1-\beta^m b_m)$$

$$-\frac{e_u}{4\pi\rho^2} \cdot \frac{e_w}{(\beta^m w_m)^2} \cdot \left( \tfrac{1}{2} Y \cdot a_l b^l + \beta^l a_l \cdot (1+y^m b_m) \right)$$

$$B_u = -\frac{e_u}{4\pi\rho^2} \cdot \frac{e_w}{(\beta^m w_m)^3} \cdot \left( \tfrac{1}{2} Y \cdot w^l a_l + \beta^l a_l \cdot y_k w^k \right)$$

$$+\frac{e_u}{4\pi\rho^2} \cdot \frac{e_w}{(\beta^m w_m)^3} \cdot \left( \tfrac{1}{2} Y \cdot w^l a_l + \beta^l a_l \cdot y_k w^k \right) \cdot (\beta^m b_m)$$

$$-\frac{e_u}{4\pi\rho^2} \cdot \frac{e_w}{(\beta^m w_m)^2} \cdot \left( \tfrac{1}{2} Y \cdot a_l b^l + \beta^l a_l \cdot (1+y^m b_m) \right)$$

==**$\hat{B}_u = -\frac{e_u e_w}{4\pi\rho^2} \cdot \left( \tfrac{1}{2} Y \cdot w^l a_l \cdot L_2 + M^l_2 a_l \cdot y_k w^k \right)$**==

==**$+\frac{e_u e_w}{4\pi\rho^2} \cdot \left( \tfrac{1}{2} Y \cdot w^l a_l \cdot M^m_2 + N^{ml}_2 a_l \cdot y_k w^k \right) \cdot b_m$**==

==**$-\frac{e_u e_w}{4\pi\rho^2} \cdot \left( \tfrac{1}{2} Y \cdot a_l b^l L_1 + M^l_1 a_l \cdot (1+y^m b_m) \right)$**==

$$B_a = \frac{e_u}{4\pi\rho^2} \times \frac{e_w}{\left(\beta^m w_m\right)^3} \cdot \left(\tfrac{1}{2} Y \cdot u_l w^l + (\rho + U) \cdot y_k w^k\right) \cdot \left(1 - \beta^m b_m\right)$$

$$+ \frac{e_u}{4\pi\rho^2} \times \frac{e_w}{\left(\beta^m w_m\right)^2} \cdot \left(\tfrac{1}{2} Y \cdot u_l b^l + (\rho + U) \cdot \left(1 + y^m b_m\right)\right)$$

$$\hat{B}_a = \frac{e_u e_w}{4\pi\rho^2} \cdot \left(\tfrac{1}{2} Y \cdot u_l w^l + (\rho + U) \cdot y_k w^k\right) \cdot \left(L_2 - M_2^m b_m\right)$$

$$+ \frac{e_u e_w}{4\pi\rho^2} \cdot \left(\tfrac{1}{2} Y \cdot u_l b^l + (\rho + U) \cdot \left(1 + y^m b_m\right)\right) \cdot L_1$$

$$\hat{B}_y = \frac{1}{r_s} \int_0^{2\pi} \left(B_y + B_\chi\right) \cdot \beta^m \xi_m \, d\varphi$$

$$B_\chi = -\frac{e_u}{4\pi\rho^2} \cdot (U + 2\rho) \cdot \left(a_l \cdot w^l \cdot \frac{e_w}{\left(\beta^m w_m\right)^3} \cdot \left(1 - \beta^m b_m\right) + a_l \cdot b^l \cdot \frac{e_w}{\left(\beta^m w_m\right)^2}\right)$$

$$- \frac{e_u}{4\pi\rho^2} \cdot \left(1 - y^m a_m - 2 \cdot \beta^m a_m\right) \times$$

$$\left(\frac{e_w}{\left(\beta^m w_m\right)^3} \cdot u_k \cdot w^k \cdot \left(1 - \beta^m b_m\right) + \frac{e_w}{\left(\beta^m w_m\right)^2} \cdot u_k \cdot b^k\right)$$

$$B_\chi = -\frac{e_u}{4\pi\rho^2} \cdot (U + 2\rho) \cdot \left(a_l \cdot w^l \cdot \frac{e_w}{\left(\beta^m w_m\right)^3} \cdot \left(1 - \beta^m b_m\right) + a_l \cdot b^l \cdot \frac{e_w}{\left(\beta^m w_m\right)^2}\right)$$

$$- \frac{e_u}{4\pi\rho^2} \cdot \left(1 - y^m a_m\right) \times \left(\frac{e_w}{\left(\beta^m w_m\right)^3} \cdot u_k \cdot w^k \cdot \left(1 - \beta^m b_m\right) + \frac{e_w}{\left(\beta^m w_m\right)^2} \cdot u_k \cdot b^k\right)$$

$$+ \frac{e_u}{4\pi\rho^2} \cdot 2 \cdot \beta^m a_m \times \left(\frac{e_w}{\left(\beta^m w_m\right)^3} \cdot u_k \cdot w^k \cdot \left(1 - \beta^m b_m\right) + \frac{e_w}{\left(\beta^m w_m\right)^2} \cdot u_k \cdot b^k\right)$$

$$B_y = \frac{e_u}{4\pi\rho^2} \cdot \left(\rho \cdot a_l + u_l \cdot \left(1 - y^m a_m - \beta^m a_m\right)\right) \times$$

$$\left(\frac{e_w}{\left(\beta^m w_m\right)^3} \cdot w^l \cdot \left(1 - \beta^m b_m\right) + \frac{e_w}{\left(\beta^m w_m\right)^2} \cdot b^l\right)$$

$$B_y = \frac{e_u}{4\pi\rho^2} \cdot \left(\rho \cdot a_l + u_l \cdot \left(1 - y^m a_m\right)\right) \times \left(\frac{e_w}{\left(\beta^m w_m\right)^3} \cdot w^l \cdot \left(1 - \beta^m b_m\right) + \frac{e_w}{\left(\beta^m w_m\right)^2} \cdot b^l\right)$$

$$- \frac{e_u}{4\pi\rho^2} \cdot \left(u_l \cdot \beta^m a_m\right) \times \left(\frac{e_w}{\left(\beta^m w_m\right)^3} \cdot w^l \cdot \left(1 - \beta^m b_m\right) + \frac{e_w}{\left(\beta^m w_m\right)^2} \cdot b^l\right)$$

$$\hat{B}_y = -\frac{e_u e_w}{4\pi\rho^2} \cdot (U + 2\rho) \cdot \left(a_l \cdot w^l \cdot \left(L_2 - M_2^m b_m\right) + a_l \cdot b^l \cdot L_1\right)$$

$$- \frac{e_u e_w}{4\pi\rho^2} \cdot \left(1 - y^m a_m\right) \times \left(u_k \cdot w^k \cdot \left(L_2 - M_2^m b_m\right) + u_k \cdot b^k \cdot L_1\right)$$

$$+ \frac{e_u e_w}{4\pi\rho^2} \cdot 2 \cdot a_m \times \left(u_k \cdot w^k \cdot \left(M_2^m - N_2^{nm} b_n\right) + M_1^m \cdot u_k \cdot b^k\right)$$

$$+ \frac{e_u e_w}{4\pi\rho^2} \cdot \left(\rho \cdot a_l + u_l \cdot \left(1 - y^m a_m\right)\right) \times \left(w^l \cdot \left(L_2 - M_2^m b_m\right) + L_1 \cdot b^l\right)$$

$$- \frac{e_u e_w}{4\pi\rho^2} \cdot \left(u_l \cdot a_m\right) \times \left(w^l \cdot \left(M_2^m - N_2^{nm} b_n\right) + M_1^m \cdot b^l\right)$$

$$\boxed{\hat{B}_y = -\frac{e_u e_w}{4\pi\rho^2} \cdot (U + \rho) \cdot \left(a_l \cdot w^l \cdot \left(L_2 - M_2^m b_m\right) + a_l \cdot b^l \cdot L_1\right)}$$

$$\boxed{+ \frac{e_u e_w}{4\pi\rho^2} \cdot a_m \times \left(u_k \cdot w^k \cdot \left(M_2^m - N_2^{nm} b_n\right) + M_1^m \cdot u_k \cdot b^k\right)}$$

$$B_y = \frac{e_u}{4\pi\rho^2} \cdot \left(\rho \cdot a_l + u_l \cdot \left(1 - y^m a_m\right)\right) \times \left(\frac{e_w}{\left(\beta^m w_m\right)^3} \cdot w^l \cdot \left(1 - \beta^m b_m\right) + \frac{e_w}{\left(\beta^m w_m\right)^2} \cdot b^l\right)$$

$$- \frac{e_u}{4\pi\rho^2} \cdot \left(u_l \cdot \beta^m a_m\right) \times \left(\frac{e_w}{\left(\beta^m w_m\right)^3} \cdot w^l \cdot \left(1 - \beta^m b_m\right) + \frac{e_w}{\left(\beta^m w_m\right)^2} \cdot b^l\right)$$

$$B_\chi + B_y = -\frac{e_u}{4\pi\rho^2} \cdot (U + \rho) \cdot \left(a_l \cdot w^l \cdot \frac{e_w}{\left(\beta^m w_m\right)^3} \cdot \left(1 - \beta^m b_m\right) + a_l \cdot b^l \cdot \frac{e_w}{\left(\beta^m w_m\right)^2}\right)$$

$$+ \frac{e_u}{4\pi\rho^2} \cdot \beta^m a_m \times \left(\frac{e_w}{\left(\beta^m w_m\right)^3} \cdot u_k \cdot w^k \cdot \left(1 - \beta^m b_m\right) + \frac{e_w}{\left(\beta^m w_m\right)^2} \cdot u_k \cdot b^k\right)$$

Checked!

$$B_w = \frac{e_u}{4\pi\rho^2} \cdot \left(\rho \cdot a_l + u_l \cdot \left(1 - \chi^m a_m\right)\right) \times \frac{e_w}{\left(\beta^m w_m\right)^3} \cdot \beta^l \cdot \left(1 - \beta^m b_m\right)$$

$$B_w = \frac{e_u}{4\pi\rho^2} \cdot (\rho + U)\left(1 - \chi^m a_m\right) \times \frac{e_w}{\left(\beta^m w_m\right)^3} \cdot \left(1 - \beta^m b_m\right)$$

$$+ \frac{e_u}{4\pi\rho^2} \cdot \rho \cdot a_l \times \frac{e_w}{\left(\beta^m w_m\right)^3} \cdot \beta^l \cdot \left(1 - \beta^m b_m\right)$$

$$B_w = \frac{e_u}{4\pi\rho^2} \cdot (\rho + U)\left(1 - y^m a_m\right) \times \frac{e_w}{\left(\beta^m w_m\right)^3} \cdot \left(1 - \beta^m b_m\right)$$

$$- \frac{e_u}{4\pi\rho^2} \cdot U \cdot \left(\beta^l a_l\right) \times \frac{e_w}{\left(\beta^m w_m\right)^3} \cdot \left(1 - \beta^m b_m\right)$$

Checked!

$$\hat{B}_w = \frac{e_u e_w}{4\pi\rho^2} \cdot (\rho + U)\left(1 - y^m a_m\right) \cdot \left(L_2 - M_2^m b_m\right)$$

$$- \frac{e_u e_w}{4\pi\rho^2} \cdot U \cdot a_l \cdot \left(M_2^l - N_2^{lm} b_m\right)$$

$$B_b = \frac{e_u}{4\pi\rho^2} \cdot \left(\rho \cdot a_l + u_l \cdot \left(1 - \chi^m a_m\right)\right) \times \frac{e_w}{\left(\beta^m w_m\right)^2} \cdot \beta^l$$

$$B_b = \frac{e_u}{4\pi\rho^2} \cdot \left(\rho \cdot a_l + u_l \cdot \left(1 - y^m a_m - \beta^m a_m\right)\right) \times \frac{e_w}{\left(\beta^m w_m\right)^2} \cdot \beta^l$$

$$B_b = \frac{e_u}{4\pi\rho^2} \cdot \left(\rho \cdot a_l \beta^l + (\rho + U) \cdot \left(1 - y^m a_m - \beta^m a_m\right)\right) \times \frac{e_w}{\left(\beta^m w_m\right)^2}$$

$$B_b = \frac{e_u}{4\pi\rho^2} \cdot \left(-U \cdot a_l \beta^l + (\rho + U) \cdot \left(1 - y^m a_m\right)\right) \times \frac{e_w}{\left(\beta^m w_m\right)^2}$$

Checked!

$$\hat{B}_b = \frac{e_u e_w}{4\pi\rho^2} \cdot \left(-U \cdot a_l M_1^l + (\rho + U) \cdot \left(1 - y^m a_m\right) \cdot L_1\right)$$

**Дополнительные сноски**

$$T_{FG}^{ik} = \frac{1}{4\pi} \cdot \left( -G^{il}F^{k}{}_{l} - F^{il}G^{k}{}_{l} + \frac{1}{2}g^{ik}G_{lm}F^{lm} \right)$$

$$G^{ik} = \frac{e_u}{\left(\chi^m u_m\right)^3} \cdot \left(\chi^i u^k - \chi^k u^i\right) \cdot \left(1 - \chi^m a_m\right) + \frac{e_u}{\left(\chi^m u_m\right)^2} \cdot \left(\chi^i a^k - \chi^k a^i\right)$$

$$\psi_l = u_l + \left(-1 + \chi^m a_m\right) \cdot \frac{\chi_l}{\rho}$$

$$\chi^m = y^m + \beta^m$$

$$F^{ik} = \frac{e_w}{\left(\beta^m w_m\right)^3} \cdot \left(\beta^i w^k - \beta^k w^i\right) \cdot \left(1 - \beta^m b_m\right) + \frac{e_w}{\left(\beta^m w_m\right)^2} \cdot \left(\beta^i b^k - \beta^k b^i\right)$$

$$T_{FG}^{ik} \cdot \psi_k = \frac{1}{4\pi} \cdot \left( -G^{i}{}_{l}F^{kl} - F^{il}G^{k}{}_{l} + \frac{1}{2}g^{ik}G_{lm}F^{lm} \right) \cdot \left(u_k + \left(-1 + \chi^m a_m\right) \cdot \frac{\chi_k}{\rho}\right)$$

$$G^{i}{}_{l} = \frac{e_u}{\left(\chi^m u_m\right)^3} \cdot \left(\chi^i u_l - \chi_l u^i\right) \cdot \left(1 - \chi^m a_m\right) + \frac{e_u}{\left(\chi^m u_m\right)^2} \cdot \left(\chi^i a_l - \chi_l a^i\right)$$

$$T_{FG}^{ik} \cdot \psi_k = -\frac{1}{4\pi} \cdot \left( \frac{e_u}{\left(\chi^m u_m\right)^3} \cdot \left(\chi^i u_l - \chi_l u^i\right) \cdot \left(1 - \chi^m a_m\right) + \frac{e_u}{\left(\chi^m u_m\right)^2} \cdot \left(\chi^i a_l - \chi_l a^i\right) \right) \cdot F^{kl}\psi_k$$

$$-\frac{1}{4\pi} \cdot F^{il} \left( \frac{e_u}{\left(\chi^m u_m\right)^3} \cdot \left(\chi^k u_l - \chi_l u^k\right) \cdot \left(1 - \chi^m a_m\right) + \frac{e_u}{\left(\chi^m u_m\right)^2} \cdot \left(\chi^k a_l - \chi_l a^k\right) \right) \cdot \psi_k$$

$$+\frac{1}{8\pi} \cdot g^{ik} \left( \frac{e_u}{\left(\chi^m u_m\right)^3} \cdot \left(\chi_m u_l - \chi_l u_m\right) \cdot \left(1 - \chi^m a_m\right) + \frac{e_u}{\left(\chi^m u_m\right)^2} \cdot \left(\chi_m a_l - \chi_l a_m\right) \right) F^{ml} \cdot \psi_k$$

$$T_{FG}^{ik} \cdot \psi_k = -\frac{1}{4\pi} \cdot \left( \frac{e_u}{\rho^3} \cdot \left(\chi^i u_l - \chi_l u^i\right) \cdot \left(1 - \chi^m a_m\right) + \frac{e_u}{\rho^2} \cdot \left(\chi^i a_l - \chi_l a^i\right) \right) \cdot F^{kl}\psi_k$$

$$-\frac{1}{4\pi} \cdot F^{il} \left( \frac{e_u}{\rho^3} \cdot \left(\chi^k u_l - \chi_l u^k\right) \cdot \left(1 - \chi^m a_m\right) + \frac{e_u}{\rho^2} \cdot \left(\chi^k a_l - \chi_l a^k\right) \right) \cdot \psi_k$$

$$+\frac{1}{8\pi} \cdot g^{ik} \left( \frac{e_u}{\rho^3} \cdot \left(\chi_m u_l - \chi_l u_m\right) \cdot \left(1 - \chi^m a_m\right) + \frac{e_u}{\rho^2} \cdot \left(\chi_m a_l - \chi_l a_m\right) \right) F^{ml} \cdot \psi_k$$

$$T_{FG}^{ik} \cdot \psi_k = -\frac{1}{4\pi} \cdot \left( \frac{e_u}{\rho^3} \cdot \left( \chi^i u_l - \chi_l u^i \right) \cdot \left( 1 - \chi^m a_m \right) + \frac{e_u}{\rho^2} \cdot \left( \chi^i a_l - \chi_l a^i \right) \right) \cdot F^{kl} \cdot u_k$$

$$- \frac{1}{4\pi} \cdot \left( \frac{e_u}{\rho^3} \cdot \left( \chi^i u_l - \chi_l u^i \right) \cdot \left( 1 - \chi^m a_m \right) + \frac{e_u}{\rho^2} \cdot \left( \chi^i a_l - \chi_l a^i \right) \right) \cdot F^{kl} \cdot \left( -1 + \chi^m a_m \right) \cdot \frac{\chi_k}{\rho}$$

$$- \frac{1}{4\pi} \cdot F^{il} \left( \frac{e_u}{\rho^3} \cdot \left( \chi^k u_l - \chi_l u^k \right) \cdot \left( 1 - \chi^m a_m \right) + \frac{e_u}{\rho^2} \cdot \left( \chi^k a_l - \chi_l a^k \right) \right) \cdot u_k$$

$$- \frac{1}{4\pi} \cdot F^{il} \left( \frac{e_u}{\rho^3} \cdot \left( \chi^k u_l - \chi_l u^k \right) \cdot \left( 1 - \chi^m a_m \right) + \frac{e_u}{\rho^2} \cdot \left( \chi^k a_l - \chi_l a^k \right) \right) \cdot \left( -1 + \chi^m a_m \right) \cdot \frac{\chi_k}{\rho}$$

$$+ \frac{1}{8\pi} \cdot \left( \frac{e_u}{\rho^3} \cdot \left( \chi_m u_l - \chi_l u_m \right) \cdot \left( 1 - \chi^m a_m \right) + \frac{e_u}{\rho^2} \cdot \left( \chi_m a_l - \chi_l a_m \right) \right) F^{ml} \cdot \left( u^i + \left( -1 + \chi^m a_m \right) \cdot \frac{\chi^i}{\rho} \right)$$

$$T_{FG}^{ik} \cdot \psi_k = -\frac{1}{4\pi} \cdot \left( \frac{e_u}{\rho^3} \cdot \left( -\chi_l u^i \right) \cdot \left( 1 - \chi^m a_m \right) + \frac{e_u}{\rho^2} \cdot \left( \chi^i a_l - \chi_l a^i \right) \right) \cdot F^{kl} \cdot u_k$$

$$- \frac{1}{4\pi} \cdot \left( \frac{e_u}{\rho^3} \cdot \left( \chi^i u_l \right) \cdot \left( 1 - \chi^m a_m \right) + \frac{e_u}{\rho^2} \cdot \left( \chi^i a_l \right) \right) \cdot F^{kl} \cdot \left( -1 + \chi^m a_m \right) \cdot \frac{\chi_k}{\rho}$$

$$- \frac{1}{4\pi} \cdot F^{il} \left( \frac{e_u}{\rho^3} \cdot \left( \rho u_l - \chi_l \right) \cdot \left( 1 - \chi^m a_m \right) + \frac{e_u}{\rho^2} \cdot \left( \rho a_l \right) \right)$$

$$- \frac{1}{4\pi} \cdot F^{il} \left( \frac{e_u}{\rho^3} \cdot \left( -\chi_l \rho \right) \cdot \left( 1 - \chi^m a_m \right) + \frac{e_u}{\rho^2} \cdot \left( -\chi_l a^k \chi_k \right) \right) \cdot \left( -1 + \chi^m a_m \right) \cdot \frac{1}{\rho}$$

$$+ \frac{1}{4\pi} \cdot \left( \frac{e_u}{\rho^3} \cdot \left( \chi_m u_l \right) \cdot \left( 1 - \chi^m a_m \right) + \frac{e_u}{\rho^2} \cdot \left( \chi_m a_l \right) \right) F^{ml} \cdot \left( u^i + \left( -1 + \chi^m a_m \right) \cdot \frac{\chi^i}{\rho} \right)$$

$$T_{FG}^{ik} \cdot \psi_k = -\frac{1}{4\pi} \cdot \left( \frac{e_u}{\rho^3} \cdot \left( -\chi_l u^i \right) \cdot \left( 1 - \chi^m a_m \right) + \frac{e_u}{\rho^2} \cdot \left( \chi^i a_l - \chi_l a^i \right) \right) \cdot F^{kl} \cdot u_k$$

$$- \frac{1}{4\pi} \cdot \left( \frac{e_u}{\rho^3} \cdot \left( \chi^i u_l \right) \cdot \left( 1 - \chi^m a_m \right) + \frac{e_u}{\rho^2} \cdot \left( \chi^i a_l \right) \right) \cdot F^{kl} \cdot \left( -1 + \chi^m a_m \right) \cdot \frac{\chi_k}{\rho}$$

$$- \frac{1}{4\pi} \cdot F^{il} \left( \frac{e_u}{\rho^3} \cdot \left( \rho u_l - \chi_l \right) \cdot \left( 1 - \chi^m a_m \right) + \frac{e_u}{\rho^2} \cdot \left( \rho a_l \right) \right)$$

$$- \frac{1}{4\pi} \cdot F^{il} \left( \frac{e_u}{\rho^3} \cdot \left( -\chi_l \rho \right) \cdot \left( 1 - \chi^m a_m \right) + \frac{e_u}{\rho^2} \cdot \left( -\chi_l a^k \chi_k \right) \right) \cdot \left( -1 + \chi^m a_m \right) \cdot \frac{1}{\rho}$$

$$+ \frac{1}{4\pi} \cdot \left( \frac{e_u}{\rho^3} \cdot \left( \chi_m u_l \right) \cdot \left( 1 - \chi^m a_m \right) + \frac{e_u}{\rho^2} \cdot \left( \chi_m a_l \right) \right) F^{ml} \cdot \left( u^i + \left( -1 + \chi^m a_m \right) \cdot \frac{\chi^i}{\rho} \right)$$

$$T_{FG}^{ik} \cdot \psi_k = A_u u^i + A_\chi \chi^i + A_a a^i + F^{il} X_l$$

$$A_u = -\frac{1}{4\pi} \cdot \left( \frac{e_u}{\rho^3} \cdot (-\chi_l) \cdot (1 - \chi^m a_m) \right) \cdot F^{kl} \cdot u_k$$

$$+ \frac{1}{4\pi} \cdot \left( \frac{e_u}{\rho^3} \cdot (\chi_m u_l) \cdot (1 - \chi^m a_m) + \frac{e_u}{\rho^2} \cdot (\chi_m a_l) \right) F^{ml}$$

$$A_\chi = -\frac{1}{4\pi} \cdot \left( \frac{e_u}{\rho^2} \cdot a_l \right) \cdot F^{kl} \cdot u_k$$

$$- \frac{1}{4\pi} \cdot \left( \frac{e_u}{\rho^3} \cdot u_l \cdot (1 - \chi^m a_m) + \frac{e_u}{\rho^2} \cdot a_l \right) \cdot F^{kl} \cdot (-1 + \chi^m a_m) \cdot \frac{\chi_k}{\rho}$$

$$+ \frac{1}{4\pi} \cdot \left( \frac{e_u}{\rho^3} \cdot (\chi_m u_l) \cdot (1 - \chi^m a_m) + \frac{e_u}{\rho^2} \cdot (\chi_m a_l) \right) F^{ml} \cdot (-1 + \chi^m a_m) \cdot \frac{1}{\rho}$$

$$A_a = -\frac{1}{4\pi} \cdot \left( \frac{e_u}{\rho^2} \cdot (-\chi_l a^i) \right) \cdot F^{kl} \cdot u_k$$

$$X_l = -\frac{1}{4\pi} \cdot \left( \frac{e_u}{\rho^3} \cdot (\rho u_l - \chi_l) \cdot (1 - \chi^m a_m) + \frac{e_u}{\rho^2} \cdot (\rho a_l) \right)$$

$$- \frac{1}{4\pi} \cdot \left( \frac{e_u}{\rho^3} \cdot (-\chi_l \rho) \cdot (1 - \chi^m a_m) + \frac{e_u}{\rho^2} \cdot (-\chi_l a^k \chi_k) \right) \cdot (-1 + \chi^m a_m) \cdot \frac{1}{\rho}$$

$$A_u = -\frac{1}{4\pi} \cdot \left( \frac{e_u}{\rho^3} \cdot (-\chi_l) \cdot (1 - \chi^m a_m) \right) \cdot F^{kl} \cdot u_k$$

$$+ \frac{1}{4\pi} \cdot \left( \frac{e_u}{\rho^3} \cdot (\chi_m u_l) \cdot (1 - \chi^m a_m) + \frac{e_u}{\rho^2} \cdot (\chi_m a_l) \right) F^{ml}$$

$$A_u = \frac{1}{4\pi} \cdot \frac{e_u}{\rho^3} \cdot \left( \chi_l \cdot F^{kl} \cdot u_k + \chi_m u_l F^{ml} \right) \cdot (1 - \chi^m a_m)$$

$$+ \frac{1}{4\pi} \cdot \left( \frac{e_u}{\rho^2} \cdot (\chi_m a_l) \right) F^{ml}$$

$$A_u = \frac{e_u}{4\pi \rho^2} \cdot \chi_m a_l \cdot F^{ml}$$

$$A_\chi = -\frac{1}{4\pi} \cdot \left( \frac{e_u}{\rho^2} \cdot a_l \right) \cdot F^{kl} \cdot u_k$$

$$- \frac{1}{4\pi} \cdot \left( \frac{e_u}{\rho^3} \cdot u_l \cdot (1 - \chi^m a_m) + \frac{e_u}{\rho^2} \cdot a_l \right) \cdot F^{kl} \cdot (-1 + \chi^m a_m) \cdot \frac{\chi_k}{\rho}$$

$$+ \frac{1}{4\pi} \cdot \left( \frac{e_u}{\rho^3} \cdot (\chi_m u_l) \cdot (1 - \chi^m a_m) + \frac{e_u}{\rho^2} \cdot (\chi_m a_l) \right) F^{ml} \cdot (-1 + \chi^m a_m) \cdot \frac{1}{\rho}$$

$$A_\chi = -\frac{1}{4\pi}\cdot\left(\frac{e_u}{\rho^2}\cdot a_l\right)\cdot F^{kl}\cdot u_k - \frac{1}{4\pi}\cdot\frac{e_u}{\rho^2}\cdot\left((a_l)\cdot F^{kl}\cdot\frac{\chi_k}{\rho} - ((\chi_m a_l))F^{ml}\cdot\frac{1}{\rho}\right)\cdot(-1+\chi^m a_m)$$

$$-\frac{1}{4\pi}\cdot\left(\frac{e_u}{\rho^3}\cdot u_l\cdot(1-\chi^m a_m)\right)\cdot F^{kl}\cdot(-1+\chi^m a_m)\cdot\frac{\chi_k}{\rho}$$

$$+\frac{1}{4\pi}\cdot\left(\frac{e_u}{\rho^3}\cdot(\chi_k u_l)\cdot(1-\chi^m a_m)\right)F^{kl}\cdot(-1+\chi^m a_m)\cdot\frac{1}{\rho}$$

$$A_\chi = -\frac{1}{4\pi}\cdot\left(\frac{e_u}{\rho^2}\cdot a_l\right)\cdot F^{kl}\cdot u_k - \frac{1}{4\pi}\cdot\frac{e_u}{\rho^2}\cdot\left(a_l\cdot F^{kl}\cdot\frac{\chi_k}{\rho} - \chi_m a_l F^{ml}\cdot\frac{1}{\rho}\right)\cdot(-1+\chi^m a_m)$$

$$-\frac{1}{4\pi}\cdot\frac{e_u}{\rho^3}\cdot\left(u_l\cdot F^{kl}\cdot\frac{\chi_k}{\rho} - \chi_k u_l F^{kl}\cdot\frac{1}{\rho}\right)\cdot(1-\chi^m a_m)\cdot(-1+\chi^m a_m)$$

$$A_\chi = -\frac{1}{4\pi}\cdot\left(\frac{e_u}{\rho^2}\cdot a_l\right)\cdot F^{kl}\cdot u_k$$

$$A_a = -\frac{1}{4\pi}\cdot\frac{e_u}{\rho^2}\cdot F^{kl}\cdot\chi_k\cdot u_l$$

$$X_l = -\frac{1}{4\pi}\cdot\left(\frac{e_u}{\rho^3}\cdot(\rho u_l - \chi_l)\cdot(1-\chi^m a_m) + \frac{e_u}{\rho^2}\cdot(\rho a_l)\right)$$

$$-\frac{1}{4\pi}\cdot\left(\frac{e_u}{\rho^3}\cdot(-\chi_l\rho)\cdot(1-\chi^m a_m) + \frac{e_u}{\rho^2}\cdot(-\chi_l a^k\chi_k)\right)\cdot(-1+\chi^m a_m)\cdot\frac{1}{\rho}$$

$$X_l = -\frac{1}{4\pi}\cdot\frac{e_u}{\rho}\cdot a_l - \frac{1}{4\pi}\cdot\frac{e_u}{\rho^3}\cdot(\rho u_l - \chi_l)\cdot(1-\chi^m a_m)$$

$$-\frac{1}{4\pi}\cdot\left(\frac{e_u}{\rho^2}\cdot(-1+\chi^m a_m) + \frac{e_u}{\rho^2}\cdot(-a^k\chi_k)\right)\cdot\chi_l\cdot(-1+\chi^m a_m)\cdot\frac{1}{\rho}$$

$$X_l = -\frac{1}{4\pi}\cdot\frac{e_u}{\rho}\cdot a_l - \frac{1}{4\pi}\cdot\frac{e_u}{\rho^3}\cdot(\rho u_l - \chi_l)\cdot(1-\chi^m a_m) - \frac{1}{4\pi}\cdot\frac{e_u}{\rho^3}\cdot\chi_l\cdot(1-\chi^m a_m)$$

$$X_l = -\frac{1}{4\pi}\cdot\frac{e_u}{\rho}\cdot a_l - \frac{1}{4\pi}\cdot\frac{e_u}{\rho^2}\cdot u_l\cdot(1-\chi^m a_m)$$

$$T_{FG}^{ik}\cdot\psi_k = \frac{e_u}{4\pi\rho^2}\cdot\chi_k a_l\cdot F^{kl}\cdot u^i - \frac{e_u}{4\pi\rho^2}\cdot a_l\cdot F^{kl}\cdot u_k\chi^i - \frac{e_u}{4\pi\rho^2}\cdot F^{kl}\cdot\chi_k\cdot u_l\cdot a^i$$

$$-F^{il}\left(\frac{1}{4\pi}\cdot\frac{e_u}{\rho}\cdot a_l + \frac{1}{4\pi}\cdot\frac{e_u}{\rho^2}\cdot u_l\cdot(1-\chi^m a_m)\right)$$

$$\boxed{T_{FG}^{ik}\cdot\psi_k = \frac{e_u}{4\pi\rho^2}\cdot\left(\chi_k\cdot a_l\cdot u^i - a_l\cdot u_k\cdot\chi^i - \chi_k\cdot u_l\cdot a^i\right)\cdot F^{kl}}$$

$$\boxed{-\frac{e_u}{4\pi\rho^2}\cdot F^{il}\left(\rho\cdot a_l + u_l\cdot(1-\chi^m a_m)\right)}$$

$$\oiint_S \chi^i \chi^k \chi^l \cdot d\mathbf{S} = 8\pi\rho^5 \cdot u^i u^k u^l - \frac{4\pi\rho^5}{3} \cdot \left(g^{ik} \cdot u^l + g^{il} \cdot u^k + g^{kl} \cdot u^i\right)$$

$$\oiint_S \chi^i \chi^k \chi^l \chi^m \cdot d\mathbf{S} = A \cdot u^i u^k u^l u^m + B \cdot \left(g^{ik} g^{lm} + g^{il} g^{km} + g^{im} g^{kl}\right)$$

$$+ C \cdot \left(g^{ik} \cdot u^l u^m + g^{il} \cdot u^k u^m + g^{kl} \cdot u^i u^m + g^{im} \cdot u^k u^l + g^{km} \cdot u^i u^l + g^{lm} \cdot u^i u^k\right)$$

$$\oiint_S \chi^i \chi^k \chi^l \chi^m \cdot d\mathbf{S} \cdot g_{lm} = A \cdot u^i u^k + B \cdot \left(g^{ik} 4 + g^{ik} + g^{ik}\right)$$

$$+ C \cdot \left(g^{ik} + u^k u^i + u^i u^k + u^k u^i + u^i u^k + 4 \cdot u^i u^k\right) = 0$$

$$A + 8C = 0$$
$$6B + C = 0$$

$$\oiint_S \chi^i \chi^k \chi^l \chi^m u_m \cdot d\mathbf{S} = A \cdot u^i u^k u^l + B \cdot \left(g^{ik} u^l + g^{il} u^k + u^i g^{kl}\right)$$

$$+ C \cdot \left(g^{ik} \cdot u^l + g^{il} \cdot u^k + g^{kl} \cdot u^i + u^i \cdot u^k u^l + u^k \cdot u^i u^l + u^l \cdot u^i u^k\right)$$

$$\oiint_S \chi^i \chi^k \chi^l \chi^m u_m \cdot d\mathbf{S} = (A + 3C) \cdot u^i u^k u^l + (B + C) \cdot \left(g^{ik} u^l + g^{il} u^k + u^i g^{kl}\right)$$

$$A + 3C = 8\pi\rho^6$$

$$B + C = -\frac{4\pi}{3}\rho^6$$

$$6B + C = -\frac{4\pi}{3}\rho^6 + 5B = 0$$

$$B = \frac{4\pi}{15}\rho^6$$

$$C = -\frac{8\pi}{5}\rho^6$$

$$A = \frac{64\pi}{5}\rho^6$$

$$\oiint_S \chi^i \chi^k \chi^l \chi^m \cdot d\mathbf{S} = \frac{64\pi}{5}\rho^6 \cdot u^i u^k u^l u^m + \frac{4\pi}{15}\rho^6 \cdot \left(g^{ik} g^{lm} + g^{il} g^{km} + g^{im} g^{kl}\right)$$

$$- \frac{8\pi}{5}\rho^6 \cdot \left(g^{ik} \cdot u^l u^m + g^{il} \cdot u^k u^m + g^{kl} \cdot u^i u^m + g^{im} \cdot u^k u^l + g^{km} \cdot u^i u^l + g^{lm} \cdot u^i u^k\right)$$

$$\oiint_S \chi^0 \chi^0 \chi^0 \chi^0 \cdot d\mathbf{S} = \frac{64\pi}{5}\rho^6 + \frac{4\pi}{15}\rho^6 \cdot (1+1+1) - \frac{8\pi}{5}\rho^6 \cdot 6$$

$$= 4\pi\rho^6 \cdot \left(\frac{16}{5} + \frac{1}{5} - \frac{12}{5}\right) = 4\pi\rho^6$$

$$F^{il} = \frac{e_w}{\left(\beta^m w_m\right)^3} \cdot \left(\beta^i w^l - \beta^l w^i\right) \cdot \left(1 - \beta^m b_m\right) + \frac{e_w}{\left(\beta^m w_m\right)^2} \cdot \left(\beta^i b^l - \beta^l b^i\right)$$

$$F^{kl} = \frac{e_w}{\left(\beta^m w_m\right)^3} \cdot \left(\beta^k w^l - \beta^l w^k\right) \cdot \left(1 - \beta^m b_m\right) + \frac{e_w}{\left(\beta^m w_m\right)^2} \cdot \left(\beta^k b^l - \beta^l b^k\right)$$

$$T_{FG}^{ik} \cdot \psi_k = \frac{e_u}{4\pi\rho^2} \cdot \left(\chi_k \cdot a_l \cdot u^i - a_l \cdot u_k \cdot \chi^i - \chi_k \cdot u_l \cdot a^i\right) \times$$

$$\left(\frac{e_w}{\left(\beta^m w_m\right)^3} \cdot \left(\beta^k w^l - \beta^l w^k\right) \cdot \left(1 - \beta^m b_m\right) + \frac{e_w}{\left(\beta^m w_m\right)^2} \cdot \left(\beta^k b^l - \beta^l b^k\right)\right)$$

$$-\frac{e_u}{4\pi\rho^2} \cdot \left(\rho \cdot a_l + u_l \cdot \left(1 - \chi^m a_m\right)\right) \times$$

$$\left(\frac{e_w}{\left(\beta^m w_m\right)^3} \cdot \left(\beta^i w^l - \beta^l w^i\right) \cdot \left(1 - \beta^m b_m\right) + \frac{e_w}{\left(\beta^m w_m\right)^2} \cdot \left(\beta^i b^l - \beta^l b^i\right)\right)$$

$$T_{FG}^{ik} \cdot \psi_k = \frac{e_u}{4\pi\rho^2} \cdot \left(\chi_k \cdot a_l \cdot u^i - a_l \cdot u_k \cdot \chi^i - \chi_k \cdot u_l \cdot a^i\right) \times$$

$$\left(\frac{e_w}{\left(\beta^m w_m\right)^3} \cdot \left(\beta^k w^l - \beta^l w^k\right) \cdot \left(1 - \beta^m b_m\right) + \frac{e_w}{\left(\beta^m w_m\right)^2} \cdot \left(\beta^k b^l - \beta^l b^k\right)\right)$$

$$-\frac{e_u}{4\pi\rho^2} \cdot \left(\rho \cdot a_l + u_l \cdot \left(1 - \chi^m a_m\right)\right) \times$$

$$\left(\frac{e_w}{\left(\beta^m w_m\right)^3} \cdot \left(\beta^i w^l - \beta^l w^i\right) \cdot \left(1 - \beta^m b_m\right) + \frac{e_w}{\left(\beta^m w_m\right)^2} \cdot \left(\beta^i b^l - \beta^l b^i\right)\right)$$

$$T_{FG}^{ik} \cdot \psi_k = B_u u^i + B_a a^i + B_\chi \chi^i + B_y y^i + B_w w^i + B_b b^i$$

$$B_u = \frac{e_u}{4\pi\rho^2} \cdot \left(\chi_k \cdot a_l\right) \times \frac{e_w}{\left(\beta^m w_m\right)^3} \cdot \left(\beta^k w^l - \beta^l w^k\right) \cdot \left(1 - \beta^m b_m\right)$$

$$+ \frac{e_u}{4\pi\rho^2} \cdot \left(\chi_k \cdot a_l\right) \times \frac{e_w}{\left(\beta^m w_m\right)^2} \cdot \left(\beta^k b^l - \beta^l b^k\right)$$

$$B_a = -\frac{e_u}{4\pi\rho^2} \cdot \chi_k \cdot u_l \times \left(\frac{e_w}{\left(\beta^m w_m\right)^3} \cdot \left(\beta^k w^l - \beta^l w^k\right) \cdot \left(1 - \beta^m b_m\right)\right)$$

$$- \frac{e_u}{4\pi\rho^2} \cdot \chi_k \cdot u_l \times \left(\frac{e_w}{\left(\beta^m w_m\right)^2} \cdot \left(\beta^k b^l - \beta^l b^k\right)\right)$$

$$B_\chi = -\frac{e_u}{4\pi\rho^2} \cdot a_l \cdot u_k \times \frac{e_w}{\left(\beta^m w_m\right)^3} \cdot \left(\beta^k w^l - \beta^l w^k\right) \cdot \left(1 - \beta^m b_m\right)$$

$$-\frac{e_u}{4\pi\rho^2} \cdot a_l \cdot u_k \times \frac{e_w}{\left(\beta^m w_m\right)^2} \cdot \left(\beta^k b^l - \beta^l b^k\right)$$

$$-\frac{e_u}{4\pi\rho^2} \cdot \left(\rho \cdot a_l + u_l \cdot \left(1 - \chi^m a_m\right)\right) \times \left(\frac{e_w}{\left(\beta^m w_m\right)^3} \cdot w^l \cdot \left(1 - \beta^m b_m\right) + \frac{e_w}{\left(\beta^m w_m\right)^2} \cdot b^l\right)$$

$$B_y = \frac{e_u}{4\pi\rho^2} \cdot \left(\rho \cdot a_l + u_l \cdot \left(1 - \chi^m a_m\right)\right) \times \left(\frac{e_w}{\left(\beta^m w_m\right)^3} \cdot w^l \cdot \left(1 - \beta^m b_m\right) + \frac{e_w}{\left(\beta^m w_m\right)^2} \cdot b^l\right)$$

$$B_w = \frac{e_u}{4\pi\rho^2} \cdot \left(\rho \cdot a_l + u_l \cdot \left(1 - \chi^m a_m\right)\right) \times \frac{e_w}{\left(\beta^m w_m\right)^3} \cdot \beta^l \cdot \left(1 - \beta^m b_m\right)$$

$$B_b = \frac{e_u}{4\pi\rho^2} \cdot \left(\rho \cdot a_l + u_l \cdot \left(1 - \chi^m a_m\right)\right) \times \frac{e_w}{\left(\beta^m w_m\right)^2} \cdot \beta^l$$

**Checked!**

=====================-================-================-==============

$$B_\chi = -\frac{e_u}{4\pi\rho^2} \cdot a_l \cdot u_k \times \frac{e_w}{\left(\beta^m w_m\right)^3} \cdot \left(\beta^k w^l - \beta^l w^k\right) \cdot \left(1 - \beta^m b_m\right)$$

$$-\frac{e_u}{4\pi\rho^2} \cdot a_l \cdot u_k \times \frac{e_w}{\left(\beta^m w_m\right)^2} \cdot \left(\beta^k b^l - \beta^l b^k\right)$$

$$-\frac{e_u}{4\pi\rho^2} \cdot \rho \cdot a_l \times \left(\frac{e_w}{\left(\beta^m w_m\right)^3} \cdot w^l \cdot \left(1 - \beta^m b_m\right) + \frac{e_w}{\left(\beta^m w_m\right)^2} \cdot b^l\right)$$

$$-\frac{e_u}{4\pi\rho^2} \cdot u_l \cdot \left(1 - \chi^m a_m\right) \times \left(\frac{e_w}{\left(\beta^m w_m\right)^3} \cdot w^l \cdot \left(1 - \beta^m b_m\right) + \frac{e_w}{\left(\beta^m w_m\right)^2} \cdot b^l\right)$$

$$B_\chi = -\frac{e_u}{4\pi\rho^2} \cdot a_l \cdot w^l \cdot \frac{e_w}{\left(\beta^m w_m\right)^3} \cdot \left(u_k \cdot \beta^k \cdot \left(1 - \beta^m b_m\right) + \rho \cdot \left(1 - \beta^m b_m\right)\right)$$

$$+\frac{e_u}{4\pi\rho^2} \cdot a_l \beta^l \cdot \left(\frac{e_w}{\left(\beta^m w_m\right)^3} \cdot u_k w^k \cdot \left(1 - \beta^m b_m\right) + \frac{e_w}{\left(\beta^m w_m\right)^2} \cdot u_k b^k\right)$$

$$-\frac{e_u}{4\pi\rho^2} \cdot a_l \cdot b^l \cdot \frac{e_w}{\left(\beta^m w_m\right)^2} \cdot \left(\rho + u_k \beta^k\right)$$

$$-\frac{e_u}{4\pi\rho^2} \cdot \left(1 - \chi^m a_m\right) \times \left(\frac{e_w}{\left(\beta^m w_m\right)^3} \cdot u_l \cdot w^k \cdot \left(1 - \beta^m b_m\right) + \frac{e_w}{\left(\beta^m w_m\right)^2} \cdot u_k \cdot b^k\right)$$

$$B_\chi = -\frac{e_u}{4\pi\rho^2} \cdot a_l \cdot w^l \cdot \frac{e_w}{\left(\beta^m w_m\right)^3} \cdot \left(u_k \cdot \beta^k + \rho\right) \cdot \left(1 - \beta^m b_m\right)$$

$$-\frac{e_u}{4\pi\rho^2} \cdot a_l \cdot b^l \cdot \frac{e_w}{\left(\beta^m w_m\right)^2} \cdot \left(\rho + u_k \beta^k\right)$$

$$-\frac{e_u}{4\pi\rho^2} \cdot \left(1 - \chi^m a_m - \beta^m a_m\right) \times \left(\frac{e_w}{\left(\beta^m w_m\right)^3} \cdot u_k \cdot w^k \cdot \left(1 - \beta^m b_m\right) + \frac{e_w}{\left(\beta^m w_m\right)^2} \cdot u_k \cdot b^k\right)$$

$$B_\chi = -\frac{e_u}{4\pi\rho^2} \cdot \left(u_k \cdot \beta^k + \rho\right) \cdot \left( a_l \cdot w^l \cdot \frac{e_w}{\left(\beta^m w_m\right)^3} \cdot \left(1 - \beta^m b_m\right) + a_l \cdot b^l \cdot \frac{e_w}{\left(\beta^m w_m\right)^2}\right)$$

$$-\frac{e_u}{4\pi\rho^2} \cdot \left(1 - \chi^m a_m - \beta^m a_m\right) \times \left(\frac{e_w}{\left(\beta^m w_m\right)^3} \cdot u_k \cdot w^k \cdot \left(1 - \beta^m b_m\right) + \frac{e_w}{\left(\beta^m w_m\right)^2} \cdot u_k \cdot b^k\right)$$

$$B_\chi = -\frac{e_u}{4\pi\rho^2} \cdot \left(U + 2\rho\right) \cdot \left( a_l \cdot w^l \cdot \frac{e_w}{\left(\beta^m w_m\right)^3} \cdot \left(1 - \beta^m b_m\right) + a_l \cdot b^l \cdot \frac{e_w}{\left(\beta^m w_m\right)^2}\right)$$

$$-\frac{e_u}{4\pi\rho^2} \cdot \left(1 - y^m a_m - 2\cdot\beta^m a_m\right) \times \left(\frac{e_w}{\left(\beta^m w_m\right)^3} \cdot u_k \cdot w^k \cdot \left(1 - \beta^m b_m\right) + \frac{e_w}{\left(\beta^m w_m\right)^2} \cdot u_k \cdot b^k\right)$$

**Checked!**

$$B_a = -\frac{e_u}{4\pi\rho^2} \times \left(\frac{e_w}{\left(\beta^m w_m\right)^3} \cdot \left(\chi_k \beta^k \cdot u_l w^l - u_l \beta^l \cdot \chi_k w^k\right) \cdot \left(1 - \beta^m b_m\right)\right)$$

$$-\frac{e_u}{4\pi\rho^2} \times \left(\frac{e_w}{\left(\beta^m w_m\right)^2} \cdot \left(\chi_k \beta^k \cdot u_l b^l - u_l \beta^l \cdot \chi_k b^k\right)\right)$$

$$B_a = -\frac{e_u}{4\pi\rho^2} \times \left(\frac{e_w}{\left(\beta^m w_m\right)^3} \cdot \left(\left(-\tfrac{1}{2}Y\right) \cdot u_l w^l - \left(\rho + U\right) \cdot \left(\beta_k + y_k\right) w^k\right) \cdot \left(1 - \beta^m b_m\right)\right)$$

$$-\frac{e_u}{4\pi\rho^2} \times \left(\frac{e_w}{\left(\beta^m w_m\right)^2} \cdot \left(\left(-\tfrac{1}{2}Y\right) \cdot u_l b^l - \left(\rho + U\right) \cdot \chi_k b^k\right)\right)$$

$$B_a = -\frac{e_u}{4\pi\rho^2} \times \left(\frac{e_w}{\left(\beta^m w_m\right)^3} \cdot \left(\left(-\tfrac{1}{2}Y\right) \cdot u_l w^l - \left(\rho + U\right) \cdot y_k w^k\right) \cdot \left(1 - \beta^m b_m\right)\right)$$

$$-\frac{e_u}{4\pi\rho^2} \times \frac{e_w}{\left(\beta^m w_m\right)^2} \cdot \left(\left(\left(-\tfrac{1}{2}Y\right) \cdot u_l b^l - \left(\rho + U\right) \cdot \chi_k b^k\right) - \left(\rho + U\right) \cdot \left(1 - \beta^m b_m\right)\right)$$

$$B_a = \frac{e_u}{4\pi\rho^2} \times \left( \frac{e_w}{\left(\beta^m w_m\right)^3} \cdot \left(\tfrac{1}{2}Y \cdot u_l w^l + \left(\rho+U\right) \cdot y_k w^k\right) \cdot \left(1-\beta^m b_m\right) \right)$$

$$+ \frac{e_u}{4\pi\rho^2} \times \frac{e_w}{\left(\beta^m w_m\right)^2} \cdot \left(\tfrac{1}{2}Y \cdot u_l b^l + \left(\rho+U\right) \cdot \left(1 + \chi_k b^k - \beta^m b_m\right)\right)$$

$$\boxed{B_a = \frac{e_u}{4\pi\rho^2} \times \frac{e_w}{\left(\beta^m w_m\right)^3} \cdot \left(\tfrac{1}{2}Y \cdot u_l w^l + \left(\rho+U\right) \cdot y_k w^k\right) \cdot \left(1-\beta^m b_m\right)}$$

$$\boxed{+ \frac{e_u}{4\pi\rho^2} \times \frac{e_w}{\left(\beta^m w_m\right)^2} \cdot \left(\tfrac{1}{2}Y \cdot u_l b^l + \left(\rho+U\right) \cdot \left(1 + y^m b_m\right)\right)}$$

**Checked!**

$$B_u = \frac{e_u}{4\pi\rho^2} \cdot \frac{e_w}{\left(\beta^m w_m\right)^3} \cdot \left(\chi_k \beta^k w^l \cdot a_l - \beta^l \cdot a_l \chi_k w^k\right) \cdot \left(1-\beta^m b_m\right)$$

$$+ \frac{e_u}{4\pi\rho^2} \cdot \frac{e_w}{\left(\beta^m w_m\right)^2} \cdot \left(\chi_k \beta^k \cdot a_l b^l - \beta^l a_l \cdot \chi_k b^k\right)$$

$$B_u = \frac{e_u}{4\pi\rho^2} \cdot \frac{e_w}{\left(\beta^m w_m\right)^3} \cdot \left(\left(-\tfrac{1}{2}Y\right) \cdot w^l a_l - \beta^l a_l \cdot y_k w^k - \beta^l a_l \cdot \beta_k w^k\right) \cdot \left(1-\beta^m b_m\right)$$

$$+ \frac{e_u}{4\pi\rho^2} \cdot \frac{e_w}{\left(\beta^m w_m\right)^2} \cdot \left(\left(-\tfrac{1}{2}Y\right) \cdot a_l b^l - \beta^l a_l \cdot \chi_k b^k\right)$$

$$B_u = \frac{e_u}{4\pi\rho^2} \cdot \frac{e_w}{\left(\beta^m w_m\right)^3} \cdot \left(\left(-\tfrac{1}{2}Y\right) \cdot w^l a_l - \beta^l a_l \cdot y_k w^k\right) \cdot \left(1-\beta^m b_m\right)$$

$$+ \frac{e_u}{4\pi\rho^2} \cdot \frac{e_w}{\left(\beta^m w_m\right)^2} \cdot \left(-\tfrac{1}{2}Y \cdot a_l b^l - \beta^l a_l \cdot \chi_k b^k - \beta^l a_l \cdot \left(1-\beta^m b_m\right)\right)$$

$$B_u = \frac{e_u}{4\pi\rho^2} \cdot \frac{e_w}{\left(\beta^m w_m\right)^3} \cdot \left(\left(-\tfrac{1}{2}Y\right) \cdot w^l a_l - \beta^l a_l \cdot y_k w^k\right) \cdot \left(1-\beta^m b_m\right)$$

$$+ \frac{e_u}{4\pi\rho^2} \cdot \frac{e_w}{\left(\beta^m w_m\right)^2} \cdot \left(-\tfrac{1}{2}Y \cdot a_l b^l - \beta^l a_l \cdot \left(1 + y^m b_m\right)\right)$$

$$\boxed{B_u = -\frac{e_u}{4\pi\rho^2} \cdot \frac{e_w}{\left(\beta^m w_m\right)^3} \cdot \left(\tfrac{1}{2}Y \cdot w^l a_l + \beta^l a_l \cdot y_k w^k\right) \cdot \left(1-\beta^m b_m\right)}$$

$$\boxed{- \frac{e_u}{4\pi\rho^2} \cdot \frac{e_w}{\left(\beta^m w_m\right)^2} \cdot \left(\tfrac{1}{2}Y \cdot a_l b^l + \beta^l a_l \cdot \left(1 + y^m b_m\right)\right)}$$

**Checked!**

$$T_{FF}^{ik}\psi_k = \frac{1}{4\pi} \cdot \left(-F^{il}F^k{}_l + \frac{1}{4}g^{ik}F_{lm}F^{lm}\right) \cdot \left(u_k + \left(-1 + \chi^m a_m\right) \cdot \frac{\chi_k}{\rho}\right)$$

$$F^{ik} = \frac{e'}{\left(\beta^m w_m\right)^3} \cdot \left(\beta^i w^k - \beta^k w^i\right) \cdot \left(1 - \beta^m b_m\right) + \frac{e'}{\left(\beta^m w_m\right)^2} \cdot \left(\beta^i b^k - \beta^k b^i\right)$$

$$F_{pk} = \frac{e'}{\left(\beta^m w_m\right)^3} \cdot \left(\beta_p w_k - \beta_k w_p\right) \cdot \left(1 - \beta^m b_m\right) + \frac{e'}{\left(\beta^m w_m\right)^2} \cdot \left(\beta_p b_k - \beta_k b_p\right)$$

$$F^{ik}F_{pk} = \left(\frac{e'}{\left(\beta^m w_m\right)^3} \cdot \left(\beta^i w^k - \beta^k w^i\right) \cdot \left(1 - \beta^m b_m\right) + \frac{e'}{\left(\beta^m w_m\right)^2} \cdot \left(\beta^i b^k - \beta^k b^i\right)\right) \times$$
$$\frac{e'}{\left(\beta^m w_m\right)^3} \cdot \left(\beta_p w_k - \beta_k w_p\right) \cdot \left(1 - \beta^m b_m\right)$$
$$+ \left(\frac{e'}{\left(\beta^m w_m\right)^3} \cdot \left(\beta^i w^k - \beta^k w^i\right) \cdot \left(1 - \beta^m b_m\right) + \frac{e'}{\left(\beta^m w_m\right)^2} \cdot \left(\beta^i b^k - \beta^k b^i\right)\right) \times$$
$$\frac{e'}{\left(\beta^m w_m\right)^2} \cdot \left(\beta_p b_k - \beta_k b_p\right)$$

$$F^{ik}F_{pk} = \left(\frac{e'}{\left(\beta^m w_m\right)^3} \cdot \left(\beta^i w^k - \beta^k w^i\right) \cdot \left(1 - \beta^m b_m\right)\right) \times \frac{e'}{\left(\beta^m w_m\right)^3} \cdot \left(\beta_p w_k - \beta_k w_p\right) \cdot \left(1 - \beta^m b_m\right)$$
$$\left(\frac{e'}{\left(\beta^m w_m\right)^2} \cdot \left(\beta^i b^k - \beta^k b^i\right)\right) \times \frac{e'}{\left(\beta^m w_m\right)^3} \cdot \left(\beta_p w_k - \beta_k w_p\right) \cdot \left(1 - \beta^m b_m\right)$$
$$+ \left(\frac{e'}{\left(\beta^m w_m\right)^3} \cdot \left(\beta^i w^k - \beta^k w^i\right) \cdot \left(1 - \beta^m b_m\right)\right) \times \frac{e'}{\left(\beta^m w_m\right)^2} \cdot \left(\beta_p b_k - \beta_k b_p\right)$$
$$+ \left(\frac{e'}{\left(\beta^m w_m\right)^2} \cdot \left(\beta^i b^k - \beta^k b^i\right)\right) \times \frac{e'}{\left(\beta^m w_m\right)^2} \cdot \left(\beta_p b_k - \beta_k b_p\right)$$

$$F^{ik}F_{pk} = \frac{e'^2}{\left(\beta^m w_m\right)^6} \cdot \left(1-\beta^m b_m\right)^2 \cdot \left(\beta^i w^k - \beta^k w^i\right) \cdot \left(\beta_p w_k - \beta_k w_p\right)$$

$$+ \frac{e'^2}{\left(\beta^m w_m\right)^5} \cdot \left(1-\beta^m b_m\right) \cdot \left(\beta^i b^k - \beta^k b^i\right) \cdot \left(\beta_p w_k - \beta_k w_p\right)$$

$$+ \frac{e'^2}{\left(\beta^m w_m\right)^5} \cdot \left(1-\beta^m b_m\right) \cdot \left(\beta^i w^k - \beta^k w^i\right) \cdot \left(\beta_p b_k - \beta_k b_p\right)$$

$$+ \frac{e'^2}{\left(\beta^m w_m\right)^4} \cdot \left(\beta^i b^k - \beta^k b^i\right) \cdot \left(\beta_p b_k - \beta_k b_p\right)$$

$$F^{ik}F_{pk} = \frac{e'^2}{\left(\beta^m w_m\right)^6} \cdot \left(1-\beta^m b_m\right)^2 \cdot \left(\beta^i w^k \beta_p w_k - \beta^i w^k \beta_k w_p - \beta^k w^i \beta_p w_k + \beta^k w^i \beta_k w_p\right)$$

$$+ \frac{e'^2}{\left(\beta^m w_m\right)^5} \cdot \left(1-\beta^m b_m\right) \cdot \left(-\beta^i b^k \cdot \beta_k w_p - \beta^k b^i \cdot \beta_p w_k\right)$$

$$+ \frac{e'^2}{\left(\beta^m w_m\right)^5} \cdot \left(1-\beta^m b_m\right) \cdot \left(-\beta^i w^k \cdot \beta_k b_p - \beta^k w^i \cdot \beta_p b_k\right)$$

$$+ \frac{e'^2}{\left(\beta^m w_m\right)^4} \cdot \left(\beta^i b^k \cdot \beta_p b_k - \beta^i b^k \cdot \beta_k b_p - \beta^k b^i \cdot \beta_p b_k\right)$$

$$F^{ik}F^p{}_k = \frac{e'^2}{\left(\beta^m w_m\right)^6} \cdot \left(1-\beta^m b_m\right)^2 \cdot \left(\beta^i \beta^p - \beta^i w^k \beta_k w^p - \beta^k w^i \beta^p w_k\right)$$

$$+ \frac{e'^2}{\left(\beta^m w_m\right)^5} \cdot \left(1-\beta^m b_m\right) \cdot \left(-\beta^i b^k \cdot \beta_k w^p - \beta^k b^i \cdot \beta^p w_k\right)$$

$$+ \frac{e'^2}{\left(\beta^m w_m\right)^5} \cdot \left(1-\beta^m b_m\right) \cdot \left(-\beta^i w^k \cdot \beta_k b^p - \beta^k w^i \cdot \beta^p b_k\right)$$

$$+ \frac{e'^2}{\left(\beta^m w_m\right)^4} \cdot \left(\beta^i b^k \cdot \beta^p b_k - \beta^i b^k \cdot \beta_k b^p - \beta^k b^i \cdot \beta^p b_k\right)$$

$$F^{ik}F^p{}_k = \frac{e'^2}{(\beta^m w_m)^6} \cdot (1-\beta^m b_m)^2 \cdot \beta^i \beta^p - \frac{e'^2}{(\beta^m w_m)^5} \cdot (1-\beta^m b_m)^2 \cdot (\beta^i w^p + w^i \beta^p)$$

$$- \frac{e'^2}{(\beta^m w_m)^5} \cdot (1-\beta^m b_m) \cdot \beta^i w^p \cdot b^k \beta_k - \frac{e'^2}{(\beta^m w_m)^4} \cdot (1-\beta^m b_m) \cdot b^i \cdot \beta^p$$

$$- \frac{e'^2}{(\beta^m w_m)^4} \cdot (1-\beta^m b_m) \cdot (\beta^i b^p) - \frac{e'^2}{(\beta^m w_m)^5} \cdot (1-\beta^m b_m) \cdot (\beta^p w^i \cdot \beta^k b_k)$$

$$+ \frac{e'^2}{(\beta^m w_m)^4} \cdot \beta^i \beta^p b^k b_k - \frac{e'^2}{(\beta^m w_m)^4} \cdot b^k \beta_k \cdot (\beta^i b^p + b^i \cdot \beta^p)$$

$$F^{ik}F^p{}_k = \frac{e'^2}{(\beta^m w_m)^6} \cdot (1-\beta^m b_m)^2 \cdot \beta^i \beta^p - \frac{e'^2}{(\beta^m w_m)^5} \cdot (1-\beta^m b_m)^2 \cdot (\beta^i w^p + w^i \beta^p)$$

$$- \frac{e'^2}{(\beta^m w_m)^5} \cdot (1-\beta^m b_m) \cdot (\beta^i w^p + \beta^p w^i) \cdot b^k \beta_k$$

$$- \frac{e'^2}{(\beta^m w_m)^4} \cdot (\beta^i b^p + b^i \cdot \beta^p) + \frac{e'^2}{(\beta^m w_m)^4} \cdot \beta^i \beta^p b^k b_k$$

$$\boxed{F^{ik}F^p{}_k = \frac{e'^2}{(\beta^m w_m)^6} \cdot (1-\beta^m b_m)^2 \cdot \beta^i \beta^p - \frac{e'^2}{(\beta^m w_m)^5} \cdot (1-\beta^m b_m) \cdot (\beta^i w^p + w^i \beta^p) \\ - \frac{e'^2}{(\beta^m w_m)^4} \cdot (\beta^i b^p + b^i \cdot \beta^p) + \frac{e'^2}{(\beta^m w_m)^4} \cdot \beta^i \beta^p b^k b_k}$$

$$F^{pk}F_{pk} = \frac{e'^2}{(\beta^m w_m)^6} \cdot (1-\beta^m b_m)^2 \cdot (\beta^p w^k \beta_p w_k - \beta^p w^k \beta_k w_p - \beta^k w^p \beta_p w_k + \beta^k w^p \beta_k w_p)$$

$$+ \frac{e'^2}{(\beta^m w_m)^5} \cdot (1-\beta^m b_m) \cdot (-\beta^p b^k \cdot \beta_k w_p - \beta^k b^p \cdot \beta_p w_k)$$

$$+ \frac{e'^2}{(\beta^m w_m)^5} \cdot (1-\beta^m b_m) \cdot (-\beta^p w^k \cdot \beta_k b_p - \beta^k w^p \cdot \beta_p b_k)$$

$$+ \frac{e'^2}{(\beta^m w_m)^4} \cdot (\beta^p b^k \cdot \beta_p b_k - \beta^p b^k \cdot \beta_k b_p - \beta^k b^p \cdot \beta_p b_k)$$

$$F^{pk}F_{pk} = -\frac{2e'^2}{\left(\beta^m w_m\right)^4} \cdot \left(1 - \beta^m b_m\right)^2$$

$$-\frac{2e'^2}{\left(\beta^m w_m\right)^4} \cdot \left(1 - \beta^m b_m\right) \cdot \left(b^k \beta_k\right)$$

$$-\frac{2e'^2}{\left(\beta^m w_m\right)^4} \cdot \left(1 - \beta^m b_m\right) \cdot \left(\beta^p b_p\right)$$

$$-\frac{2e'^2}{\left(\beta^m w_m\right)^4} \cdot \left(\beta^p b_p\right)^2$$

$$F^{pk}F_{pk} = -\frac{2e'^2}{\left(\beta^m w_m\right)^4} \cdot \left(1 - \beta^m b_m\right)\left(1 - \beta^m b_m + b^k \beta_k\right)$$

$$-\frac{2e'^2}{\left(\beta^m w_m\right)^4} \cdot \left(1 - \beta^m b_m + \beta^p b_p\right) \cdot \left(\beta^p b_p\right)$$

$$F^{pk}F_{pk} = -\frac{2e'^2}{\left(\beta^m w_m\right)^4} \cdot \left(1 - \beta^m b_m\right) - \frac{2e'^2}{\left(\beta^m w_m\right)^4} \cdot \left(\beta^p b_p\right)$$

$$\boxed{F^{pk}F_{pk} = -\frac{2e'^2}{\left(\beta^m w_m\right)^4}}$$

$$T^{ik}_{FF}\psi_k = \frac{1}{4\pi} \cdot \left(-F^{il}F^k{}_l + \frac{1}{4}g^{ik}F_{lm}F^{lm}\right) \cdot \left(u_k + \left(-1 + \chi^m a_m\right) \cdot \frac{\chi_k}{\rho}\right)$$

$$T^{ip}_{FF}\psi_p = \frac{1}{4\pi} \cdot \left(-F^{il}F^p{}_l + \frac{1}{4}g^{ip}F_{lm}F^{lm}\right) \cdot \psi_p$$

$$T^{ip}_{FF}\psi_p = \frac{1}{4\pi} \cdot \left(-\frac{e'^2}{\left(\beta^m w_m\right)^6} \cdot \left(1 - \beta^m b_m\right)^2 \cdot \beta^i \beta^p + \frac{e'^2}{\left(\beta^m w_m\right)^5} \cdot \left(1 - \beta^m b_m\right) \cdot \left(\beta^i w^p + w^i \beta^p\right)\right) \cdot \psi_p$$

$$\frac{1}{4\pi} \cdot \left(+\frac{e'^2}{\left(\beta^m w_m\right)^4} \cdot \left(\beta^i b^p + b^i \cdot \beta^p\right) - \frac{e'^2}{\left(\beta^m w_m\right)^4} \cdot \beta^i \beta^p b^k b_k - \frac{1}{4}g^{ip}\frac{2e'^2}{\left(\beta^m w_m\right)^4}\right) \cdot \psi_p$$

$$\boxed{\begin{aligned}T^{ip}_{FF}\psi_p &= \frac{e'^2}{4\pi} \cdot \left(-\frac{1}{\left(\beta^m w_m\right)^6} \cdot \left(1 - \beta^m b_m\right)^2 \cdot \beta^i \beta^p + \frac{1}{\left(\beta^m w_m\right)^5} \cdot \left(1 - \beta^m b_m\right) \cdot \left(\beta^i w^p + w^i \beta^p\right)\right) \cdot \psi_p \\ &\quad + \frac{e'^2}{4\pi} \cdot \left(\frac{1}{\left(\beta^m w_m\right)^4} \cdot \left(\beta^i b^p + b^i \cdot \beta^p\right) - \frac{1}{\left(\beta^m w_m\right)^4} \cdot \beta^i \beta^p b^k b_k - \frac{1}{2}g^{ip}\frac{1}{\left(\beta^m w_m\right)^4}\right) \cdot \psi_p\end{aligned}}$$

$$^e X_u{}^i = \int_0^{2\pi} d\varphi \cdot \frac{\beta^i}{r_s} \cdot \left( \frac{w^p}{(\beta^m w_m)^4} \cdot (1 - \beta^m b_m) + \frac{b^p}{(\beta^m w_m)^3} - \frac{\beta^p \cdot (1 - \beta^m b_m)^2}{(\beta^m w_m)^5} - \frac{\beta^p \cdot b^k b_k}{(\beta^m w_m)^3} \right) \cdot u_p$$

$$X_\rho{}^i = -\int_0^{2\pi} d\varphi \cdot \frac{\beta^i}{r_s} \cdot \left( \frac{w^p}{(\beta^m w_m)^4} \cdot (1 - \beta^m b_m) + \frac{b^p}{(\beta^m w_m)^3} - \frac{\beta^p \cdot (1 - \beta^m b_m)^2}{(\beta^m w_m)^5} - \frac{\beta^p \cdot b^k b_k}{(\beta^m w_m)^3} \right) \frac{\chi_p}{\rho}$$

$$X_a{}^{im} = \int_0^{2\pi} d\varphi \cdot \frac{\beta^i}{r_s} \cdot \left( \frac{w^p}{(\beta^m w_m)^4} \cdot (1 - \beta^m b_m) + \frac{b^p}{(\beta^m w_m)^3} - \frac{\beta^p \cdot (1 - \beta^m b_m)^2}{(\beta^m w_m)^5} - \frac{\beta^p \cdot b^k b_k}{(\beta^m w_m)^3} \right) \frac{\chi_p \chi^m}{\rho}$$

$$X_u{}^i = \int_0^{2\pi} d\varphi \cdot \frac{\beta^i}{r_s} \cdot \left( \frac{w^p}{(\beta^m w_m)^4} \cdot (1 - \beta^m b_m) + \frac{b^p}{(\beta^m w_m)^3} \right) \cdot u_p$$
$$- \int_0^{2\pi} d\varphi \cdot \frac{\beta^i}{r_s} \cdot \left( \frac{(1 - \beta^m b_m)^2}{(\beta^m w_m)^5} + \frac{b^k b_k}{(\beta^m w_m)^3} \right) \cdot \beta^p u_p$$

$$X_u{}^i = \int_0^{2\pi} d\varphi \cdot \frac{1}{r_s} \cdot \left( \frac{w^p u_p}{(\beta^m w_m)^4} \cdot (\beta^i - \beta^i \beta^m b_m) + \frac{\beta^i \cdot b^p u_p}{(\beta^m w_m)^3} \right)$$
$$- \int_0^{2\pi} d\varphi \cdot \frac{\beta^i}{r_s} \cdot \left( \frac{(1 - 2 \cdot \beta^m b_m + \beta^m b_m \cdot \beta^n b_n)}{(\beta^m w_m)^5} + \frac{b^k b_k}{(\beta^m w_m)^3} \right) \cdot (\rho - y^p u_p)$$

$$\boxed{X_u{}^i = w^p u_p \cdot (M_4^i - N_4^{im} b_m) + M_3^i \cdot b^p u_p - (M_5^i - 2 \cdot N_5^{im} b_m + P_5^{imn} b_m b_n + b^k b_k \cdot M_3^i) \cdot (\rho - y^p u_p)}$$

$$X_\rho{}^i = -\int_0^{2\pi} d\varphi \cdot \frac{\beta^i}{r_s} \cdot \left( \frac{w^p \cdot (y_p + \beta_p)}{(\beta^m w_m)^4} \cdot (1 - \beta^m b_m) + \frac{b^p \cdot (y_p + \beta_p)}{(\beta^m w_m)^3} \right) \frac{1}{\rho}$$
$$- \int_0^{2\pi} d\varphi \cdot \frac{\beta^i}{r_s} \cdot \left( -\frac{(1 - \beta^m b_m)^2}{(\beta^m w_m)^5} - \frac{b^k b_k}{(\beta^m w_m)^3} \right) \frac{(-y^p + \chi^p) \chi_p}{\rho}$$

$$X_\rho{}^i = -\int_0^{2\pi} d\varphi \cdot \frac{\beta^i}{r_s} \cdot \left( \frac{w^p \cdot y_p}{(\beta^m w_m)^4} \cdot (1 - \beta^m b_m) + \frac{1}{(\beta^m w_m)^3} \cdot (1 - \beta^m b_m) + \frac{b^p \cdot y_p}{(\beta^m w_m)^3} + \frac{b^p \cdot \beta_p}{(\beta^m w_m)^3} \right) \frac{1}{\rho}$$
$$- \int_0^{2\pi} d\varphi \cdot \frac{\beta^i}{r_s} \cdot \left( \frac{(1 - \beta^m b_m)^2}{(\beta^m w_m)^5} + \frac{b^k b_k}{(\beta^m w_m)^3} \right) \frac{y^p y_p}{2\rho}$$

$$X_\rho{}^i = -\int_0^{2\pi} d\varphi \cdot \frac{\beta^i}{r_s} \cdot \left( \frac{w^p \cdot y_p}{(\beta^m w_m)^4} \cdot (1 - \beta^m b_m) + \frac{1}{(\beta^m w_m)^3} + \frac{b^p \cdot y_p}{(\beta^m w_m)^3} \right) \frac{1}{\rho}$$
$$- \int_0^{2\pi} d\varphi \cdot \frac{\beta^i}{r_s} \cdot \left( \frac{(1 - 2 \cdot \beta^m b_m + \beta^m b_m \cdot \beta^n b_n)}{(\beta^m w_m)^5} + \frac{b^k b_k}{(\beta^m w_m)^3} \right) \frac{y^p y_p}{2\rho}$$

$$X_\rho^{\ i} = -\left(w^p \cdot y_p \cdot \left(M_4^i - N_4^{im} b_m\right) + \left(1 + b^p \cdot y_p\right) \cdot M_3^i\right)\frac{1}{\rho}$$

$$-\left(M_5^i - 2 \cdot N_5^{im} b_m + P_5^{imn} b_m b_n + b^k b_k \cdot M_3^i\right) \cdot \frac{y^p y_p}{2\rho}$$

$$X_a^{\ im} = \int_0^{2\pi} d\varphi \cdot \frac{\beta^i}{r_s} \cdot \left(\frac{w^p}{(\beta^m w_m)^4} \cdot (1 - \beta^m b_m) + \frac{b^p}{(\beta^m w_m)^3}\right)\frac{\chi_p \chi^m}{\rho}$$

$$-\int_0^{2\pi} d\varphi \cdot \frac{\beta^i}{r_s} \cdot \left(\frac{\beta^p \cdot (1 - \beta^m b_m)^2}{(\beta^m w_m)^5} + \frac{\beta^p \cdot b^k b_k}{(\beta^m w_m)^3}\right)\frac{\chi_p \chi^m}{\rho}$$

$$X_a^{\ im} = \int_0^{2\pi} d\varphi \cdot \frac{\beta^i}{r_s} \cdot \left(\frac{w^p (\beta_p + y_p)}{(\beta^m w_m)^4} \cdot (1 - \beta^m b_m) + \frac{b^p (\beta_p + y_p)}{(\beta^m w_m)^3}\right)\frac{\chi^m}{\rho}$$

$$-\int_0^{2\pi} d\varphi \cdot \frac{\beta^i}{r_s} \cdot \left(\frac{(1 - \beta^m b_m)^2}{(\beta^m w_m)^5} + \frac{b^k b_k}{(\beta^m w_m)^3}\right)\frac{(-y_p + \chi_p)\chi^p \chi^m}{\rho}$$

$$X_a^{\ im} = \int_0^{2\pi} d\varphi \cdot \frac{\beta^i}{r_s} \cdot \left(\frac{w^p (\beta_p + y_p)}{(\beta^m w_m)^4} \cdot (1 - \beta^m b_m) + \frac{b^p (\beta_p + y_p)}{(\beta^m w_m)^3}\right)\frac{\chi^m}{\rho}$$

$$-\int_0^{2\pi} d\varphi \cdot \frac{\beta^i}{r_s} \cdot \left(\frac{(1 - \beta^m b_m)^2}{(\beta^m w_m)^5} + \frac{b^k b_k}{(\beta^m w_m)^3}\right)\frac{(-y_p + \chi_p)\chi^p \chi^m}{\rho}$$

$$X_a^{\ im} = \int_0^{2\pi} d\varphi \cdot \frac{\beta^i}{r_s} \cdot \left(\frac{w^p y_p}{(\beta^m w_m)^4} \cdot (1 - \beta^m b_m) + \frac{1}{(\beta^m w_m)^3} \cdot (1 - \beta^m b_m) + \frac{b^p y_p}{(\beta^m w_m)^3} + \frac{b^p \beta_p}{(\beta^m w_m)^3}\right)\frac{\chi^m}{\rho}$$

$$-\int_0^{2\pi} d\varphi \cdot \frac{\beta^i}{r_s} \cdot \left(\frac{(1 - \beta^m b_m)^2}{(\beta^m w_m)^5} + \frac{b^k b_k}{(\beta^m w_m)^3}\right)\frac{(-y_p + \chi_p)\chi^p \chi^m}{\rho}$$

$$X_a^{\ im} = \int_0^{2\pi} d\varphi \cdot \frac{\beta^i}{r_s} \cdot \left(\frac{w^p y_p}{(\beta^m w_m)^4} \cdot (1 - \beta^q b_q) + \frac{1 + b^p y_p}{(\beta^m w_m)^3}\right)\frac{\chi^m}{\rho}$$

$$+\int_0^{2\pi} d\varphi \cdot \frac{\beta^i}{r_s} \cdot \left(\frac{(1 - 2 \cdot \beta^q b_q + \beta^q b_q \cdot \beta^n b_n)}{(\beta^m w_m)^5} + \frac{b^k b_k}{(\beta^m w_m)^3}\right)\frac{y_p y^p \cdot \chi^m}{2\rho}$$

$$X_a^{\,im} = \int_0^{2\pi} d\varphi \cdot \frac{1}{r_s} \cdot \left( \frac{w^p y_p}{(\beta^m w_m)^4} \cdot \beta^i \cdot (\beta^m + y^m) - \frac{w^p y_p}{(\beta^m w_m)^4} \cdot \beta^i \cdot (\beta^m + y^m) \beta^q b_q \right) \cdot \frac{1}{\rho}$$

$$+ \int_0^{2\pi} d\varphi \cdot \frac{1}{r_s} \cdot \left( \frac{1 + b^p y_p}{(\beta^m w_m)^3} \cdot \beta^i \cdot (\beta^m + y^m) \right) \cdot \frac{1}{\rho}$$

$$+ \int_0^{2\pi} d\varphi \cdot \frac{1}{r_s} \cdot \frac{(1 - 2 \cdot \beta^q b_q + \beta^q b_q \cdot \beta^n b_n)}{(\beta^m w_m)^5} \cdot \beta^i \cdot (\beta^m + y^m) \cdot \frac{y_p y^p}{2\rho}$$

$$+ \int_0^{2\pi} d\varphi \cdot \frac{1}{r_s} \cdot \frac{b^k b_k}{(\beta^m w_m)^3} \cdot \beta^i \cdot (\beta^m + y^m) \cdot \frac{y_p y^p}{2\rho}$$

$$\boxed{\begin{aligned}X_a^{\,im} &= \frac{w^p y_p}{\rho} \cdot (N_4^{im} + M_4^i y^m) + \frac{w^p y_p}{\rho} \cdot (P_4^{imq} + y^m N_4^{iq}) b_q + \frac{1 + b^p y_p}{\rho} \cdot (N_3^{im} + y^m M_3^i) \\ &+ (N_5^{im} + y^m M_5^i - 2 \cdot (P_5^{imq} + y^m N_5^{iq}) b_q) \cdot \frac{y_p y^p}{2\rho} \\ &+ ((Q_5^{imqk} + y^m P_5^{iqk}) b_q b_k + (N_3^{im} + y^m M_3^i) \cdot b^k b_k) \cdot \frac{y_p y^p}{2\rho} \end{aligned}}$$

$$^f X\_w = \int_0^{2\pi} d\varphi \cdot \frac{1}{r_s} \cdot \frac{\beta^l}{(\beta^m w_m)^4} \cdot (1 - \beta^m b_m) \cdot \left( u_l + (-1 + \chi^m a_m) \cdot \frac{\chi_l}{\rho} \right)$$

$$X\_w_u = \int_0^{2\pi} d\varphi \cdot \frac{1}{r_s} \cdot \frac{\beta^l u_l}{(\beta^m w_m)^4} \cdot (1 - \beta^m b_m)$$

$$\boxed{X\_w_u = (\rho - y^l u_l) \cdot (L_4 - M_4^m b_m)}$$

$$X\_w_\rho = -\int_0^{2\pi} d\varphi \cdot \frac{1}{r_s} \cdot \frac{\beta^l}{(\beta^m w_m)^4} \cdot (1 - \beta^m b_m) \cdot \frac{\chi_l}{\rho}$$

$$\boxed{X\_w_\rho = (L_4 - M_4^m b_m) \cdot \frac{y^l y_l}{2\rho}}$$

$$X\_w_a^m = -\int_0^{2\pi} d\varphi \cdot \frac{1}{r_s} \cdot \frac{1}{(\beta^m w_m)^4} \cdot (1 - \beta^p b_p) \cdot \frac{y^l y_l \chi^m}{2\rho}$$

$$X\_w_a^m = -\int_0^{2\pi} d\varphi \cdot \frac{1}{r_s} \cdot \frac{1}{(\beta^m w_m)^4} \left( (\beta^m + y^m) - (\beta^m \beta^p + y^m \beta^p) \cdot b_p \right) \cdot \frac{y^l y_l}{2\rho}$$

$$X\_w_a^m = -\left(\left(M_4^m + y^m \cdot L_4\right) - \left(N_4^{mp} + y^m M_4^p\right) \cdot b_p\right) \cdot \frac{y^l y_l}{2\rho}$$

$$X\_b = \int_0^{2\pi} d\varphi \cdot \frac{1}{r_s} \cdot \frac{\beta^l}{(\beta^m w_m)^3} \left(u_l + \left(-1 + \chi^m a_m\right) \cdot \frac{\chi_l}{\rho}\right)$$

$$X\_b_u = \left(\rho - y^l u_l\right) \cdot L_3$$

$$X\_b_\rho = \frac{y^l y_l}{2\rho} \cdot L_3$$

$$X\_b_a^m = -\int_0^{2\pi} d\varphi \cdot \frac{1}{r_s} \cdot \frac{1}{(\beta^m w_m)^3} \left(\left(y^m + \beta^m\right) \cdot \frac{y^l y_l}{2\rho}\right)$$

$$X\_b_a^m = -\left(y^m \cdot L_3 + M_3^m\right) \cdot \frac{y^l y_l}{2\rho}$$

$$B^i = -\frac{1}{2} \cdot \int_0^{2\pi} d\varphi \cdot \frac{1}{r_s} \cdot \frac{g^{ip}}{(\beta^m w_m)^3} \psi_p$$

$$B^i = -\frac{1}{2} \cdot \int_0^{2\pi} d\varphi \cdot \frac{1}{r_s} \cdot \frac{1}{(\beta^m w_m)^3} \left(u^i + \left(-1 + \chi^m a_m\right) \cdot \frac{\chi^i}{\rho}\right)$$

$$B_u^{\,i} = -\frac{1}{2} \cdot L_3 \cdot u^i$$

$$B^i = \frac{1}{2} \cdot \int_0^{2\pi} d\varphi \cdot \frac{1}{r_s} \cdot \frac{1}{(\beta^m w_m)^3} \frac{\chi^i}{\rho}$$

$$B_\rho^{\,i} = \frac{1}{2\rho} \cdot \left(M_3^i + y^i \cdot L_3\right)$$

$$B_a^{\,i} = -\frac{1}{2} \cdot \int_0^{2\pi} d\varphi \cdot \frac{1}{r_s} \cdot \frac{1}{(\beta^m w_m)^3} \cdot \frac{\chi^m \chi^i}{\rho}$$

$$B_a^{\,i} = -\frac{1}{2} \cdot \int_0^{2\pi} d\varphi \cdot \frac{1}{r_s} \cdot \frac{1}{(\beta^m w_m)^3} \cdot \frac{\beta^m \beta^i + y^m \beta^i + \beta^m y^i + y^m y^i}{\rho}$$

$$B_a^{\,i} = -\frac{1}{2\rho} \cdot \left(N_3^{mi} + y^m M_3^i + y^i M_3^m + y^m y^i L_3\right)$$

$$A = \frac{1}{\sqrt{-e^{iklm}u_k y_l \xi_m \cdot e_{ipqr}u^p y^q \xi^r}}$$

$$e^{ikls} \cdot e_{pqrs} = -\delta_p^i \cdot \left(\delta_q^k \delta_r^l - \delta_q^l \delta_r^k\right) + \delta_p^k \cdot \left(\delta_q^i \delta_r^l - \delta_q^l \delta_r^i\right) + \delta_p^l \cdot \left(\delta_q^k \delta_r^i - \delta_q^i \delta_r^k\right)$$

$$e^{iklm}u_k y_l \xi_m \cdot e_{ipqr}u^p y^q \xi^r = e^{ikls}u_i y_k \xi_l \cdot e_{pqrs}u^p y^q \xi^r$$

$$e^{iklm}u_k y_l \xi_m \cdot e_{ipqr}u^p y^q \xi^r = u_i y_k \xi_l \cdot u^p y^q \xi^r \left(-\delta_p^i \cdot \left(\delta_q^k \delta_r^l - \delta_q^l \delta_r^k\right) + \delta_p^k \cdot \left(\delta_q^i \delta_r^l - \delta_q^l \delta_r^i\right) + \delta_p^l \cdot \left(\delta_q^k \delta_r^i - \delta_q^i \delta_r^k\right)\right)$$

$$e^{iklm}u_k y_l \xi_m \cdot e_{ipqr}u^p y^q \xi^r = u_i y_k \xi_l \cdot u^p y^q \xi^r \left(-\delta_p^i \cdot \left(\delta_q^k \delta_r^l - \delta_q^l \delta_r^k\right)\right)$$
$$+ u_i y_k \xi_l \cdot u^p y^q \xi^r \left(\delta_p^k \cdot \left(\delta_q^i \delta_r^l - \delta_q^l \delta_r^i\right)\right)$$
$$+ u_i y_k \xi_l \cdot u^p y^q \xi^r \left(\delta_p^l \cdot \left(\delta_q^k \delta_r^i - \delta_q^i \delta_r^k\right)\right)$$

$$e^{iklm}u_k y_l \xi_m \cdot e_{ipqr}u^p y^q \xi^r = -y_k \xi_l \cdot y^k \xi^l + y_k \xi_l \cdot y^l \xi^k$$
$$-U\left(u_i \xi_l \cdot y^i \xi^l - u_i \xi_l \cdot y^l \xi^i\right)$$
$$+ \xi_p u^p \left(u_i y_k \cdot y^k \xi^i - u_i y_k \cdot y^i \xi^k\right)$$

$$e^{iklm}u_k y_l \xi_m \cdot e_{ipqr}u^p y^q \xi^r = \left(U^2 - Y\right) \cdot \xi^m \xi_m + y^n \xi_m \cdot y^m \xi_n$$
$$-U\left(-u^m \xi_n \cdot y^n \xi_m\right)$$
$$+ \xi_m u^m \left(u^n Y \xi_n + U y^n \xi_n\right)$$

$$e^{iklm}u_k y_l \xi_m \cdot e_{ipqr}u^p y^q \xi^r = \left(U^2 - Y\right) \cdot \xi^m \xi_m + \left(y^n \cdot y^m + U \cdot u^m y^n + U u^m y^n + Y u^m u^n\right) \cdot \xi_m \xi_n$$

$$A = \frac{1}{\sqrt{-\left(\left(u_k y^k\right)^2 - y_k y^k\right) \cdot \xi^m \xi_m - \left(y^n \cdot y^m - 2 \cdot u_k y^k \cdot u^m y^n + Y u^m u^n\right) \cdot \xi_m \xi_n}}$$

Пусть $\xi^n = \{1, 0, 0, 0\}$

$$e^{iklm}u_k y_l \xi_m \cdot e_{ipqr}u^p y^q \xi^r = \left(U^2 - Y\right) + \left(y^0 \cdot y^0 + 2 \cdot U \cdot u^0 y^0 + Y u^0 u^0\right)$$

$$e^{iklm}u_k y_l \xi_m \cdot e_{ipqr}u^p y^q \xi^r = \left(\left(u^0 y^0 - \vec{u}\vec{y}\right)^2 + |\vec{y}|^2\right) - 2 \cdot \left(u^0 y^0 - \vec{u}\vec{y}\right) \cdot u^0 y^0 + \left(y^0 y^0 - |\vec{y}|^2\right) u^0 u^0$$

$$e^{iklm}u_k y_l \xi_m \cdot e_{ipqr}u^p y^q \xi^r = -2 \cdot \left(u^0 y^0 \cdot \vec{u}\vec{y}\right) + \left(\vec{u}\vec{y}\right)^2 + |\vec{y}|^2 - 2 \cdot \left(-\vec{u}\vec{y}\right) \cdot u^0 y^0 + \left(-|\vec{y}|^2\right) u^0 u^0$$

$$e^{iklm}u_k y_l \xi_m \cdot e_{ipqr}u^p y^q \xi^r = \left(\vec{u}\vec{y}\right)^2 - |\vec{y}|^2 \left(u^0 u^0 - 1\right)$$

$$e^{iklm}u_k y_l \xi_m \cdot e_{ipqr}u^p y^q \xi^r = \left(\vec{u}\vec{y}\right)^2 - |\vec{y}|^2 |\vec{u}|^2 < 0$$

$$\tilde{\Theta}^{nm} = -\left(\left(u_k y^k\right)^2 - y_k y^k\right) \cdot g^{nm} - \left(y^n \cdot y^m - u_k y^k \cdot \left(u^m y^n + u^n y^m\right) + y_k y^k \cdot u^m u^n\right)$$

$$\tilde{\Theta}^{nm} = -\left(\left(u_k y^k\right)^2 - y_k y^k\right) \cdot g^{nm} - \left(y^n \cdot y^m - u_k y^k \cdot \left(u^m y^n + u^n y^m\right) + y_k y^k \cdot u^m u^n\right)$$

$$\tilde{\Theta}^{ik} = A_g g^{ik} + A_\lambda \lambda^i \lambda^k + A_\mu \mu^i \mu^k + A_{\lambda\mu} \left( \lambda^i \mu^k + \lambda^k \mu^i \right)$$

$$\tilde{\Theta}^{ik} \lambda_k = -r_s^2 \lambda^i$$

$$\tilde{\Theta}^{ik} \mu_k = -r_s^2 \mu^i$$

$$A_{\lambda\mu} = 0$$

$$A_g - A_\lambda = -r_s^2$$

$$A_g - A_\mu = -r_s^2$$

Рассмотрим след:

$$\tilde{\Theta}^{nm} g_{nm} = -4 \cdot r_s^2 - \left( Y + U \cdot (-2U) + Y \right) = -4 \cdot r_s^2 - 2\left( -U^2 + Y \right) = -2 \cdot r_s^2$$

$$-2r_s^2 = 4A_g - A_\lambda - A_\mu$$

$$A_g = 0$$

$$A_\lambda = A_\mu = r_s^2$$

$$\boxed{\tilde{\Theta}^{ik} = r_s^2 \cdot \left( \lambda^i \lambda^k + \mu^i \mu^k \right)}$$

$$\tilde{\Theta}^{nm} = -\left( \left( u_k y^k \right)^2 - y_k y^k \right) \cdot g^{nm} - \left( y^n \cdot y^m - u_k y^k \cdot \left( u^m y^n + u^n y^m \right) + y_k y^k \cdot u^m u^n \right)$$

$$\tilde{\Theta}^{nm} \xi_m = -\left( \left( u_k y^k \right)^2 - y_k y^k \right) \cdot \xi^n - \left( y^n \cdot y^m - u_k y^k \cdot \left( u^m y^n + u^n y^m \right) + y_k y^k \cdot u^m u^n \right) \cdot \xi_m$$

$$\tilde{\Theta}^{nm} \xi_m = A_u u^n + A_y y^n + A_\lambda \cdot \lambda^i$$

$$A_u + A_y y^i u_i = -\left( \left( u_k y^k \right)^2 - y_k y^k \right) \cdot \xi^n u_n - \left( y^n \cdot y^m - u_k y^k \cdot \left( u^m y^n + u^n y^m \right) + y_k y^k \cdot u^m u^n \right) \cdot \xi_m u_n$$

$$A_u + A_y y^i u_i = -\left( U^2 - Y \right) \cdot \xi^n u_n - \left( -U \cdot y^m + U \cdot \left( -u^m U + y^m \right) + Y \cdot u^m \right) \cdot \xi_m$$

$$A_u + A_y y^i u_i = -\left( U^2 - Y \right) \cdot \xi^n u_n - \left( -U^2 \cdot u^m + Y \cdot u^m \right) \cdot \xi_m = 0$$

$$A_u u^n y_n + A_y y^n y_n = -\left( U^2 - Y \right) \cdot \xi^n y_n - \left( Y \cdot y^m + U \cdot \left( u^m Y - U y^m \right) - Y \cdot u^m U \right) \cdot \xi_m$$

$$A_u u^n y_n + A_y y^n y_n = -\left( U^2 - Y \right) \cdot \xi^n y_n - \left( Y + U \cdot (-U) \right) y^m \cdot \xi_m = 0$$

$$A_\lambda \lambda^n \lambda_n = -\left( \left( u_k y^k \right)^2 - y_k y^k \right) \cdot \xi^n \lambda_n$$

$$A_\lambda = r_s^2 \cdot \xi^n \lambda_n = r_s \cdot \sqrt{\tilde{\Theta}^{nm} \xi_n \xi_m}$$

$$\tilde{\Theta}^{nm} \xi_m = r_s \cdot \sqrt{\tilde{\Theta}^{nm} \xi_n \xi_m} \cdot \lambda^i$$

$$Q_5^{iklm} = 2\pi \cdot \left( \frac{1}{8} \cdot \frac{\left( \Theta^{il} \Theta^{km} + \Theta^{kl} \Theta^{im} + \Theta^{ik} \Theta^{lm} \right)}{\left( \Theta^{nm} \xi_m \xi_n \right)^{\frac{5}{2}}} + \frac{35}{8} \frac{\Theta^{iq} \xi_q \Theta^{kn} \xi_n \Theta^{lp} \xi_p \Theta^{ms} \xi_s}{\left( \Theta^{nm} \xi_m \xi_n \right)^{\frac{9}{2}}} \right)$$

$$-\frac{5\pi}{4} \frac{\left( \Theta^{im} \Theta^{kq} \Theta^{lp} + \Theta^{iq} \Theta^{km} \Theta^{lp} + \Theta^{iq} \Theta^{kp} \Theta^{lm} + \Theta^{il} \Theta^{kp} \Theta^{mq} + \Theta^{kl} \Theta^{ip} \Theta^{mq} + \Theta^{ik} \Theta^{lp} \Theta^{mq} \right) \xi_q \xi_p}{\left( \Theta^{nm} \xi_m \xi_n \right)^{\frac{7}{2}}}$$

$$P_5^{ikl} = 2\pi \cdot \left( \frac{1}{8} \cdot \frac{\left( \Theta^{il} \Upsilon^k + \Theta^{kl} \Upsilon^i + \Theta^{ik} \Upsilon^l \right)}{\left( \Theta^{nm} \xi_m \xi_n \right)^{\frac{5}{2}}} + \frac{35}{8} \frac{\Theta^{iq} \xi_q \Theta^{kn} \xi_n \Theta^{lp} \xi_p \Upsilon^s \xi_s}{\left( \Theta^{nm} \xi_m \xi_n \right)^{\frac{9}{2}}} \right)$$

$$-\frac{5\pi}{4} \frac{\left( \Upsilon^i \Theta^{kq} \Theta^{lp} + \Theta^{iq} \Upsilon^k \Theta^{lp} + \Theta^{iq} \Theta^{kp} \Upsilon^l + \Theta^{il} \Theta^{kp} \Upsilon^q + \Theta^{kl} \Theta^{ip} \Upsilon^q + \Theta^{ik} \Theta^{lp} \Upsilon^q \right) \xi_q \xi_p}{\left( \Theta^{nm} \xi_m \xi_n \right)^{\frac{7}{2}}}$$

$$N_5^{ik} = 2\pi \cdot \left( \frac{1}{8} \cdot \frac{\left( \Upsilon^i \Upsilon^k + \Upsilon^k \Upsilon^i + \Theta^{ik} \cdot r_s^2 \right)}{\left( \Theta^{nm} \xi_m \xi_n \right)^{\frac{5}{2}}} + \frac{35}{8} \frac{\Theta^{iq} \xi_q \Theta^{kn} \xi_n \Upsilon^p \xi_p \Upsilon^s \xi_s}{\left( \Theta^{nm} \xi_m \xi_n \right)^{\frac{9}{2}}} \right)$$

$$-\frac{5\pi}{4} \frac{\left( \Upsilon^i \Theta^{kq} \Upsilon^p + \Theta^{iq} \Upsilon^k \Upsilon^p + \Theta^{iq} \Theta^{kp} \cdot r_s^2 + \Upsilon^i \Theta^{kp} \Upsilon^q + \Upsilon^k \Theta^{ip} \Upsilon^q + \Theta^{ik} \Upsilon^p \Upsilon^q \right) \xi_q \xi_p}{\left( \Theta^{nm} \xi_m \xi_n \right)^{\frac{7}{2}}}$$

$$\boxed{N_5^{ik} = 2\pi \cdot \left( \frac{1}{8} \cdot \frac{\left( \Upsilon^i \Upsilon^k + \Upsilon^k \Upsilon^i + \Theta^{ik} \cdot r_s^2 \right)}{\left( \Theta^{nm} \xi_m \xi_n \right)^{\frac{5}{2}}} + \frac{35}{8} \frac{\Theta^{iq} \xi_q \Theta^{kn} \xi_n \Upsilon^p \xi_p \Upsilon^s \xi_s}{\left( \Theta^{nm} \xi_m \xi_n \right)^{\frac{9}{2}}} \right) \\ -\frac{5\pi}{4} \frac{\left( 2 \cdot \Upsilon^i \Theta^{kq} \Upsilon^p + 2 \cdot \Upsilon^k \Theta^{ip} \Upsilon^q + \Theta^{iq} \Theta^{kp} \cdot r_s^2 + \Theta^{ik} \Upsilon^p \Upsilon^q \right) \xi_q \xi_p}{\left( \Theta^{nm} \xi_m \xi_n \right)^{\frac{7}{2}}}}$$

$$M_5^i = 2\pi \cdot \left( \frac{1}{8} \cdot \frac{\left( \Upsilon^i \cdot r_s^2 + \Upsilon^i \cdot r_s^2 + \Upsilon^i \cdot r_s^2 \right)}{\left( \Theta^{nm} \xi_m \xi_n \right)^{\frac{5}{2}}} + \frac{35}{8} \frac{\Theta^{iq} \xi_q \Upsilon^n \xi_n \Upsilon^p \xi_p \Upsilon^s \xi_s}{\left( \Theta^{nm} \xi_m \xi_n \right)^{\frac{9}{2}}} \right)$$

$$-\frac{5\pi}{4} \frac{\left( 2 \cdot \Upsilon^i \Upsilon^q \Upsilon^p + 2 \cdot \Theta^{ip} \Upsilon^q \cdot r_s^2 + \Theta^{iq} \Upsilon^p \cdot r_s^2 + \Upsilon^i \Upsilon^p \Upsilon^q \right) \xi_q \xi_p}{\left( \Theta^{nm} \xi_m \xi_n \right)^{\frac{7}{2}}}$$

$$M_5^i = 2\pi \cdot \left( \frac{3}{8} \cdot \frac{\Upsilon^i \cdot r_s^2}{\left( \Theta^{nm} \xi_m \xi_n \right)^{\frac{5}{2}}} + \frac{35}{8} \frac{\Theta^{iq} \xi_q \Upsilon^n \xi_n \Upsilon^p \xi_p \Upsilon^s \xi_s}{\left( \Theta^{nm} \xi_m \xi_n \right)^{\frac{9}{2}}} \right) - \frac{15\pi}{4} \frac{\left( \Upsilon^i \Upsilon^q \Upsilon^p + \Theta^{ip} \Upsilon^q \cdot r_s^2 \right) \xi_q \xi_p}{\left( \Theta^{nm} \xi_m \xi_n \right)^{\frac{7}{2}}}$$

$$M_5^i = \frac{\pi}{4} \cdot \left( \frac{3 \cdot \Upsilon^i \cdot r_s^2}{\left( \Theta^{nm} \xi_m \xi_n \right)^{\frac{5}{2}}} - \frac{15 \cdot \left( \Upsilon^i \Upsilon^q \Upsilon^p + \Theta^{ip} \Upsilon^q \cdot r_s^2 \right) \xi_q \xi_p}{\left( \Theta^{nm} \xi_m \xi_n \right)^{\frac{7}{2}}} + \frac{35 \cdot \Theta^{iq} \xi_q \Upsilon^n \xi_n \Upsilon^p \xi_p \Upsilon^s \xi_s}{\left( \Theta^{nm} \xi_m \xi_n \right)^{\frac{9}{2}}} \right)$$

$$\boxed{M_5^i = \frac{\pi}{4} \cdot \left( \frac{3 \cdot \Upsilon^i \cdot r_s^2}{\left( \Theta^{nm} \xi_m \xi_n \right)^{\frac{5}{2}}} - \frac{15 \cdot \left( \Upsilon^i \Upsilon^q \Upsilon^p + \Theta^{ip} \Upsilon^q \cdot r_s^2 \right) \xi_q \xi_p}{\left( \Theta^{nm} \xi_m \xi_n \right)^{\frac{7}{2}}} + \frac{35 \cdot \Theta^{iq} \xi_q \left( \Upsilon^n \xi_n \right)^3}{\left( \Theta^{nm} \xi_m \xi_n \right)^{\frac{9}{2}}} \right)}$$

$$N_4^{ik} = -\frac{1}{2} \cdot \frac{2\pi \cdot \left(\Upsilon^i \Theta^{kn} + \Upsilon^k \Theta^{in} + \Theta^{ik} \Upsilon^n\right) \cdot \xi_n}{\left(\Theta^{nm} \xi_m \xi_n\right)^{\frac{5}{2}}} + \frac{5}{2} \frac{2\pi \cdot \Theta^{im} \xi_m \Theta^{kn} \xi_n \cdot \Upsilon^p \xi_p}{\left(\Theta^{nm} \xi_m \xi_n\right)^{\frac{7}{2}}}$$

$$M_4^i = -\frac{1}{2} \cdot \frac{2\pi \cdot \left(\Upsilon^i \Upsilon^n + \Theta^{in} \cdot r_s^2 + \Upsilon^i \Upsilon^n\right) \cdot \xi_n}{\left(\Theta^{nm} \xi_m \xi_n\right)^{\frac{5}{2}}} + \frac{5}{2} \frac{2\pi \cdot \Theta^{im} \xi_m \cdot \left(\Upsilon^p \xi_p\right)^2}{\left(\Theta^{nm} \xi_m \xi_n\right)^{\frac{7}{2}}}$$

$$M_4^i = -\frac{1}{2} \cdot \frac{2\pi \cdot \left(2 \cdot \Upsilon^i \Upsilon^n + \Theta^{in} \cdot r_s^2\right) \cdot \xi_n}{\left(\Theta^{nm} \xi_m \xi_n\right)^{\frac{5}{2}}} + \frac{5}{2} \frac{2\pi \cdot \Theta^{im} \xi_m \cdot \left(\Upsilon^p \xi_p\right)^2}{\left(\Theta^{nm} \xi_m \xi_n\right)^{\frac{7}{2}}}$$

$$L_4 = -\frac{1}{2} \cdot \frac{2\pi \cdot \left(2 \cdot \Upsilon^n \cdot r_s^2 + \Upsilon^n \cdot r_s^2\right) \cdot \xi_n}{\left(\Theta^{nm} \xi_m \xi_n\right)^{\frac{5}{2}}} + \frac{5}{2} \frac{2\pi \cdot \Upsilon^m \xi_m \cdot \left(\Upsilon^p \xi_p\right)^2}{\left(\Theta^{nm} \xi_m \xi_n\right)^{\frac{7}{2}}}$$

$$L_4 = -3\pi \cdot \frac{\Upsilon^n \xi_n \cdot r_s^2}{\left(\Theta^{nm} \xi_m \xi_n\right)^{\frac{5}{2}}} + 5\pi \cdot \frac{\left(\Upsilon^p \xi_p\right)^3}{\left(\Theta^{nm} \xi_m \xi_n\right)^{\frac{7}{2}}}$$

$$M_3^i = \frac{3}{2} \cdot \frac{2\pi \cdot \Theta^{im} \xi_m \cdot \Upsilon^n \xi_n}{\left(\Theta^{nm} \xi_m \xi_n\right)^{\frac{5}{2}}} - \frac{1}{2} \frac{2\pi \cdot \Upsilon^i}{\left(\Theta^{nm} \xi_m \xi_n\right)^{\frac{3}{2}}}$$

$$L_3 = \frac{3}{2} \cdot \frac{2\pi \cdot \left(\Upsilon^m \xi_m\right)^2}{\left(\Theta^{nm} \xi_m \xi_n\right)^{\frac{5}{2}}} - \frac{1}{2} \frac{2\pi \cdot r_s^2}{\left(\Theta^{nm} \xi_m \xi_n\right)^{\frac{3}{2}}}$$

$$L_3 = 3\pi \cdot \frac{\left(\Upsilon^m \xi_m\right)^2}{\left(\Theta^{nm} \xi_m \xi_n\right)^{\frac{5}{2}}} - \pi \cdot \frac{r_s^2}{\left(\Theta^{nm} \xi_m \xi_n\right)^{\frac{3}{2}}}$$

[k] Кривая рис. 10.19. построена на основании рассчитанных значений, собранных в следующей таблице:

| u | 0.01 | 0.02 | 0.03 | 0.04 | 0.05 | 0.06 | 0.07 | 0.10 | 0.12 |
|---|------|------|------|------|------|------|------|------|------|
| d | 800 | 310 | 177 | 120 | 88 | 69 | 57 | 35 | 28 |
| u | 0.14 | 0.16 | 0.18 | 0.2 | 0.3 | 0.4 | 0.5 | 0.6 | 0.7 |
| d | 23.5 | 19.5 | 17.25 | 15.25 | 9.85 | 7.25 | 6.15 | 5.34 | 4.80 |
| u | 0.8 | 0.9 | 1.0 | 1.1 | 1.2 | 1.3 | 1.4 | 1.5 | 1.6 |
| d | 4.42 | 4.15 | 3.94 | 3.78 | 3.66 | 3.58 | 3.46 | 3.54 | 3.55 |
| u | 1.80 | 2.00 | 2.20 | 2.40 | 2.60 | 2.80 | 3.00 | 3.20 | 3.40 |
| d | 3.61 | 3.70 | 3.84 | 4.01 | 4.19 | 4.37 | 4.56 | 4.78 | 5.00 |
| u | 3.60 | 3.80 | 4.00 | 4.25 | 4.50 | 4.75 | 5.00 | 5.25 | 5.50 |
| d | 5.24 | 5.47 | 5.70 | 6.01 | 6.30 | 6.62 | 6.92 | 7.23 | 7.55 |
| u | 5.75 | 6.00 | 6.50 | 7.00 | 7.50 | 8.00 | 8.50 | 9.00 | 9.50 |
| d | 7.87 | 8.17 | 8.82 | 9.47 | 10.12 | 10.78 | 11.41 | 12.07 | 12.71 |
| u | 11.0 | | | | | | | | |
| d | 14.70 | | | | | | | | |

$$\Theta^{nm} u_n = \left(-(\rho+U)\cdot U + \tfrac{1}{2}Y\right)\cdot\left((\rho+U)\cdot y^m + \tfrac{1}{2}Y u^m\right)$$
$$\quad - Y\cdot\left(\tfrac{1}{4}Y + \rho\cdot U + \rho^2\right)\cdot u^m$$

$$\Theta^{nm} u_n = \left(-(\rho+U)\cdot U + \tfrac{1}{2}Y\right)\cdot(\rho+U)\cdot y^m$$
$$\quad + \left(\left(-(\rho+U)\cdot U + \tfrac{1}{2}Y\right)\cdot\tfrac{1}{2}Y - Y\cdot\left(\tfrac{1}{4}Y + \rho\cdot U + \rho^2\right)\right)\cdot u^m$$

$$\Theta^{nm} u_n = \left(-(\rho+U)\cdot U + \tfrac{1}{2}Y\right)\cdot(\rho+U)\cdot y^m$$
$$\quad - \tfrac{1}{2}(U + 2\rho)\cdot(\rho+U)\cdot Y \cdot u^m$$